\documentclass[10pt,journal,double]{IEEEtran}			
\usepackage{graphicx}
\usepackage{epsfig}
\usepackage{lipsum}
\usepackage{graphics}
\usepackage{cite}
\usepackage{amsmath}
\usepackage{amssymb}
\usepackage{amsfonts}
\usepackage{tikz}
\usepackage{pgfplots}
\usepackage{setspace}
\usepackage{color}
	\usetikzlibrary{shapes,arrows,fit,calc,positioning}
	\usetikzlibrary{decorations.shapes}
	\usepackage{multicol}
	\definecolor{colorgreen}{rgb}{0.2,0.5,0.3}
	\definecolor{colorblue}{rgb}{0.2,0.1,0.7}
	\usetikzlibrary{decorations.pathmorphing}
	\usepackage{bigints}
\usepackage{setspace}
\usepackage{multirow}
\usepackage{tablefootnote}
\usepackage[skip=2pt,font=small]{caption}
\usepackage{subcaption}
\usepackage{bm, amsmath, amssymb}
\usepackage{amsfonts}
\usepackage {amssymb,amsmath}
\usepackage{cite}
\usepackage{epstopdf}
\usepackage{booktabs}
\usepackage{float}
\usepackage{suffix}
\DeclareGraphicsExtensions{.pdf,.png,.jpg}

\DeclareMathOperator{\E}{\mathbb{E}}
\DeclareMathOperator{\D}{\textit{d}}
\DeclareMathOperator{\SSS}{\mathcal{S}}
\DeclareMathOperator{\SR}{\mathcal{R}}
\DeclareMathOperator{\SD}{\mathcal{D}}
\DeclareMathOperator{\PD}{\mathcal{P}}
\usepackage{cite}

\usepackage{tikz}
\usepackage{textcomp}
\usepackage{lipsum}
\DeclareMathOperator\erfc{erfc}
\newcommand{\A}{\mathcal{A}}
\newcommand{\DD}{\mathcal{D}}

\newcommand\pipr{\mathrel{\stackrel{\makebox[0pt]{\mbox{\small\ $\hspace{-0.0 in}\gamma_{max}\rightarrow \text{large}$}}}{\hspace{0.4 in} \approx \hspace{0.3 in} }}}

\newcommand\copyrighttext{%
	\footnotesize This work has been submitted to the IEEE for possible publication. Copyright may be transferred without prior notice, after which this version may no longer be accessible.}
\newcommand\copyrightnotice{%
	\begin{tikzpicture}[remember picture,overlay]
	\node[anchor=south,yshift=760pt] at (current page.south) {{\parbox{\dimexpr\textwidth-\fboxsep-\fboxrule\relax}{\copyrighttext}}};
	\end{tikzpicture}%
}

\begin{document}

\title{Performance of Adaptive Link Selection with Buffer-Aided Relays in Underlay Cognitive Networks}
\author{Bhupendra Kumar,~\IEEEmembership{Student Member,~IEEE}
	and Shankar Prakriya,~\IEEEmembership{Senior Member,~IEEE}
	\thanks{Bhupendra Kumar is with Bharti School of Telecommunication Technology and Management, Indian Institute of  Technology Delhi, Hauz Khas, New Delhi 110 016, India     (e-mail: bkumar0810@gmail.com).}
	\thanks{Shankar Prakriya is with the Department of Electrical Engineering and Bharti School of Telecommunication Technology and Management, Indian Institute of Technology Delhi, Hauz Khas, New Delhi 110 016, India (e-mails: shankar@ee.iitd.ac.in).}
	\thanks{A conference version with initial results \cite{Bhupendra2016} related to average rate was submitted to IEEE Sarnoff 2016.}    
	\thanks{This work was supported by Information Technology Research Academy (a unit of Media Labs Asia) through sponsored project ITRA/15(63)/Mobile/MBSSCRN/01.}
}
\maketitle
\copyrightnotice
\begin{abstract}
\textbf{
In this paper, we investigate the performance of a  three-node dual-hop cognitive radio network (CRN) with a half-duplex (HD) decode-and-forward (DF) buffer-aided relay. We derive expressions for the average rate and symbol error rate (SER) performance of an adaptive link selection based channel-aware buffer-aided relay (CABR) scheme that imposes peak-power and peak-interference constraints on the secondary nodes, and compare them with those of conventional non-buffer-aided relay (CNBR) and conventional buffer-aided relay (CBR) schemes for a delay-tolerant system.  For finite-delay systems, we analyze the performance of a modified threshold-based scheme for fixed-rate transmission, and demonstrate that use of a last-in-first-out buffer is advantageous in some situations. We bring out the trade-offs between delay, throughput and SER. Computer simulation results are presented to demonstrate accuracy of the derived expressions.\\
Keywords- Underlay Cognitive Network, Adaptive Link Selection, Buffer-Aided Relay.
}
\end{abstract}
\vspace{-0.5cm}
\section{Introduction}
\par Next Generation wireless networks are expected to support a wide variety of data services with different traffic characteristics and quality-of-service (QoS) requirements.  Besides, the surge in the number of data services has already led to spectrum scarcity. It is well known that the former problem can be alleviated by use of relays\cite{Laneman2004} while the latter can be alleviated by use of cognitive radio networks (CRN)s\cite{Haykin2005}.  In particular, underlay cognitive radio technology, in which the secondary (unlicensed) users utilize the same frequency band as the primary (licensed) network, but with transmit powers carefully controlled to limit interference caused to the primary receiver below an interference temperature limit (ITL), has shown great promise   \cite{Goldsmith2009}.  The interference constraint imposed by the primary receiver limits the secondary transmitter power, making the use of relays advantageous, Clearly, analysis of the performance of CRNs with relays is well motivated.
\par In non-cognitive cooperative networks,  protocols and techniques that improve the performance of relay networks have received attention.  With buffer-aided relays, channel-aware scheduling  was shown to improve the QoS over conventional half-duplex (HD) decode-and-forward (DF) relaying in two-hop networks  \cite{Madsen2005}\cite{Bing2008}\cite{Zlatanov2014}-\cite{Nomikos2015}. In such networks  adaptive link selection was shown to lead to significant performance improvement \cite {Zlatanov2013_1, Zlatanov2013_2}.  With exact channel state information (CSI) and infinite-sized buffer, an average rate of half of the maximum of the capacities of two links can be achieved using adaptive rate transmission \cite {Zlatanov2013_1}, and a diversity order of two can be achieved with fixed-rate transmission \cite {Zlatanov2013_2},\cite{Islam2013_2}. With outdated CSI estimates, it was shown  in \cite{Islam2015}  that diversity of one is still achievable with positive coding gain over conventional relays. Analysis of delay performance was also taken up in these works. Selection of one of several buffer-aided relays has been shown to improve performance\cite{Ikhlef2012_MMRS,Krikidis2012_2,Teh2015}.
\par Since the performance of CRN is degraded by the transmit power constraints,  development of techniques to improve the performance of relay-aided networks is of great interest. In \cite{Darabi2014}, an interference cancellation based scheme is proposed where the primary and the secondary sources pick one buffer-aided relay {\em each} for two-hop transmission. Power allocation issues are also addressed.  In \cite{Darabi2015}, a throughput-optimal adaptive link selection policy is proposed for a secondary two-hop underlay network in which the secondary node only transmits if the average or instantaneous interference power at the primary receiver is below a threshold.  In \cite{Shaqfeh2015}, an {\em overlay} secondary source (using knowledge of the primary message) maximizes its rate in a link without relays, while assisting the primary to attain its target rate.  For underlay two-hop buffer-aided relay networks with finite-sized buffers,  a sub-optimal relay selection scheme is proposed and its outage performance is analyzed \cite{Chen2014} assuming {\em only} the peak interference constraint (ignoring the peak-power constraint). With finite-sized buffers,  the outage probability of the underlay network is analyzed in \cite{Tang2016} assuming outdated CSI. It is emphasized that they simply select a link from source to relay and from relay to destination {\em without taking into consideration the channel to primary destination in their link selection procedure}.  Neither \cite{Tang2016} nor \cite{Chen2014}  derive insights into the delay performance of the system, or derive expressions for ergodic rate or symbol error rate (SER).
\par In this paper, we analyze the average rate and SER performance of a three-node dual-hop underlay CRN. We use the adaptive link selection scheme  proposed in \cite{Zlatanov2013_1} for use in the non-cognitive context, and derive expressions for both the average rate and SER in delay tolerant links. We also discuss the performance of delay-limited systems, and  assuming fixed-rate transmission, present delay analysis for the threshold based scheme presented in \cite{Islam2015}. We also introduce the concept of reversibility in the buffer due to which it is also possible to stabilize the buffer and transmit with finite system delay even if the arrival rate is more than departure rate. We also discuss trade-offs between throughput, SER and delay.
\par The rest of the paper is organized as follows. Section~\ref{sec:SysMod} discusses the buffer-aided cognitive relaying model. The relaying schemes that will be considered are discussed in Section~\ref{sec:RelSch}. Performance analysis is presented in Section~\ref{sec:PerAna}, which includes the derivation of the complementary cumulative distribution function (CCDF) of the link SNRs, as well as the expressions for the link selection probability (LSP), average rate, and SER of various schemes for a delay-tolerant system. In section~\ref{sec:DelAna}, we present the delay analysis, and relate the average rate and SER to the delay performance.  Simulation results are presented in Section~\ref{sec:SimRes} to validate the derived expressions, and to obtain insights into performance. Conclusions are presented in Section~\ref{sec:ConClu}.\\
{\em Notations}:  $\mathcal{CN}(0,\Omega)$ denotes circularly symmetric complex Gaussian distribution with zero-mean and $\Omega$ variance. $F^c_{X,Y}(\,)$, $f_{X,Y}(\,)$ and $\mathbf{\E}_{X,Y}[\,]$  denote the joint CCDF, the joint probability distribution function (PDF), and expectation w.r.t.  random variables $X$ and $Y$ respectively. $\Pr\{{\cal A}\}$ denotes the probability of an event ${\cal A}$. $\delta()$ and $(k)!!$ denote the delta function and double factorial respectively. $\erfc()$ denotes the complementary error function.
$\Gamma(a,x)$ and $E_{n}(x)$ denote the upper incomplete gamma function and  the generalized exponential integral respectively.  $Li_{2}(x) $ denotes the Euler-Dilogarithm function and is given by {\small $Li_{2}(x) =-\int\limits_{0}^{x}\frac{\ln(1-t)}{t}{\D}t=\int\limits_{0}^{\infty}\frac{t\,xe^{-t}}{1-xe^{-t}}\,{\D}t$}. 
\section{System Model}\label{sec:SysMod}
\par We consider a two-hop underlay CRN depicted in Fig.~\ref{fig:sysmod1}. The primary network consists of the primary source (not depicted in the figure) and the primary destination ($\PD$).  The secondary or unlicensed network consists of the secondary source ($\SSS$), the secondary destination ($\SD$), and a HD DF secondary relay ($\SR$) equipped with a buffer. All nodes are equipped with a single antenna. The $\SSS-\SD$ direct link is heavily shadowed. As in most works on underlay cognitive radio \cite{Lee2011}\cite{Tourki2012}, we ignore the primary signal at the secondary nodes. This assumption has been justified on information-theoretic grounds \cite{Kashyap2013}.
\vspace{-0.2cm}
\subsection{Channel Model}
\par The channel coefficients of secondary and interference links are denoted by $h_{i}\sim {\cal CN}(0,\Omega_{h_{i}})$ and $g_{i}\sim {\cal CN}(0,\Omega_{g_{i}})$ respectively, where $i\in\{s,r\}$ ($s$ and $r$ denote  $\SSS$ and $\SR$).  We assume a path-loss Rayleigh fading channel model so that  $\Omega_{h_{i}}=d_{ij}^{-\alpha}$ (with $j\in\{s,r\}$ and $j\neq i$), and $\Omega_{g_{i}}=d_{ip}^{-\alpha}$ respectively, where $\alpha$ is the path-loss exponent, $d_{ij}$ denotes the distance between secondary nodes, and $d_{sp}$ and $d_{rp}$ denote the distance of $\SSS$ and $\SR$ to the primary receiver. Zero-mean additive white Gaussian noise of  $\mathcal{N}_{o}$ variance is assumed at all terminals.  All channels undergo mutually independent, ergodic, quasi-static fading that remains fixed for one slot duration but varies independently between consecutive slots. 
\par Underlay cognitive radio nodes \cite{Hong2011} use a peak interference power (PIP) constraint - $\SSS$ and $\SR$ restrict their instantaneous transmit power so as to limit the peak interference to $\PD$ below a certain ITL ${\mathcal I}_{p}$.  We also assume a peak transmit power (PTP) constraint at $\SSS$ and $\SR$ that limits the transmit power to $P_{max}$. Define $\gamma_{p}=\mathcal{I}_{p}/\mathcal{N}_{o}$, and  system SNR as $\gamma_{max}= P_{max}/\mathcal{N}_{o}$.
 The instantaneous SNRs $\gamma_{i}$ with combined PTP and PIP constraints are given by \cite{Hong2011}: 
\begin{figure}[!h]
	\begin{center}
		\includegraphics[scale=0.6]{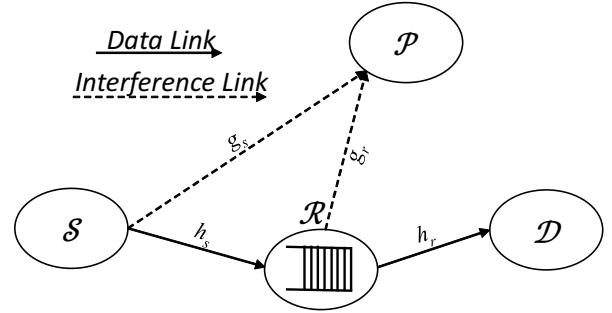}
		\caption{Three Node CRN with buffer-aided relay}
		\label{fig:sysmod1}
	\end{center}
	\vspace{-.25cm}
\end{figure}
{\small
\begin{eqnarray}\label{eqn:link_snrs}
	\gamma_{i} &=& \min\left(\gamma_{max},\dfrac{\gamma_p}{|g_{i}|^2}\right)|h_{i}|^2,\quad  i\in\{s,r\}.
\end{eqnarray}
}
The  instantaneous capacity $C_{i}$ is  $C_{i}=\log_{2}(1+\gamma_{i})$. In the very low (high) SNR region, where $\gamma_{max}$ is very small (large), the $\gamma_{max}<<\gamma_p/|g_{i}|^{2}$ ($\gamma_{max}>>\gamma_p/|g_{i}|^{2}$) event is encountered with high probability. We refer to this as the unlicensed transmitter is in PTP (PIP) or non-cognitive (fully-cognitive) regime\footnote{The terms non-cognitive and fully-cognitive are used to refer respectively to the situations when unlicensed transmitter (either $\SSS$ or $\SR$) experiences negligible and severe interference respectively.}. The link SNRs in PTP scenario are limited by peak power and modelled as  exponential random variables, whereas in PIP regime, they are limited by peak interference and modelled as the ratio of the exponential random variables.  Denote by $p_{i}$, $i\in \{s,r\}$, the probability that interference at $\PD$ is higher than ${\cal I}_{p}$ with secondary transmit power $P_{max}$:
{\small
\begin{eqnarray}
p_{i}&=&\Pr\left\{\gamma_{max}>\dfrac{\gamma_{p}}{|g_{i}|^2}\right\}= e^{-{\mu_{i}}/{\lambda_{i}}}, \quad  i\in\{s,r\},
\end{eqnarray}
}
where $\lambda_{i}=\gamma_{max}\,\Omega_{h_{i}}$ and $\mu_{i}={\gamma_{p}\Omega_{h_{i}}}/{\Omega_{g_{i}}}$ are the average values of instantaneous SNR $\gamma_{i}$ of the link $i\in\{s,r\}$ when the corresponding unlicensed transmitter (either $\SSS$ or $\SR$) is transmitting with power $P_{max}$  and ${\mathcal{I}_p}/{\Omega_{g_{i}}}$ respectively. It is clear that $\lambda_{i}$ is the average SNR in the PTP regime (non-cognitive case) whereas $\mu_{i}$ is a virtual parameter related to the PIP regime. Note that the ratio ${\mu_{i}}/{\lambda_{i}}$ tends to $\infty$ ($0$) when the node $i\in \{s,r\}$ is in the  PTP (PIP) regime. Expressions have been derived for CCDF of the link SNRs $\gamma_{i}$ of (\ref{eqn:link_snrs}) in literature \cite{Lee2011}\cite{Tourki2012}. Here for ease of exposition, we will find it convenient to write the CCDF and PDF of the link SNRs $\gamma_{i}$  in terms of $p_{i}$ as follows:
{\small
\begin{eqnarray}\label{eqn:F_Gi_s}
\begin{array}{lll}
F_{\gamma_{i}}^{c}(s)\hspace{-0.2cm} &=&\hspace{-0.2cm} e^{-{s}/\lambda_{i}}\left[1-p_{i}\left(1- \dfrac{\mu_{i}}{s+\mu_{i}}\right) \right],\\
f_{\gamma_{i}}(s) \hspace{-0.2cm}&=&\hspace{-0.2cm} \dfrac{1}{\lambda_{i}}e^{-{s}/\lambda_{i}}\left[1-p_{i}\left(1- \dfrac{\mu_{i}}{s+\mu_{i}}- \dfrac{\lambda_{i}\mu_{i}}{(s+\mu_{i})^2}\right) \right].
\end{array}{}
\end{eqnarray}} 
In the above, use of $p_{i}=0$ ($p_{i}=1$) gives the expressions valid for the PTP (PIP) regime. All the expressions, including those of the ergodic rate, SER, and delay presented in this paper are expressed in terms of $p_i$. Expressions for PTP and PIP regime can thus be obtained simply by substituting $p_i=0$ and $p_i=1$ for $i\in\{s,r\}$.
\vspace{-0.7cm}
\subsection{Relay Schemes}\label{sec:RelSch}
\par In all the relay schemes, we assume that signalling takes place in time-slots of fixed duration. We describe the CABR scheme, and then outline some conventional relay schemes for comparison of performance in the CRN context.
\subsubsection{\textbf{Channel-Aware-Buffer-Aided Relay (CABR) Scheme}}
\par For the non-cognitive scenario, Zlatanov et al. proposed a link-selection protocol applicable to two-hop signalling which is optimum in the average rate sense\cite{Zlatanov2013_1}. According to it, in every time-slot, the CABR scheme assuming an infinite-sized buffer selects either the $\SSS-\SR$ or $\SR-\SD$ link, whichever has higher capacity, while ensuring buffer stability. In this paper, we consider finite-sized  buffers in addition to infinite-sized buffers. We briefly describe the buffer dynamics in what follows. Consider a first-in first-out (FIFO) buffer of finite size $L$ bits with $B(n)$ bits\footnote{The length of buffer and amount of stored information, which is given in bits, is actually normalized w.r.t. symbols i.e. bits/symbol.} in the $n^{th}$ signalling interval\footnote{Here, we explicitly show the time-dependence of variables for clarity.}.  When $d(n)=0$ so that $i=s$, the $\SSS-\SR$ link is selected, and $C_s(n)$ bits are added to the buffer (unless this would exceed the buffer size). Similarly, when $d(n)=1$ so that  $i=r$,  the $\SR-\SD$ link is selected, and $C_r(n)$ bits are removed from the buffer (unless the buffer size becomes zero).  Denote by {\small $\overline{\mathcal{R}}_{fifo}^{CABR}$}
   the achievable rate of the FIFO buffer for the CABR scheme. Concisely, we can write:
  {\small
  	\begin{eqnarray}\label{eqn:rate_expr_fl_fifo}
  		\begin{array}{lll}
  			\hspace{-0.15cm}\overline{\mathcal{R}}_{fifo}^{CABR}=\min\Big( \mathbf{\E}_{\gamma_{s},\gamma_{r}}[(1-d(n))\min(C_{s}(n),L-B(n))]\\ \hspace{3.25cm},\hspace{-0.0cm}\mathbf{\E}_{\gamma_{s},\gamma_{r}}[d(n) \min(C_{r}(n),B(n))]\Big).
  		\end{array}
  	\end{eqnarray}
  } 
Two observations can readily be made: a)  analysis of performance of each hop suffices, and b)   the achievable rate is maximized when inflow and outflow rates are equal implying that the argument of the $\min()$ function are equal.  Link-selection presented in the non-cognitive context \cite{Zlatanov2013_1} is based on the ratio of instantaneous capacities $C_{s}$ and $C_{r}$ of $\SSS-\SR$ and $\SR-\SD$ links. Here $d(n)=0$ when $\gamma_r\leq\rho\gamma_s$ and $d(n)=1$ otherwise (the buffer size is assumed to be infinity), where $\rho$ is a parameter. Here, they choose $\rho$ so as to maximize {\small $\overline{\mathcal{R}}_{fifo}^{CABR}$} by making the inflow and outflow rates equal.
  \par For choice of $d(n)$, there are two simple options a) $d(n)$ does not depend on buffer state $B(n)$ or, b) $d(n)$ is modified whenever $B(n)=0 \text{ or }L$ (buffer is empty or full respectively). The former leads to buffer underflow ($\SR-\SD$ link is selected when $B(n)=0$) or overflow ($\SSS-\SR$ link is selected when $B(n)=L$).
  For finite-sized buffers, when $B(n)=L$  ($B(n)=0$), $d(n)$ can be forced to $1$ ($0$). However, this results in poor SER performance as can be expected intuitively. An alternative is to decide probabilistically when the buffer is empty ($B(n)=0$) or full ($B(n)=L$). We refer to this as the modified threshold based transmission protocol, and consider its use with buffers of finite size (we also analyze its performance) later in this paper. When $B(n)=0$ ($B(n)=L$) or the buffer is empty (full), we choose the $\SSS-\SR$ ($\SR - \SD$) link when $\gamma_r \leq \rho_c\gamma_s$ for appropriately chosen $\rho_c$ ($\rho_d$).  Here, we use the following link selection mechanism when $B(n)\neq 0,L$:
  {\small 
  \begin{eqnarray}\label{eqn:linkcrit}
  \hspace{0.5cm}d(n)\hspace{-0.cm}= \hspace{-0.cm}\left\{ 
  \begin{array}{ll}
  \hspace{-0.15cm}0 &\hspace{-0.15cm}  \gamma_{r}\leq\,\rho\gamma_{s}, B(n)\neq 0,L\\
  \hspace{-0.15cm}1 &\hspace{-0.15cm}  \gamma_{r}>\,\rho\gamma_{s},B(n)\neq 0,L
  \end{array} \right.\hspace{-0.2cm}.
  \end{eqnarray}
}
When $B(n)=0$ or $B(n)=L$, we use:
\vspace{-0.05cm}
{\small
\begin{eqnarray}\label{eqn:linkcrit2}   
  \hspace{0.1cm}d(n)\hspace{-0.cm}= \hspace{-0.cm}\left\{ 
  \begin{array}{l l}
  \hspace{-0.15cm}0 &\hspace{-0.15cm}  {\gamma_{r}}\leq\,\rho_{c}{\gamma_{s}},\, B(n)=0,/ {\gamma_{r}}\leq\,\rho_{d}{\gamma_{s}},\, B(n)=L\\
  \hspace{-0.15cm}1 &\hspace{-0.15cm} {\gamma_{r}}>\,\rho_{d}{\gamma_{s}},\, B(n)=L, / {\gamma_{r}}>\,\rho_{c}{\gamma_{s}},\, B(n)=0
  \end{array} \right.\hspace{-0.2cm}.
 \end{eqnarray}
}
  where $\rho$, $\rho_{c}$, and $\rho_{d}$ are positive parameters. It is obvious that when $d(n)$ does not depend on the buffer state, then $\rho=\rho_{c}=\rho_{d}$. We note that since the channels vary in every signalling interval, so do $\gamma_{r},\,\gamma_{s},\,C_{s},\,C_{r}$ and hence $d$. However for conciseness, we do not show the dependence on the time index $n$ unless explicitly required. The LSP of the $\SSS-\SR$ ( $\SR-\SD$) link, when buffer is neither empty nor full is given by $q_{s}=\Pr\{\gamma_{r}\leq \rho \gamma_{s} \}$ ($q_{r}=\Pr\{\gamma_{r}> \rho \gamma_{s} \}$). The LSPs of the $\SSS-\SR$ ( $\SR-\SD$) link when buffer is empty (full), is given by   $q_{c}=\Pr\{\gamma_{r}\leq \rho_{c} \gamma_{s} \}$ ($q_{d}=\Pr\{\gamma_{r}> \rho_{d} \gamma_{s} \}$).
  \par It is important to note that finite-size buffers are always stable. For a stable infinite-sized buffer operating at optimum rate,  the choices of $\rho_{c}$ and $\rho_{d}$ (hence $q_{c}$ and $q_{d}$) are irrelevant. Hence, for an infinite-sized buffer, $\rho$ is carefully optimized to ensure stability of the FIFO buffer i.e.{\small $\mathbf{\E}_{\gamma_{s},\gamma_{r}}[(1-d)C_{s}] \leq\mathbf{\E}_{\gamma_{s},\gamma_{r}}[d C_{r}]$}, where the achievable rate for CABR is the minimum of the inflow and outflow rate in stabilized buffer condition, i.e.
     {\small
		\begin{eqnarray}\label{eqn:rate_expr_in_len}
    	\begin{array}{lll}
    	\hspace{0.5cm}\overline{\mathcal{R}}_{fifo}^{CABR}\hspace{-0.2cm}=\min(\mathbf{\E}_{\gamma_{s},\gamma_{r}}[(1-d)C_{s}] ,\mathbf{\E}_{\gamma_{s},\gamma_{r}}[d C_{r}]).
    	\end{array}
    	\end{eqnarray}
    } 
   The achievable rate is optimal when $\rho=\rho_{opt}$ is chosen to make inflow and outflow rates equal ({\small $\mathbf{\E}_{\gamma_{s},\gamma_{r}}[(1-d)C_{s}] = \mathbf{\E}_{\gamma_{s},\gamma_{r}}[d C_{r}]$} ). Note that when {\small $\mathbf{\E}_{\gamma_{s},\gamma_{r}}[(1-d)C_{s}] < \mathbf{\E}_{\gamma_{s},\gamma_{r}}[d C_{r}]$}, or $\rho<\rho_{opt}$, the buffer underflows so that   $\Pr\{B(n)=0,d(n)=1\}$ is finite (except when $q_c=1$), which clearly decreases rate. This amounts to starving the buffer since $q_{r}>q_{s}$.  On the other hand, when {\small $\mathbf{\E}_{\gamma_{s},\gamma_{r}}[(1-d)C_{s}] > \mathbf{\E}_{\gamma_{s},\gamma_{r}}[d C_{r}]$} , $B(n)$ increases without bound and the buffer becomes unstable. In this situation, the buffer can still be stabilized by using a  Last-in First-Out (LIFO) mechanism.
   \par The pointer is placed at the end of buffer. The direction of filling the packets and the link selection method (i.e. $d=0$ implies $\SSS$ transmits), are both the same as with the FIFO buffer\footnote{Please note that conventional LIFO buffer assumes stack mechanism in which the direction of filling packets is opposite to that of FIFO.} to operate the buffer. The rate can be obtained by substituting $B'(n)=L-B(n)$ in (\ref{eqn:rate_expr_fl_fifo}), to get:
   {\small
   	\begin{eqnarray}\label{eqn:rate_expr_fl_lifo}
   	\begin{array}{lll}
   	\hspace{-0.15cm}\overline{\mathcal{R}}_{lifo}^{CABR}=\min\Big( \mathbf{\E}_{\gamma_{s},\gamma_{r}}[(1-d(n))\min(C_{s}(n),B'(n))]\\ \hspace{2.5cm},\hspace{-0.0cm}\mathbf{\E}_{\gamma_{s},\gamma_{r}}[d(n) \min(C_{r}(n),L-B'(n))]\Big).
   	\end{array}
   	\end{eqnarray}
   } 
   It is evident that when the $\SSS-\SR$ link is chosen more often, there is buffer underflow instead of an overflow. It is then obvious that  infinite-sized LIFO buffer is stable when arrival rate is more than departure rate.
   \par The following important observations are possible by comparing (\ref{eqn:rate_expr_fl_fifo}) with (\ref{eqn:rate_expr_fl_lifo}). If we exchange  $d\leftrightarrow 1-d$ and $C_{s}\leftrightarrow C_{r}$, the LIFO and FIFO models presented in (\ref{eqn:rate_expr_fl_fifo}) and (\ref{eqn:rate_expr_fl_lifo}) can be seen to be duals of each other. It is clear from (\ref{eqn:linkcrit}) and (\ref{eqn:linkcrit2}) that this exchange can be achieved as follows:
   {\small
   \begin{eqnarray*}\label{eqn:rever}
   \begin{array}{lll}
   \rho\leftrightarrow 1/\rho,&\hspace{.5cm} \rho_{c}\leftrightarrow\rho_{d},&\hspace{.9cm}\gamma_{s}\leftrightarrow\gamma_{r}.
   \end{array}
   \end{eqnarray*}
	}
  The following change of parameters are now implied:
  {\small
   \begin{eqnarray}\label{eqn:reversibility}
   \begin{array}{llll}
   \rho\rightarrow1/\rho,&\hspace{-1.5cm} \rho_{c}\leftrightarrow\rho_{d},&\hspace{-1.5cm} \lambda_{s}\leftrightarrow\lambda_{r}, &\hspace{-1.5cm} \mu_{s}\leftrightarrow\mu_{r},\\
   \text{Consequently}\,\, p_{s}\leftrightarrow p_{r},& p_{c}\leftrightarrow p_{d},& \lambda_{\rho}/{\rho}\leftrightarrow\lambda_{\rho},
   \end{array}
   \end{eqnarray}
}
   where $1/\lambda_{\rho}$ is defined as $1/\lambda_{\rho}=1/\lambda_{r}+1/
   (\rho\lambda_{s})$. Its physical significance will be brought out later. We refer to (\ref{eqn:reversibility}) as the reversibility equation. It will become apparent later that the reversibility equation can be used to obtain performance of one link (either $\SSS-\SR$ or $\SR-\SD$) given the other\footnote{In the rest of the paper, we assume FIFO relays unless mentioned otherwise. The subscript FIFO is therefore no longer user with rate and other variables.}.
\par It is assumed that the buffer of the source $\SSS$ is backlogged and always has data to transmit. We analyze the average rate and SER  performance of the CABR scheme. For analysing the average rate, it is assumed that both the source $\SSS$ and relay $\SR$ adapt their transmission rate using capacity achieving codes, and are hence capable of exploiting  CSI of corresponding forward and interference links ($|h_{i}|^{2}$ and $|g_{i}|^{2}$). For analysing the SER, it is assumed that both the source $\SSS$ and relay $\SR$ transmit at a predefined fixed rates. Without loss of generality, we assume that the two rates are equal to $R$. It is evident from (\ref{eqn:rate_expr_fl_fifo}) that an infinite-sized buffer is stabilized by selecting the statistical parameter $\rho$ to make the LSP the same ($1/2$) since $C_{s}(n)=C_{r}(n)=R$ (fixed, and independent of the channels).
\par For implementing the link-selection protocol for adaptive or fixed rate, the control node requires perfect CSI of $\SSS-\PD$ and $\SR-\PD$ (interference links) besides $\SSS-\SR$ and $\SR-\SD$ links. Though the choice depends largely on the scenario being considered, the buffer-aided relay itself can be chosen to be the control node. In this case, since the relay can estimate its interference channel as well as first and second hop channels, and $\SSS$ needs to just pass on the $\SSS-\PD$ channel gain to enable the relay $\SR$ to perform link selection.
\subsubsection{\textbf{Conventional Relay Schemes}}
\par To facilitate comparison of performance with the CABR scheme in the cognitive radio framework, we briefly describe the conventional schemes. One of the schemes is without a buffer, while the other is with a buffer. However, both the schemes use fixed scheduling, thereby imposing no excessive CSI requirement.
\vspace{-0.2cm}
\subsection*{\textbf{Conventional Non-Buffer-Aided Relay (CNBR) Scheme }}
In the simple CNBR scheme, a fixed scheduler is used with a buffer of just one packet at the relay \cite{Laneman2004}. Here, $d(n)=0$ for even $n$ and $1$ for odd $n$. 
 It is obvious that the delay incurred by this scheme is fixed (one time-slot). Due to nature of the scheduling, the capacity of the network is dominated by the capacity of the bottleneck hop. The end-to-end instantaneous SNR $\gamma$ and average rate of the CNBR scheme are given by:
 {\small
\begin{eqnarray}\label{eqn:eqn_CNBR}
\begin{array}{lll}
\hspace{1cm}\gamma &=& \min(\gamma_{s},\gamma_{r}),\\
\hspace{0.2cm}\overline{{\mathcal{R}}}^{CNBR}&=& \dfrac{1}{2}\,\mathbf{\E}_{\gamma_{s},\gamma_{r}}[\min(C_{s},C_{r})].
\end{array}
\end{eqnarray}
}
\vspace{-0.75cm}
\subsection*{\textbf{Conventional Buffer-Aided Relay (CBR) Scheme }}
Like CNBR, CBR too uses fixed scheduling \cite{Bing2008}: $d(n)=0$ for $n=0,1,\ldots,\frac{N}{2}-1$ and $d(n)=1$ for $n=\dfrac{N}{2},\ldots,N-1$ ($N$ is even). A buffer is used to store the packets.  
For the adaptive rate scenario, there is a loss in rate performance when there is asymmetry in $\SSS-\SR$ and $\SR-\SD$ links. Further,  the overall rate is limited by the rate of the bottleneck link. For the infinite-sized buffer, the achievable rate for the CBR scheme is half the minimum of capacities of individual links:
{\small
\begin{eqnarray}\label{eqn:eqn_CBR}
\overline{{\mathcal{R}}}^{CBR} &=& \dfrac{1}{2} \min(\mathbf{\E}_{\gamma_{s}}[C_{s}],\mathbf{\E}_{\gamma_{r}}[C_{r}]).
\end{eqnarray}
}
It is clear that from (\ref{eqn:eqn_CBR}) that the throughput of the CBR scheme is the same that of CNBR ($R/2$).
\begin{figure*}[!t]
	\vspace*{0.5pt}
	{\small
		\begin{subequations}\label{eqn:CCDF_SR}
			\begin{eqnarray}\label{eqn:CCDF_SR1}
			\begin{array}{lll}
			\hspace{-0.05in}F_{d,\gamma_{s}}^{c}(0,x)=(1-p_{s})\Big[e^{-x/\lambda_{s}}-(1-p_{r})\dfrac{\lambda_{\rho}}{\rho \lambda_{s}}e^{-(\rho x)/\lambda_{\rho}}\Big]+ p_{s}\dfrac{\mu_{s}}{x+\mu_{s}}\Big[e^{-x/\lambda_{s}} - \Big(1-p_{r}+ \dfrac{\mu_{r}p_{r}}{\mu_{r}-\rho\mu_{s}}\Big)e^{-(\rho x)/\lambda_{\rho}}\Big]\\
			\vspace{0.0cm}
			\hspace{1.in}+p_{s} \left(1-p_{r}+  \dfrac{\mu_{r}p_{r}}{\mu_{r}-\rho\mu_{s}}+\dfrac{\lambda_{r}\,\mu_{r}p_{r}}{(\mu_{r}-\rho\mu_{s})^2}\right)\dfrac{\rho\mu_{s}}{\lambda_{r}} \exp\left(\dfrac{\rho \mu_{s}}{\lambda_{\rho}}\right)  E_{1}\left(\dfrac{\rho x+\rho \mu_{s}}{\lambda_{\rho}}\right)\\
			\vspace{-0.175cm}
			\hspace{1.in}-p_{r}\left(1-p_{s}- \dfrac{\rho\mu_{s}p_{s}}{\mu_{r}-\rho\mu_{s}}+\dfrac{\rho\lambda_{s}\,\rho\mu_{s}\,p_{s}}{(\mu_{r}-\rho\mu_{s})^2}\right)\dfrac{\mu_{r}}{\rho\lambda_{s}} \exp\left(\dfrac{\mu_{r}}{\lambda_{\rho}}\right) E_{1}\left(\dfrac{\rho x+\mu_{r}}{\lambda_{\rho}}\right)\hspace{1.4cm} \textbf{when}\,\,\bf{\mu_{r}\neq\rho\mu_{s}},\\
			\end{array}
			\end{eqnarray}
			\hrulefill
			\vspace{-0.1cm}
			\begin{eqnarray}
			\begin{array}{lll}
			\hspace{-0.05in}F_{d,\gamma_{s}}^{c}(0,x) \hspace{-0. in}= (1-p_{s})\Big[e^{-x/\lambda_{s}}-(1-p_{r})\dfrac{\lambda_{\rho}}{\rho \lambda_{s}}e^{-(\rho x)/\lambda_{\rho}}\;\Big]-\dfrac{p_{s}p_{r}}{2}  e^{-(\rho x)/\lambda_{\rho}}\left(\dfrac{\mu_{s}}{x+\mu_{s}}\right)^2 \\
			\vspace{-0.cm}
			\hspace{1.5 cm} +\
			\dfrac{p_{s}\mu_{s}}{x+\mu_{s}}\Big[e^{-x/\lambda_{s}}-(1-p_{r})e^{-(\rho x)/\lambda_{\rho}}+\dfrac{p_{r}}{2}\left(\dfrac{\mu_{r}}{\lambda_{r}}-\,\dfrac{\mu_{s}}{\lambda_{s}}\right)e^{-(\rho x)/\lambda_{\rho}} \Big]
			\\
			\vspace{-0.175cm}
			\hspace{1.5 cm} +\ \Big[p_{s}(1-p_{r})\dfrac{\mu_{r}}{\lambda_{r}}-p_{r}(1-p_{s})\dfrac{\mu_{s}}{\lambda_{s}}+\dfrac{p_{s}p_{r}}{2}\left(\dfrac{\mu_{s}^2}{\lambda_{s}^2}-\dfrac{\mu_{r}^2}{\lambda_{r}^2}\right)\Big]\exp\left(\dfrac{\mu_{r}}{\lambda_{\rho}}\right) E_{1}\left(\dfrac{\rho x+\mu_{r}}{\lambda_{\rho}}\right)\hspace{1.1cm} \textbf{when}\,\,\bf{\mu_{r}=\rho\mu_{s}}.\label{eqn:CCDF_SR2}\\
			\end{array}
			\end{eqnarray}
		\end{subequations}\vspace{-0.1cm}
		\hrulefill
	}
	\hrulefill
\end{figure*}
{\small
	\begin{table*}[t]
		\hspace{.35cm}\caption{Joint CCDF of $\SSS-\SR$ and $\SR-\SD$ links with link LSP (note that (\ref{eqn:reversibility}) has been applied)}
		\footnotesize
		\hspace{.35cm}
		\renewcommand{\arraystretch}{1.2}
		\label{tab:CCDF_reversibility} 
		\begin{minipage}{\textwidth}
			\begin{tabular}{|p{.15cm}|p{.15cm}|p{7.1cm}|p{.15cm}|p{.15cm}|p{ 7.1cm}|}
				\hline\hline
				$p_{s}$ & $p_{r}$& $F_{d,\gamma_{s}}^{c}(0,x)$ &$p_{r}$ & $p_{s}$& $F_{d,\gamma_{r}}^c(1,x)$ \\ \hline\hline
				0 & 0 & $e^{-x/\lambda_{s}}\left\{1-\dfrac{\lambda_{\rho}}{\rho \lambda_{s}}e^{-{\rho x}/\lambda_{r}}\right\}.$   &0 & 0 & $e^{-x/\lambda_{r}}\left\{1-\dfrac{\lambda_{\rho}}{\lambda_{r}}e^{-{ x}/(\rho\lambda_{s})}\right\}.$  \\\hline
				0&  1 & $e^{-x/\lambda_{s}}\left\{1-\dfrac{\mu_{r}}{\rho\lambda_{s}}\exp\left(\dfrac{\mu_{r}}{\rho\lambda_{s}}\right)E_{1}\left(\dfrac{\rho x+\mu_{r}}{\rho\lambda_{s}}\right)\right\}.$   &0&  1 & $e^{-x/\lambda_{r}}\left\{1-\dfrac{\rho\mu_{s}}{\lambda_{r}}\exp\left(\dfrac{\rho\mu_{s}}{\lambda_{r}}\right)E_{1}\left(\dfrac{x+\rho\mu_{s}}{\lambda_{r}}\right)\right\}.$  \\\hline
				1& 0 & $\dfrac{\mu_{s}}{x+\mu_{s}}\left\{1-\exp\left(\dfrac{\rho\mu_{s}}{\lambda_{r}}\right)E_{2}\left(\dfrac{\rho x+\rho\mu_{s}}{\lambda_{r}}\right)\right\}.$   &1& 0 &$\dfrac{\mu_{r}}{x+\mu_{r}}\left\{1-\exp\left(\dfrac{\mu_{r}}{\rho\lambda_{s}}\right)E_{2}\left(\dfrac{x+\mu_{r}}{\rho\lambda_{s}}\right)\right\}.$\\\hline
				\vspace{0.1cm} 1 &\vspace{0.1cm} 1
				&$-\dfrac{\mu_{s}}{x+\mu_{s}}\left\{\dfrac{\rho\mu_{s}}{\mu_{r}-\rho \mu_{s}}\right\}-\dfrac{\rho\mu_{s}\mu_{r}}{(\mu_{r}-\rho \mu_{s})^2}\ln\dfrac{\rho x + \rho \mu_{s}}{\rho x + \mu_{r}},$   
				&\vspace{0.1cm} 1 &\vspace{0.1cm} 1 
				&$\dfrac{\mu_{r}}{x+\mu_{r}}\left\{\dfrac{\mu_{r}}{\mu_{r}-\rho \mu_{s}}\right\}-\dfrac{\rho\mu_{s}\mu_{r}}{(\mu_{r}-\rho \mu_{s})^2}\ln\dfrac{x + \mu_{r}}{x + \rho\mu_{s}},$\\\cline{3-3}\cline{6-6}
				&
				&\vspace{-0.2cm}$\textbf{or}\quad \dfrac{\mu_{s}}{x+\mu_{s}}-\dfrac{1}{2}\left(\dfrac{\mu_{s}}{x+\mu_{s}}\right)^2\quad \textbf{when}\,\,\bf{\mu_{r}=\rho\mu_{s}}.$
				&&
				&\vspace{-0.2cm}$\textbf{or}\quad \dfrac{\mu_{r}}{x+\mu_{r}}-\dfrac{1}{2}\left(\dfrac{\mu_{r}}{x+\mu_{r}}\right)^2\quad when \quad\textbf{when}\,\,\bf{\mu_{r}=\rho\mu_{s}}.$	 \\\hline\hline
			\end{tabular}
			\vspace{-0.175cm}
			\label{tab:CCDF_LSP} 
		\end{minipage}
	\end{table*}
} 
\section{Rate and SER Performance Analysis}\label{sec:PerAna}
\par In this section, we derive expressions for the average rate and high-SNR SER of the CABR scheme in underlay cognitive radio network assuming both peak interference and peak power constraints at $\SSS$ are $\SR$. We also present these expressions for conventional schemes to facilitate comparison.  Assuming adaptive rate transmission,  we first derive an expression for the average ergodic rate. Then, for fixed-rate transmission, we derive the approximate closed-form expressions of the SER. Since we adopt the CCDF based approach to obtain the expressions for average rate and SER, we first derive these joint CCDF expressions. In all the derived expressions, we can use $p_{i}=0$ and $p_{i}=1$ to obtain expressions for the PTP and PIP cases.
\subsection{\textbf{Complementary Cumulative Distribution Function}}
We first define an integral $\mathcal{I}_{n}(\mu,\lambda;x)$ as follows \cite[eq. (3.353.1)]{gradshteyn2007}:
{\footnotesize\begin{eqnarray}\label{eqn:I_n_mu_lambda_x}
\hspace{-0.4cm}\mathcal{I}_{n}(\mu,\lambda;x)\hspace{-0.3cm}&=&\hspace{-0.35cm}\int\limits_{x}^{\infty}\hspace{-0.1cm}\dfrac{\mu^{n-1}e^{-{s}/\lambda}}{(s+\mu)^n} {\D}s=\hspace{-0.1cm}\left(\dfrac{\mu}{x+\mu}\right)^{n-1}\hspace{-0.5cm} \exp\left(\dfrac{\mu}{\lambda}\right) E_{n}\left(\dfrac{x+\mu}{\lambda}\right)\hspace{-0.1cm}.
\end{eqnarray}} 
We also define a variant of $\mathcal{I}_{n}(\mu,\lambda;x)$ which finds application in average rate analysis as follows\cite[eq. (3.353.2)]{gradshteyn2007}:
{\footnotesize\begin{eqnarray}
\label{eqn:I_n_mu_lambda}
\hspace{-0.3cm}\mathcal{I}_{n}(\mu,\lambda)\hspace{-0.3cm}&=&\hspace{-0.3cm}\mathcal{I}_{n}(\mu,\lambda;0)=\hspace{-0.cm}\int\limits_{0}^{\infty}\hspace{-0.cm}\dfrac{\mu^{n-1}\,e^{-{s}/\lambda}}{(s+\mu)^n} {\D}s=\exp\left(\dfrac{\mu}{\lambda}\right) E_{n}\left(\dfrac{\mu}{\lambda}\right)\hspace{-0.1cm}.
\end{eqnarray}} 
{\bf CABR Scheme}\\
For the CABR scheme, we present expressions for joint CCDF of instantaneous SNR with link-selection parameter $d$ of $\SSS-\SR$ and $\SR-\SD$  links  ($F^c_{d,\gamma_{s}}(0,x)$ and $F^c_{d,\gamma_{r}}(1,x)$)  separately.  We note that for the $\SSS-\SR$ link, joint CCDF $F^c_{d,\gamma_{s}}(0,x)$ is given by:
{\small\begin{eqnarray}\label{eqn:jointccdfbasic}
\begin{array}{lll}
\hspace{-0.25cm}F^c_{d,\gamma_{s}}(0,x)\hspace{-0.1 in}&=&\hspace{-0.25cm} \Pr\{\gamma_{s}>x\}-\Pr\left\{{\gamma_{r}}/{\rho}>\gamma_{s}>x\right\},\\
&=&
F^c_{\gamma_{s}}(x)-\displaystyle\int\limits_{x}^{\infty}F^c_{\gamma_{r}}(\rho s)f_{\gamma_{s}}(s) {\D}s.
\end{array}
\end{eqnarray}} 
It is shown in Appendix A that using (\ref{eqn:F_Gi_s}) and  some manipulations, $F^c_{d,\gamma_{s}}(0,x)$ is given by (\ref{eqn:CCDF_SR}), where $\lambda_{\rho}$ is given by harmonic mean of $\rho\lambda_{s}$ and $\lambda_{r}$ i.e.  $1/\lambda_{\rho}=1/(\rho\lambda_{s})+1/\lambda_{r}$ (which confirms  the fact that for non-cognitive system, the link selection policy given by (\ref{eqn:linkcrit}) establishes a virtual $\SSS-\SR$ link with average SNR $\rho\lambda_{s}$)\cite{Islam2013_2}. 
We note that expression (\ref{eqn:CCDF_SR2}) derived in Appendix A applies instead of (\ref{eqn:CCDF_SR1}) when $\mu_{r}=\rho\mu_{s}$. We further note that when $\mu_{r}=\rho\mu_{s}$, relation $\dfrac{\rho\mu_{s}}{\lambda_{\rho}}=\dfrac{\mu_{r}}{\lambda_{\rho}}=\dfrac{\mu_{s}}{\lambda_{s}}+\dfrac{\mu_{r}}{\lambda_{r}}$ holds, and is frequently used in subsequent analysis.
\par It can be readily seen from  (\ref{eqn:jointccdfbasic}) that the LSP $q_s$ of the $\SSS-\SR$ link is given by $q_s=F_{d,\gamma_s}^{c}(0,0)$.    
The expression for $F_{d,\gamma_{r}}^c(1,x)$ is analogous to $F_{d,\gamma_{s}}^c(0,x)$. Also, It can be verified that $F_{d,\gamma_{r}}^c(1,x)$ can also be obtained from (\ref{eqn:CCDF_SR}) using reversibility relation (\ref{eqn:reversibility}). It is therefore omitted. However, the CCDFs of $\SSS-\SR$ and $\SR-\SD$ links for some special cases have been extracted from (\ref{eqn:CCDF_SR}) and reversibility (\ref{eqn:reversibility}), and are listed in Table~\ref{tab:CCDF_reversibility}. Note that reversibility changes the position of $p_{s}$ and $p_{r}$ and exchanges $d$ with  $1-d$ in the Table~\ref{tab:CCDF_reversibility}.\\
\vspace{0.5cm}
\\{\bf CNBR and CBR Schemes}
\par In CNBR, unlike CABR, the rates can be selected based on the rates of the two hops since signalling takes place in two consecutive time-slots over which the channels remain the same. It follows from (\ref{eqn:eqn_CNBR}) that the CCDF of end-to-end SNR for CNBR scheme can be obtained using: $F_{\gamma}^{c,CNBR}(x) = F_{\gamma_{s}}^c(x) F_{\gamma_{r}}^c(x)$ and (\ref{eqn:F_Gi_s}).
\vspace{-0.125cm} 
\subsection{\textbf{Average Rate}}\label{sec:AvgRate}
\par In this subsection, we evaluate the average rate for various relaying schemes.  We first define an important integral $\mathcal{J}(\mu,\lambda)$  as follows:
{\footnotesize\begin{eqnarray}
\label{eqn:J_mu_lambda}
\hspace{-0.4cm}\mathcal{J}(\mu,\lambda)\hspace{-0.3cm}&=& \hspace{-0.4cm} \displaystyle\int\limits_{0}^{\infty}\dfrac{\ln(1+x)e^{-{x}/\lambda}}{x+\mu} {\D}x=\exp\left(\dfrac{\mu}{\lambda}\right)   \displaystyle\int\limits_{0}^{\infty}\dfrac{E_{1}\left(\frac{x+\mu}{\lambda}\right)}{1+x} {\D}x.
\end{eqnarray}} 
The integrals in (\ref{eqn:I_n_mu_lambda})  and (\ref{eqn:J_mu_lambda}) are useful in average rate analysis. The average rate is evaluated using $f_{d,\gamma_{i}}(d,x)=-{\D}F_{d,\gamma_{i}}^{c}(d,x)$, and integration by parts as follows:

{\small\begin{eqnarray}
\begin{array}{lll}\label{eqn:rate_basic}
\hspace{-0.1cm}\overline{{\mathcal{R}}}_{i}\hspace{-0.1in}&=&\hspace{-0.3cm}-\displaystyle\int\limits_{0}^{\infty}\hspace{-0.05in} \log_{2}(1+x)\,{\D}F_{d,\gamma_{i}}^{c}(d,x)=\hspace{-0.01cm} \dfrac{1}{\ln(2)}\displaystyle\int\limits_{0}^{\infty}\hspace{-0.04in} \dfrac{F^{c}_{d,\gamma_{i}}(d,x)}{(1+x)}\,{\D}x,
\end{array}
\end{eqnarray}} 
where $d$ takes value of $0$ or $1$ depending on whether $i=s$ or $i=r$. In what follows, we derive expressions for average rate of the CABR and conventional schemes. We initially assume the infinite-sized buffer, hence the probability of buffer overflow is zero. We also assume that probability of buffer underflow is negligible. For the rate to be maximum in a balanced buffer, the average number of bits $\mathbf{\E}_{\gamma_{s},\gamma_{r}}[(1-d)C_{s}]$ entering the buffer (inflow) should be equal to the number of bits $\mathbf{\E}_{\gamma_{s},\gamma_{r}}[d C_{r}]$ leaving the buffer (outflow). To enable this, $\rho$ has to be chosen to be $\rho_{opt}$. Clearly, the average rate of the CABR scheme can be written from (\ref{eqn:rate_expr_in_len}) as:
{\small
\begin{equation}\label{eqn:rate_expr}
\overline{{\mathcal{R}}}^{CABR}\hspace{-0.3cm}=\mathbf{\E}_{\gamma_{s},\gamma_{r}}[d C_{r}]=\mathbf{\E}_{\gamma_{s},\gamma_{r}}[(1-d)C_{s}]\,\, \text{for }\rho=\rho_{opt},
\end{equation}
}
where substituting (\ref{eqn:CCDF_SR}) in (\ref{eqn:rate_basic}), the effective average rate of the $\SSS-\SR$ link $\overline{\mathcal{R}}_{s}^{CABR}$ is given by (\ref{eqn:AR_CABR_SR}), where $\mathcal{I}_{n}(\mu,\lambda)$ and $\mathcal{J}(\mu,\lambda)$ are defined in (\ref{eqn:I_n_mu_lambda}) and (\ref{eqn:J_mu_lambda}). Now the effective average rate of the $\SR-\SD$ link $\overline{\mathcal{R}}_{r}^{CABR}$ can be written directly from (\ref{eqn:AR_CABR_SR})  by using reversibility specified by (\ref{eqn:reversibility}), and is therefore omitted for brevity.
\\
\\{\bf CNBR and CBR Schemes}\\
In both the CNBR and CBR schemes, $\SSS-\SR$ and $\SR-\SD$ links are selected equally. Hence the average rate for CNBR is given by:
{\small\begin{eqnarray}
	\label{eqn:General_ER}
	\hspace{-0.35cm}\overline{{\mathcal{R}}}\hspace{-0.35cm}  &=&\hspace{-0.35cm} -\dfrac{1}{2}\hspace{-0.1cm}\displaystyle\int\limits_{0}^{\infty} \hspace{-0.1cm}\log_{2}(1+x)\,{\D}F_{\gamma}^{c}(x)= \dfrac{1}{2\ln(2)}  \int\limits_{0}^{\infty} \dfrac{F^c_{\gamma}(x)}{1+x}\,{\D}x.
	\end{eqnarray}} 
 Substituting for the CCDF of $\gamma$, it can be seen from (\ref{eqn:eqn_CNBR}) that $F_{\gamma}^{c,CNBR}(x) = F_{\gamma_{s}}^c(x) F_{\gamma_{r}}^c(x)$.  Using (\ref{eqn:F_Gi_s}),  we obtain the end-to-end average rate for CNBR  as in  (\ref{eqn:AR_CNBR}), where $\lambda$ is defined as harmonic mean of $\lambda_{s}$ and $\lambda_{r}$ i.e.  $1/\lambda=1/\lambda_{s}+1/\lambda_{r}$. In the PIP regime, under symmetric link conditions  when $\mu_{r}=\mu_{s}$ (ratio of $\SSS-\SR$ distance to $\SSS-\PD$ distance is the same as the ratio of $\SR-\SD$ distance to $\SR-\PD$ distance), (\ref{eqn:AR_CNBR1}) is invalid, and the average rate is given by (\ref{eqn:AR_CNBR2}). The average rate of CBR is obtained using (\ref{eqn:eqn_CBR}), and is given by (\ref{eqn:AR_CBR}).
\begin{figure*}[!t]
	\vspace*{-0.5pt}
	{\small
		\begin{subequations}\label{eqn:AR_CABR_SR}
			\begin{eqnarray}\label{eqn:AR_CABR_SR1}
			\begin{array}{lll}
			\hspace{-0.05in}\overline{\mathcal{R}}_{s}^{CABR}=\mathbf{\E}_{\gamma_{s},\gamma_{r}}[(1-d)C_{s}]\hspace{-0.0in}  =\dfrac{1}{\ln(2)}\Bigg[ 
			(1-p_{s}) \Big[\mathcal{I}_{1}(1,\lambda_{s})-(1-p_{r})\dfrac{\lambda_{\rho}}{\rho \lambda_{s}}\mathcal{I}_{1}\left(1,\dfrac{\lambda_{\rho}}{\rho}\right)\;\Big]+p_{s}\dfrac{\mu_{s}}{\mu_{s}-1}\\
			\vspace{-0.cm}
			\hspace{-.0 cm} \times\ 
			\Big[\mathcal{I}_{1}(1,\lambda_{s})-\mathcal{I}_{1}(\mu_{s},\lambda_{s})-\left(1-p_{r}+ \dfrac{\mu_{r}p_{r}}{\mu_{r}-\rho\mu_{s}}\right)\left\{\mathcal{I}_{1}\left(1,\dfrac{\lambda_{\rho}}{\rho}\right)-\mathcal{I}_{1}\left(\mu_{s},\dfrac{\lambda_{\rho}}{\rho}\right)\right\}\Big] +\dfrac{\rho\mu_{s}p_{s}}{\lambda_{r}}\mathcal{J}\left(\mu_{s},\dfrac{\lambda_{\rho}}{\rho}\right) \\
			\vspace{-0.175cm}
			\hspace{-.0 cm} \times\ 
			\left(1-p_{r}+ \dfrac{\mu_{r}p_{r}}{\mu_{r}-\rho\mu_{s}}+\dfrac{\lambda_{r}\mu_{r}p_{r}}{(\mu_{r}-\rho\mu_{s})^2}\right) -\dfrac{\mu_{r}p_{r}}{\rho\lambda_{s}}\mathcal{J}\left(\dfrac{\mu_{r}}{\rho},\dfrac{\lambda_{\rho}}{\rho}\right)\left(1-p_{s}- \dfrac{\rho\mu_{s}p_{s}}{\mu_{r}-\rho\mu_{s}}+\dfrac{\rho\lambda_{s}\,\rho\mu_{s}\,p_{s}}{(\mu_{r}-\rho\mu_{s})^2}\right)\Bigg]\hspace{.0cm} \textbf{when}\,\,\bf{\mu_{r}\neq\rho\mu_{s}},
			\end{array}
			\end{eqnarray}
			\hrulefill
			\begin{eqnarray}\label{eqn:AR_CABR_SR2}
			\begin{array}{lll}
			\hspace{-0.05in}\overline{\mathcal{R}}_{s}^{CABR}=\mathbf{\E}_{\gamma_{s},\gamma_{r}}[(1-d)C_{s}]  =\dfrac{1}{\ln(2)}\Bigg[
			(1-p_{s}) \Big[\mathcal{I}_{1}(1,\lambda_{s})-(1-p_{r})\dfrac{\lambda_{\rho}}{\rho \lambda_{s}}\mathcal{I}_{1}\left(1,\dfrac{\lambda_{\rho}}{\rho}\right)\;\Big]+\dfrac{\mu_{s}p_{s}}{\mu_{s}-1}  \Big\{\mathcal{I}_{1}(1,\lambda_{s})-\mathcal{I}_{1}(\mu_{s},\lambda_{s})\Big\}\\
			\vspace{-0.cm}
			\hspace{.0cm} -\ 
			\Big[\left\{p_{s}(1-p_{r})-\dfrac{p_{s}p_{r}}{2}\left(\dfrac{\mu_{r}}{\lambda_{r}}-\,\dfrac{\mu_{s}}{\lambda_{s}}\right)\right\}\dfrac{\mu_{s}}{\mu_{s}-1}+\dfrac{p_{s}p_{r}}{2}\left(\dfrac{\mu_{s}}{\mu_{s}-1}\right)^2\Big]\bigg\{\mathcal{I}_{1}\left(1,\dfrac{\lambda_{\rho}}{\rho}\right)-\mathcal{I}_{1}\left(\mu_{s},\dfrac{\lambda_{\rho}}{\rho}\right)\bigg\}\\
			\vspace{-0.175cm}
			\hspace{.0cm} +\ 
			\dfrac{p_{s}p_{r}}{2}\dfrac{\mu_{s}}{\mu_{s}-1}  \mathcal{I}_{2}\left(\mu_{s},\dfrac{\lambda_{\rho}}{\rho}\right)+\Big[p_{s}(1-p_{r})\dfrac{\mu_{r}}{\lambda_{r}}-p_{r}(1-p_{s})\dfrac{\mu_{s}}{\lambda_{s}}+\dfrac{p_{s}p_{r}}{2}\left(\dfrac{\mu_{s}^2}{\lambda_{s}^2}-\dfrac{\mu_{r}^2}{\lambda_{r}^2}\right)\Big] \mathcal{J}\Big(\mu_{s},\dfrac{\lambda_{\rho}}{\rho}\Big)\Bigg]\hspace{0.3in} \textbf{when}\,\,\bf{\mu_{r}=\rho\mu_{s}}.\\
			\end{array}
			\end{eqnarray}
		\end{subequations}

		\hrulefill
		\begin{subequations}\label{eqn:AR_CNBR}
			\begin{eqnarray}
			\hspace{-2.5cm}\begin{array}{lll}
			\hspace{2.05cm}\overline{{\mathcal{R}}}^{CNBR}  = \dfrac{1}{2\ln(2)}   \Big[(1-p_{s})(1-p_{r})\mathcal{I}_{1}(1,\lambda)+\dfrac{p_{s}\mu_{s}}{\mu_{s}-1}\left(1-p_{r}+p_{r}\dfrac{\mu_{r}}{\mu_{r}-\mu_{s}}\right)\Big\{ \mathcal{I}_{1}(1,\lambda)-\mathcal{I}_{1}(\mu_{s},\lambda)\Big\}  \\
			\vspace{-0.175cm}\hspace{8.cm}
			  +\ \dfrac{p_{r}\mu_{r}}{\mu_{r}-1}\left(1-p_{s}-p_{s}\dfrac{\mu_{s}}{\mu_{r}-\mu_{s}}\right)\Big\{ \mathcal{I}_{1}(1,\lambda)-\mathcal{I}_{1}(\mu_{r},\lambda)\Big\}\Big]\hspace{.2cm} \textbf{when}\,\,\bf{\mu_{r}\neq\mu_{s}},\label{eqn:AR_CNBR1}\\
			\end{array}
			\end{eqnarray}
			\hrulefill
			\begin{eqnarray}
			\hspace{-.4cm}\begin{array}{lll}
			\overline{{\mathcal{R}}}^{CNBR}  = \dfrac{1}{2\ln(2)}  \Big[ (1-p_{s})(1-p_{r})\mathcal{I}_{1}(1,\lambda)-p_{s}p_{r}\dfrac{\mu_{s}}{\mu_{s}-1} \mathcal{I}_{2}\left(\mu_{s},\lambda\right)+  \Big[\Big\{p_{s}(1-p_{r})+p_{r}(1-p_{s})\Big\}\dfrac{\mu_{s}}{\mu_{s}-1}\\
			\vspace{-0.175cm}\hspace{7.6cm}
			+\ 
			p_{s}p_{r}\left(\dfrac{\mu_{s}}{\mu_{s}-1}\right)^2\Big]\Big\{ \mathcal{I}_{1}(1,\lambda)-\mathcal{I}_{1}(\mu_{s},\lambda)\Big\}\Big]\hspace{.3cm} \textbf{when}\,\,\bf{\mu_{r}=\mu_{s}}.\label{eqn:AR_CNBR2}\\
			\end{array}
			\end{eqnarray}
		\end{subequations}
		\hrulefill
		\begin{eqnarray}
		\begin{array}{lll}
		\hspace{-2.3cm}\overline{{\mathcal{R}}}^{CBR}  = \dfrac{1}{2\ln(2)}  \min \Big((1-p_{s}) \mathcal{I}_{1}(1,\lambda_{s})+\dfrac{p_{s}\mu_{s}}{\mu_{s}-1}
		\Big\{\mathcal{I}_{1}(1,\lambda_{s})-\mathcal{I}_{1}(\mu_{s},\lambda_{s})\Big\}\\\vspace{-0.175cm}
		&& \hspace{-4.5cm}
		,(1-p_{r}) \mathcal{I}_{1}(1,\lambda_{r})+\dfrac{p_{r}\mu_{r}}{\mu_{r}-1}
		\Big\{\mathcal{I}_{1}(1,\lambda_{r})-\mathcal{I}_{1}(\mu_{r},\lambda_{r})\Big\}\Big).\label{eqn:AR_CBR}
		\end{array}
		\end{eqnarray}
		\hrulefill
	}
\end{figure*}
\subsection{\textbf{Symbol Error Rate (SER)}}\label{sec:SER}
In this subsection, we evaluate the SER for various relay schemes. We first define two important integrals $\mathcal{K}(\mu,\lambda)$\cite[eq. (3.363.2)]{gradshteyn2007}\cite[eq. (7.4.9)]{Abramowitz1965} and $\mathcal{L}(\mu,\lambda)$ for SER as follows:
{\footnotesize\begin{eqnarray}
\mathcal{K}(\mu,\lambda)&=&\displaystyle\int\limits_{0}^{\infty}\sqrt{\dfrac{\eta}{2\pi w}}\dfrac{\mu\,e^{-\left(1+\dfrac{2}{\eta\lambda}\right)\dfrac{\eta w}{2}}}{(w+\mu)} {\D}w,\nonumber\\\label{eqn:K_mu_lambda}
&=& \sqrt{\dfrac{\pi\eta \mu}{2}}\exp\left(\dfrac{\eta \mu}{2}+\dfrac{ \mu}{\lambda}\right)\erfc\left(\sqrt{\dfrac{\eta \mu}{2}+\dfrac{ \mu}{\lambda}}\,\right).
\end{eqnarray}}
\vspace{-0.5cm}
{\footnotesize\begin{eqnarray}
	\label{eqn:L_mu_lambda}
	\mathcal{L}(\mu,\lambda)&=&\exp\left(\dfrac{\mu}{\lambda}\right)  \displaystyle\int\limits_{0}^{\infty}\sqrt{\dfrac{\eta}{2\pi w}}e^{-(\eta w)/2}E_{1}\left(\dfrac{w+\mu}{\lambda}\right) {\D}w.
	\end{eqnarray}} 
Integral $\mathcal{L}(\mu,\lambda)$ cannot be expressed in closed-form but can be approximated for high SNR. For the CABR scheme where scheduling of packets is not deterministic, as well as for the CBR scheme, it is difficult to get a closed-form expression for the end-to-end SER performance. It is assumed here that packets decoded in error are placed in the buffer. At medium and high SNRs where the packet decoding error probability is small,  the end-to-end SER is bounded tightly by the sum of individual SER of $\SSS-\SR$ and $\SR-\SD$ link as follows\cite{Ikhlef2012_MMRS}:
\begin{eqnarray}\label{eqn:TotalSER}
\overline{\mathcal{P}}&\leq& \overline{\mathcal{P}}_{s}+\overline{\mathcal{P}}_{r}.
\end{eqnarray}
{\bf CABR Scheme}\\
For the CABR scheme, we use the CCDF approach to derive expressions for SER. Joint CCDF expressions in (\ref{eqn:CCDF_SR}) are applicable in this case.  The SER $\overline{\mathcal{P}}_{i}$  is given by \cite{Islam2015}:
{\small\begin{equation}\label{eq:SER_erfc}
\hspace{-0.28cm}\overline{\mathcal{P}}_{i}\approx \dfrac{\varphi_{i}}{2}\mathbf{\E_{\gamma_{i}}}\left[\erfc\left(\sqrt{\dfrac{\eta_{i}\gamma_{i}}{2}}\right)\right]\hspace{-0.1cm}=\hspace{-0.1cm}\dfrac{\varphi_{i}}{2}\int\limits_{0}^{\infty}\hspace{-0.1cm}\erfc\left(\hspace{-0.05cm}\sqrt{\dfrac{\eta_{i}\,s}{2}}\right)f_{\gamma_{i}}(s) {\D}s,
\end{equation}} 
where $\erfc(x)$ denotes the complementary error function, and $\eta_{i}$ and $\varphi_{i}$ are the modulation parameters. For simplicity, we assume equal transmission rates, and use of the same modulation scheme at $\SSS$ and $\SR$ i.e. $R_{s}=R_{r}=R$, $\eta_{i}=\eta$, and $\varphi_{i}=\varphi$.\\
Using the relation { $\erfc(x)=\frac{\Gamma(1/2,x^2)}{\Gamma(1/2)}$} \cite[eq. (6.5.17)]{Abramowitz1965}, we get, {\small $\erfc\left(\sqrt{\dfrac{\eta\gamma_{i}}{2}}\right) =  \dfrac{\Gamma(1/2,\eta\gamma_{i}/2)}{\Gamma(1/2)}=F_{w}^{c}(\gamma_{i})$}, where $\Gamma(n,x)$ is the incomplete gamma function, and $F_{w}^{c}(x)$ is the CCDF of a Gamma distributed random variable $w$ with PDF:
{\small
\begin{equation}\label{eqn:w_pdf}
f_{w}\left(w\right) = \sqrt{\dfrac{\eta}{2\pi w}} e^{-{\eta w}/2}.
\end{equation}
}
Using {\small $\erfc\left(\sqrt{\dfrac{\eta\gamma_{i}}{2}}\right) =F_{w}^{c}(\gamma_{i})$} in (\ref{eq:SER_erfc}), and exploiting (\ref{eqn:w_pdf}), the SER can be written as:
{\small\begin{eqnarray}
\hspace{-0.cm}\overline{\mathcal{P}}_{i}\hspace{-0.25cm}&=&\hspace{-0.25cm} \dfrac{\varphi_{i}}{2}\displaystyle\int\limits_{0}^{\infty}F_{w}^{c}(s)f_{\gamma_{i}}(s) {\D}s \overset{\ell}{=}\dfrac{\varphi_{i}}{2}\int\limits_{0}^{\infty}F_{\gamma_{i}}(s)f_{w}(s) {\D}s,\nonumber\\\label{eqn:General_SER}
&=&\hspace{-0.05cm}
	\hspace{-0.2cm}\dfrac{\varphi}{2}\mathbf{\E}_{w}\left[ F_{\gamma_{i}}(w)\right]=\dfrac{\varphi}{2}\displaystyle\int\limits_{0}^{\infty}\sqrt{\dfrac{\eta}{2\pi w}} e^{-{\eta w}/2}F_{\gamma_{i}}(w){\D}w,
	\end{eqnarray}} 
where equality $\ell$ follows using integration by parts. In the CABR scheme, $\SSS$ transmits when the corresponding link-selection criteria of (\ref{eqn:linkcrit}) is fulfilled.  We focus on SER of the $\SSS-\SR$ link. The SER of the $\SR-\SD$ link can once again be generated from that of the $\SSS-\SR$ link using reversibility relation given in (\ref{eqn:reversibility}). The conditional SER of the $\SSS-\SR$ link  given that it is selected is written using (\ref{eqn:General_SER}) as:
 {\footnotesize\begin{eqnarray}\label{eqn:P_obs}
\begin{array}{lll}
\hspace{-0.0cm}\overline{\mathcal{P}}^{CABR}_{s}\hspace{-0.3cm} &=&\hspace{-0.35cm}\dfrac{\varphi}{2}{\mathbf{\E}_{w}[F_{\gamma_{s}\hspace{-0.0cm}|d =0}(w)]}\hspace{-0.1cm}=\hspace{-0.1cm} \hspace{-0.cm}\dfrac{\varphi}{2}\mathbf{\E}_{w}\hspace{-0.cm}\left[\Pr\{\gamma_{s}<w|(\rho\gamma_{s}>\gamma_{r})\}\right],\\
&\overset{m}{=}&\hspace{-0.2cm}\dfrac{\varphi}{2q_{s}}\mathbf{\E}_{w}[F_{d,\gamma_{s}}(0,w)]\overset{l}{=} \dfrac{\varphi}{2q_{s}}\hspace{-0.15cm}\left[{q_{s}-\mathbf{\E}_{w}[F_{d,\gamma_{s}}^c(0,w)]}\right],
\end{array}
\end{eqnarray}}
\begin{figure*}[!t]
	\vspace*{0.5pt}
	{\small
		\begin{subequations}\label{eqn:ECCDF_SR}
			\begin{eqnarray}\label{eqn:ECCDF_SR1}
			\begin{array}{lll}
			\hspace{-0.1in}\mathbf{\E}_{w}[F_{d,\gamma_{s}}^c(0,w)]\hspace{-0.0in}= (1-p_{s})\left[\sqrt{\dfrac{\eta\lambda_{s}}{\eta\lambda_{s}+2}}-(1-p_{r})\dfrac{\lambda_{\rho}}{\rho \lambda_{s}}\sqrt{\dfrac{\eta\lambda_{\rho}}{\eta\lambda_{\rho}+2\rho}}\;\right]+p_{s}\mathcal{K}(\mu_{s},\lambda_{s})-p_{s}\left(1-p_{r}+ \dfrac{\mu_{r}p_{r}}{\mu_{r}-\rho\mu_{s}}\right)\mathcal{K}(\mu_{s},\lambda_{\rho}/\rho)\\
			\vspace{0.cm}
			\hspace{0.5 cm} +\
			\dfrac{\rho\mu_{s}p_{s}}{\lambda_{r}}\left(1-p_{r}+ \dfrac{\mu_{r}p_{r}}{\mu_{r}-\rho\mu_{s}}+\dfrac{\lambda_{r}\,\mu_{r}p_{r}}{(\mu_{r}-\rho\mu_{s})^2}\right)\mathcal{L}\left(\mu_{s},\dfrac{\lambda_{\rho}}{\rho}\right)  - \dfrac{\mu_{r}p_{r}}{\rho\lambda_{s}}\left(1-p_{s}- \dfrac{\rho\mu_{s}p_{s}}{\mu_{r}-\rho\mu_{s}}+\dfrac{\rho\lambda_{s}\,\rho\mu_{s}\,p_{s}}{(\mu_{r}-\rho\mu_{s})^2}\right) \mathcal{L}\left(\dfrac{\mu_{r}}{\rho},\dfrac{\lambda_{\rho}}{\rho}\right)\\
			\vspace{-0.175cm}
			\hspace{15. cm}  \bf{\textbf{when}\,\mu_{r}\neq\rho\mu_{s}},
			\end{array}
			\end{eqnarray}
			\hrulefill
			\vspace{-0.1cm}
			\begin{eqnarray}\label{eqn:ECCDF_SR2}
			\begin{array}{lll}
			\vspace{0.02cm}
			\hspace{-.1in}\mathbf{\E}_{w}[F_{d,\gamma_{s}}^c(0,w)] =
			(1-p_{s})\left[\sqrt{\dfrac{\eta\lambda_{s}}{\eta\lambda_{s}+2}}-(1-p_{r})\dfrac{\lambda_{\rho}}{\rho \lambda_{s}}\sqrt{\dfrac{\eta\lambda_{\rho}}{\eta\lambda_{\rho}+2\rho}}\;\right]+p_{s}\mathcal{K}(\mu_{s},\lambda_{s})-\dfrac{p_{s}p_{r}}{4}  \eta\mu_{s}\sqrt{1+\dfrac{2 \rho}{\eta\lambda_{\rho}}}\\
			\vspace{0.02cm}
			\hspace{.5 cm} -\ 
			p_{s}\left[1-p_{r}\left(\dfrac{\mu_{r}}{\lambda_{r}}+\dfrac{\eta\mu_{s}+3}{4}\right)\right]\mathcal{K}\left(\mu_{s},\dfrac{\lambda_{\rho}}{\rho}\right)  + \Bigg[p_{s}(1-p_{r})\dfrac{\mu_{r}}{\lambda_{r}}-p_{r}(1-p_{s})\dfrac{\mu_{s}}{\lambda_{s}}+\dfrac{p_{s}p_{r}}{2}\left(\dfrac{\mu_{s}^2}{\lambda_{s}^2}-\dfrac{\mu_{r}^2}{\lambda_{r}^2}\right)\Bigg]\mathcal{L}\left(\dfrac{\mu_{r}}{\rho},\dfrac{\lambda_{\rho}}{\rho}\right) \\
			\vspace{-0.175cm}
			\hspace{15. cm}  \bf{\textbf{when}\,\mu_{r}=\rho\mu_{s}}.
			\end{array}
			\end{eqnarray}
		\end{subequations}
		\hrulefill
	}
	\hrulefill
\end{figure*} 
where $F_{\gamma_{s}|d =0\,}(w)$ and $F_{d,\gamma_{s}}(0,x)$ are the conditional and joint distributions of $\gamma_s$ and $d$ respectively. Equality $m$ follows using Bayes rule and (\ref{eqn:linkcrit}), and equality $l$ follows from the fact that $F_{d,\gamma_{s}}(0,x)= F_{d,\gamma_{s}}(0,\infty)-F^c_{d,\gamma_{s}}(0,x)=q_{s}-F^c_{d,\gamma_{s}}(0,x)$.  $\mathbf{\E}_{w}[F_{d,\gamma_{s}}^c(0,w)]$ is evaluated by averaging (\ref{eqn:CCDF_SR}) over (\ref{eqn:w_pdf}) and is given by (\ref{eqn:ECCDF_SR}). We note that the LSP $q_{s}$ can be obtained from $(\ref{eqn:CCDF_SR})$ using the relation $q_{s}=F_{d,\gamma_{s}}^c(0,0)$, and that $\mathcal{K}(\mu,\lambda)$ and $\mathcal{L}(\mu,\lambda)$ in (\ref{eqn:ECCDF_SR}) are as defined in (\ref{eqn:K_mu_lambda}) and (\ref{eqn:L_mu_lambda}) respectively. An expression for SER $\overline{\mathcal{P}}^{CABR}_{r} $ of the $\SR-\SD$ link  can be obtained using reversibility, and is therefore omitted.

{\bf CNBR and CBR Schemes}\vspace{.1cm}
\par The SER ${\small \overline{\mathcal{P}}_{i}^{CNBR} }$ of the CNBR scheme is obtained 
by substituting (\ref{eqn:F_Gi_s}) in (\ref{eqn:General_SER}), and is given by:
 {\small
\begin{eqnarray}\label{eqn:SER_s}
\overline{\mathcal{P}}_{i}^{CNBR}  =\dfrac{\varphi}{2}\Bigg[1  -(1-p_{i})\sqrt{\dfrac{\eta\lambda_{i}}{\eta\lambda_{i}+2}}-p_{i}\mathcal{K}(\mu_{i},\lambda_{i})\Bigg],
\end{eqnarray}} 
for $i\in\{s,r\}$, where $\mathcal{K}(\mu,\lambda)$ is as defined in  (\ref{eqn:K_mu_lambda}).  It can be readily verified from (\ref{eqn:P_obs}) and (\ref{eqn:ECCDF_SR}) that when $\rho\rightarrow\infty\,(q_{s}\rightarrow 1)$, $\overline{\mathcal{P}}_{s}^{CABR}
\rightarrow\overline{\mathcal{P}}_{s}^{CNBR}$, which is quite intuitive since $\SSS-\SR$ link in the CABR scheme does not get selected according to channel condition any more. Similarly, $\overline{\mathcal{P}}_{r}^{CABR}\rightarrow
\overline{\mathcal{P}}_{r}^{CNBR}$ for $\rho\rightarrow 0$. It is obvious that when the link is in the PIP regime ($\gamma_{max}$ is large, making $\lambda_i\rightarrow \infty$),  ${\small\overline{\mathcal{P}}_{i}^{CNBR}={\varphi}/{2}\,(1-\mathcal{K}(\mu_{i},\infty))} $\footnote{$\mathcal{K}(\mu_{i},\infty)$ can be approximated asymptotically when interference is not very severe i.e. for large $\mu_{i}$,\cite[eq. (7.1.23)]{Abramowitz1965}.}  By substituting (\ref{eqn:SER_s}) into  (\ref{eqn:TotalSER}), we can obtain the gross SER of the CNBR scheme. As noted already, the SER performance of CBR is not superior to that of  CNBR due to the absence of any greedy link-selection mechanism.
\begin{table*} 
	\caption{CCDF, Expectation of CCDF, Link Selection Probability, Asymptotic SER and Average Rate}
	\footnotesize
	\centering
	\label{tab:CCDF_E_LSP_ASER_AR}
	\renewcommand{\arraystretch}{1.2}
	\begin{minipage}{\textwidth}
		\begin{tabular}{|p{.15cm}|p{.15cm}|p{7.525cm}|p{ 8.5cm}|}
			\hline\hline
			$p_{s}$ & $p_{r}$&\hspace{2cm} $F_{d,\gamma_{s}}^{c}(0,x)\,(\text{in}\,(\ref{eqn:CCDF_SR}))$ &\hspace{2cm}$\mathbf{\E}_{w}[F_{d,\gamma_{s}}^c(0,w)]\,(\text{in}\,(\ref{eqn:ECCDF_SR}))$  \\ \hline\hline
			0 & 0 & $e^{-x/\lambda_{s}}\left\{1-\dfrac{\lambda_{\rho}}{\rho \lambda_{s}}e^{-{\rho x}/\lambda_{r}}\right\}.$   & $\sqrt{\dfrac{\eta\lambda_{s}}{\eta\lambda_{s}+2}}-\dfrac{\lambda_{\rho}}{\rho \lambda_{s}}\sqrt{\dfrac{\eta\lambda_{\rho}}{\eta\lambda_{\rho}+2\rho}}\;.$  \\\hline
			0&  1 & $e^{-x/\lambda_{s}}\left\{1-\dfrac{\mu_{r}}{\rho\lambda_{s}}\exp\left(\dfrac{\mu_{r}}{\rho\lambda_{s}}\right)E_{1}\left(\dfrac{\rho x+\mu_{r}}{\rho\lambda_{s}}\right)\right\}.$   & $\sqrt{\dfrac{\eta\lambda_{s}}{\eta\lambda_{s}+2}}-\dfrac{\mu_{r}}{\rho\lambda_{s}}\mathcal{L}\left(\dfrac{\mu_{r}}{\rho},\lambda_{s}\right).$  \\\hline
			1& 0 & $\dfrac{\mu_{s}}{x+\mu_{s}}\left\{1-\exp\left(\dfrac{\rho\mu_{s}}{\lambda_{r}}\right)E_{2}\left(\dfrac{\rho x+\rho\mu_{s}}{\lambda_{r}}\right)\right\}.$   & $\mathcal{K}(\mu_{s},\infty)-\mathcal{K}\left(\mu_{s},\dfrac{\lambda_{r}}{\rho}\right)+\dfrac{\rho\mu_{s}}{\lambda_{r}}\mathcal{L}\left(\mu_{s},\dfrac{\lambda_{r}}{\rho}\right).$\\\hline
			1& 1 & $-\dfrac{\mu_{s}}{x+\mu_{s}}\left\{\dfrac{\rho\mu_{s}}{\mu_{r}-\rho \mu_{s}}\right\}-\dfrac{\rho\mu_{s}\mu_{r}}{(\mu_{r}-\rho \mu_{s})^2}\ln\dfrac{\rho x + \rho \mu_{s}}{\rho x + \mu_{r}},$   &$\dfrac{-\rho\mu_{s}}{\mu_{r}-\rho \mu_{s}}\mathcal{K}(\mu_{s},\infty)+\dfrac{\rho\mu_{s}\mu_{r}}{(\mu_{r}-\rho \mu_{s})^2}\Big\{\mathcal{L}(\mu_{s},\infty)-\mathcal{L}\left(\dfrac{\mu_{r}}{\rho},\infty\right)\Big\},$\\\cline{3-4}
			&&\vspace{-0.2cm}$\textbf{or}\quad \dfrac{\mu_{s}}{x+\mu_{s}}-\dfrac{1}{2}\left(\dfrac{\mu_{s}}{x+\mu_{s}}\right)^2\quad \textbf{when}\,\,\bf{\mu_{r}=\rho\mu_{s}}.$&\vspace{-.15cm}$\textbf{or}\quad -\dfrac{\eta\mu_{s}}{4}+\Big(\dfrac{\eta\mu_{s}}{4}+\dfrac{3}{4}\Big)\mathcal{K}(\mu_{s},\infty)\quad \textbf{when}\,\,\bf{\mu_{r}=\rho\mu_{s}}.$	 \\\hline
		\end{tabular}
	\end{minipage}
	\begin{minipage}{\textwidth}
		\begin{tabular}{|p{.15cm}|p{.15cm}|p{4.2cm}|p{2.9cm}|p{8.525cm}|}
			\hline\hline
			$p_{s}$ & $p_{r}$ &\vspace{-0.1cm}\hspace{0cm} $q_{s}=F_{d,\gamma_{s}}^c(0,0)\,(\text{using}\,(\ref{eqn:CCDF_SR}))$&\vspace{-0.15cm}$\overline{\mathcal{P}}^{CABR}_{s,asym}\,\approx\,(\text{in}\,(\ref{eqn:PBUFASYM}))$&\vspace{-0.1cm}\hspace{2.5cm} $\overline{R}_{s,asym}^{CABR}\,(\text{in}\,(\ref{eqn:AR_CABR_SR}))$ \\ \hline\hline
			0 & 0 &\vspace{-0.25cm} \hspace{.75cm} $\dfrac{\lambda_{\rho}}{\lambda_{r}}=\dfrac{\rho\lambda_{s}}{\rho\lambda_{s}+\lambda_{r}}.$&\vspace{-0.15cm}$\dfrac{3 \varphi}{4\eta^2}\dfrac{\rho}{q_{s}\lambda_{s}\lambda_{r}}$&\vspace{-0.25cm} $\left\{\mathcal{I}_{1}(1,\lambda_{s})-\dfrac{\lambda_{\rho}}{\rho \lambda_{s}}\mathcal{I}_{1}\left(1,\dfrac{\lambda_{\rho}}{\rho}\right)\right\}.$    \\\hline
			0&  1 &\vspace{-0.2cm}\hspace{0.3cm}$\exp\left(\dfrac{\mu_{r}}{\rho\lambda_{s}}\right)E_{2}\left(\dfrac{\mu_{r}}{\rho\lambda_{s}}\right).$&\vspace{-0.15cm}$\dfrac{3 \varphi}{4\eta^2}\dfrac{\rho}{q_{s}\lambda_{s}\mu_{r}}$&\vspace{-0.1cm} $\mathcal{I}_{1}(1,\lambda_{s})-\dfrac{\mu_{r}}{\rho\lambda_{s}}\mathcal{J}\left(\dfrac{\mu_{r}}{\rho},\lambda_{s}\right).$   \\\hline
			1& 0    &\vspace{-0.1cm}\hspace{0.1cm}$\dfrac{\rho\mu_{s}}{\lambda_{r}}\exp\left(\dfrac{\rho\mu_{s}}{\lambda_{r}}\right)E_{1}\left(\dfrac{\rho\mu_{s}}{\lambda_{r}}\right).$&\vspace{-0.15cm}$\dfrac{3 \varphi}{4\eta^2}\dfrac{\rho}{q_{s}\mu_{s}\lambda_{r}}$&\vspace{-0.1cm} $\Big[\log_{2}\mu_{s}-\mathcal{I}_{1}\left(1,\dfrac{\lambda_{r}}{\rho}\right)+\mathcal{I}_{1}\left(\mu_{s},\dfrac{\lambda_{r}}{\rho}\right)\Big]+\dfrac{\rho\mu_{s}}{\lambda_{r}}\mathcal{J}\left(\mu_{s},\dfrac{\lambda_{r}}{\rho}\right).$ \\\hline
			1& 1    &\vspace{-0.2cm} \hspace{-0.2cm}$\dfrac{-\rho\mu_{s}}{\mu_{r}-\rho \mu_{s}}-\dfrac{\rho\mu_{s}\mu_{r}}{(\mu_{r}-\rho \mu_{s})^2}\ln\dfrac{\rho \mu_{s}}{\mu_{r}},$ &&\vspace{-0.2cm} $  \hspace{-0.15cm}\dfrac{-\rho\mu_{s}}{\mu_{r}-\rho\mu_{s}}\dfrac{\mu_{s}\log_{2}(\mu_{s})}{\mu_{s}-1}\ \hspace{-0.1cm}+
			\dfrac{\rho\mu_{s}\mu_{r}\log_{2}(e)}{(\mu_{r}-\rho\mu_{s})^2}\Big[\hspace{-0.05cm}Li_{2}(1-\mu_{s})-Li_{2}\left(1-\dfrac{\mu_{r}}{\rho}\right)\hspace{-0.05cm}\Big]$\\\cline{3-3}\cline{5-5}
			&&\vspace{-0.1cm}$\textbf{or}\quad \dfrac{1}{2}\quad \textbf{when}\,\,\bf{\mu_{r}=\rho\mu_{s}}.$&\vspace{-0.65cm}$\dfrac{3 \varphi}{4\eta^2}\dfrac{\rho}{q_{s}\mu_{s}\mu_{r}}$&\vspace{-.1cm}$\textbf{or}\quad \dfrac{0.5\mu_{s}}{\mu_{s}-1}\Big[\log_{2}(e)+\left(\frac{\mu_{s}-2}{\mu_{s}-1}\right)\log_{2}(\mu_{s})\Big]\quad \textbf{when}\,\,\bf{\mu_{r}=\rho\mu_{s}}.$	 \\\hline\hline
		\end{tabular}
		\label{tab:CCDF_LSP} 
	\end{minipage}
\end{table*}
\subsection*{\textbf{SER at High SNR - CABR}}
\par Unfortunately, it is difficult to gain useful insights on the influence of system parameters on SER performance from the exact SER expressions of the CABR scheme presented above. It is well known that the SER exhibits a floor at high SNRs in underlay cognitive radio because of the interference constraint. The SER asymptotes are therefore of importance. For the CABR scheme, it is difficult to derive the SER asymptote directly from the exact SER expression. It is shown in Appendix B that by first approximating the joint CCDF (\ref{eqn:CCDF_SR1}) and (\ref{eqn:CCDF_SR2}), it is possible to derive the following simple form for the SER of the $\SSS-\SR$ link at high SNRs:
{\small\begin{eqnarray}\label{eqn:PBUFASYM_SR}
\begin{array}{lll}
\hspace{-0.1cm}\overline{\mathcal{P}}^{CABR}_{s}\hspace{-0.35cm} &\pipr&\hspace{-0.2cm} \dfrac{3 \varphi}{4\eta^2} \dfrac{\rho }{q_{s}}\left(\dfrac{1}{\lambda_{s}}+\dfrac{p_{s}}{\mu_{s}}\right)\left(\dfrac{1}{\lambda_{r}}+\dfrac{p_{r}}{\mu_{r}}\right),\\
\hspace{-0.1cm}\overline{\mathcal{P}}^{CABR}_{r}\hspace{-0.35cm} &\pipr&\hspace{-0.2cm} \dfrac{3 \varphi}{4\eta^2} \dfrac{1 }{\rho q_{r}}\left(\dfrac{1}{\lambda_{r}}+\dfrac{p_{r}}{\mu_{r}}\right)\left(\dfrac{1}{\lambda_{s}}+\dfrac{p_{s}}{\mu_{s}}\right),
\end{array}
\end{eqnarray}} 
where the asymptotic SER $\overline{\mathcal{P}}^{CABR}_{r}$ of the $\SR-\SD$  link is written using reversibility relation (\ref{eqn:reversibility}). Substituting expressions for $\overline{\mathcal{P}}^{CABR}_{s}$ and $\overline{\mathcal{P}}^{CABR}_{r}$ in (\ref{eqn:TotalSER}), we get:
{\small\begin{eqnarray}\label{eqn:PBUFASYM}
\hspace{-0.cm}\overline{\mathcal{P}}^{CABR}\hspace{-0.2cm} \pipr\hspace{-0.05cm} \dfrac{3 \varphi}{4\eta^2}\hspace{-0.1cm} \left[\dfrac{\rho }{q_{s}}+\dfrac{1}{\rho \,q_{r}}\right]\hspace{-0.125cm}\left(\hspace{-0.075cm}\dfrac{1}{\lambda_{s}}+\dfrac{p_{s}}{\mu_{s}}\hspace{-0.075cm}\right)\hspace{-0.125cm}\left(\hspace{-0.1cm}\dfrac{1}{\lambda_{r}}+\dfrac{p_{r}}{\mu_{r}}\hspace{-0.075cm}\right).
\end{eqnarray}} 
Now for stabilizing the buffer, $\rho$ is chosen optimally using (\ref{eqn:rate_expr}) such that the inflow and outflow rates are equal. Since we are assuming equal fixed-rate signalling in both the links ($R_{s}=R_{r}=R$), the LSPs of both the links are equal i.e. $q_{s}=q_{r}=0.5$ when $\rho=\rho_{opt}$. It is obvious that for optimum $\rho$, the average rate is always $0.5R$. Now, $\rho_{opt}$ can be numerically evaluated using  $q_s = F_{d,\gamma_s}^{c}(0,0)$ from (\ref{eqn:CCDF_SR}) by making $q_{s}=0.5$ (closed form expressions are difficult to obtain). Fortunately, in some important special cases, a closed-form expression for $\rho_{opt}$ can be found.  Table~\ref{tab:CCDF_E_LSP_ASER_AR} lists the expressions for $q_s$ for various combinations of $p_s$ and $p_r$. It can be readily inferred that when $\SSS$ and $\SR$ both in either in PTP or in PIP regime, $\rho_{opt}$ is given by:
{\small
\begin{equation}\label{eqn:link_sel_pr_half}
\rho_{opt} =  \left\{ 
\begin{array}{l l}
\lambda_{r}/\lambda_{s} & \quad \text{In PTP regime, i.e.\,$p_{s},p_{r}=0$}  \\
\mu_{r}/\mu_{s} & \quad \text{In PIP regime, i.e.\,$p_{s},p_{r}=1$}
\end{array} \right..
\end{equation}
}	
Substituting the above equation in (\ref{eqn:PBUFASYM}), the SER for the CABR scheme is written for $\rho=\rho_{opt}$ as:
{\small\begin{eqnarray}\label{eqn:SER_ASYM_CABR}
\overline{\mathcal{P}}^{CABR}\hspace{-0.09cm} \pipr\hspace{-0.14cm}\left\{ 
\begin{array}{l l}
\dfrac{3 \varphi}{2\eta^2} \left(\dfrac{1}{\lambda_{s}^2}+\dfrac{1}{\lambda_{r}^2}\right) & \hspace{-0.2cm}\text{In PTP regime,}\\
\dfrac{3 \varphi}{2\eta^2} \left(\dfrac{1}{\mu_{s}^2}+\dfrac{1}{\mu_{r}^2}\right)   & \hspace{-0.2cm}  \text{In PIP regime.}
\end{array} \right.
\end{eqnarray}} 
{\bf  CNBR and CBR Schemes}\\
Using steps similar to those outlined in Appendix B, the expression for SER of the $\SSS-\SR$ link at high SNR for the CNBR scheme can be written as:
{\small
\begin{eqnarray}\label{eqn:PCNBRASYM_SR}
\begin{array}{lll}
\overline{\mathcal{P}}^{CNBR}_{s} \hspace{-0.25cm}&\pipr&\hspace{-0.25cm} \dfrac{\varphi}{2\eta}\left(\dfrac{1}{\lambda_{s}}+\dfrac{p_{s}}{\mu_{s}}\right),\\
\overline{\mathcal{P}}^{CNBR}_{r} \hspace{-0.25cm}&\pipr&\hspace{-0.25cm} \dfrac{\varphi}{2\eta}\left(\dfrac{1}{\lambda_{r}}+\dfrac{p_{r}}{\mu_{r}}\right),
\end{array}
\end{eqnarray}
}
where ${\small \overline{\mathcal{P}}^{CNBR}_{r}}$ is obtained from  ${\small \overline{\mathcal{P}}^{CNBR}_{s}}$ using reversibility. Proof is omitted for brevity. Substituting the above SERs in (\ref{eqn:TotalSER}), and noting that SER performance of the CBR and CNBR schemes is the same, we can write:
{\small\begin{eqnarray}
\begin{array}{lllll}\label{eqn:PCNBRASYM}
\overline{\mathcal{P}}^{CBR} \hspace{-0.2cm}&=&\hspace{-0.2cm}
\overline{\mathcal{P}}^{CNBR}\hspace{-0.2cm} &\approx
& \hspace{-0.2cm}\dfrac{\varphi}{2\eta} \left[\left(\dfrac{1}{\lambda_{s}}+\dfrac{p_{s}}{\mu_{s}}\right)+\left(\dfrac{1}{\lambda_{r}}+\dfrac{p_{r}}{\mu_{r}}\right)\right].
\end{array}
\end{eqnarray}} 
From the above, the high SNR SER performance in the PTP and PIP regimes is given by:
{\small\begin{eqnarray}\label{eqn:SER_ASYM_CBR}
\begin{array}{lllll}
\hspace{-0.14cm}\overline{\mathcal{P}}^{CNBR}\hspace{-0.14cm} \pipr\hspace{-0.14cm}\left\{ 
\begin{array}{l l}
\dfrac{\varphi}{2\eta} \left(\dfrac{1}{\lambda_{s}}+\dfrac{1}{\lambda_{r}}\right)& \hspace{-0.2cm}\text{In PTP regime,}\\
\dfrac{\varphi}{2\eta} \left(\dfrac{1}{\mu_{s}}+\dfrac{1}{\mu_{r}}\right)   & \hspace{-0.2cm}  \text{In PIP regime.}
\end{array} \right.
\end{array}
\end{eqnarray}} 
Equations (\ref{eqn:SER_ASYM_CABR}) and (\ref{eqn:SER_ASYM_CBR}) bring out the important fact that the SER of the CABR scheme exhibits a slope of $2$ versus average SNR in the PTP regime, whereas that of the CNBR scheme has a slope of $1$\cite{Zlatanov2013_1}. 
\section{Delay Analysis}~\label{sec:DelAna}
\par The analysis presented for ergodic rate and SER in Sections~\ref{sec:AvgRate} and \ref{sec:SER} is based on the assumption of infinite-sized buffers. Rapidly improving technology has made systems with large memory feasible. In any case, the analysis does serve to bound performance of practical systems. However, it has been established that adaptive link-selection with relay of infinite-sized buffer results in infinite queuing delay when $\rho=\rho_{opt}$ as in (\ref{eqn:rate_expr})\cite{Zlatanov2013_1}.  
As described in \cite{Zlatanov2013_1}, \cite{Islam2015}, \cite{Islam2013_2}, the average queuing delay can be made finite by either limiting the size of the buffer or  by starving it. We can starve the buffer as noted earlier by choosing $\rho<\rho_{opt}$ (this amounts to choosing the second hop more often). When we restrict the size of the buffer, the probability of buffer overflow increases.   Both of these options therefore result in throughput loss.
\subsection{Adaptive Rate Transmission} 
\par  We perform delay analysis for the case when the source $\SSS$ and the relay $\SR$\footnote{We assume FIFO buffer here. Analysis for LIFO buffer follows from reversibility in (\ref{eqn:reversibility}).} employ adaptive rate transmission techniques. We first define an integral $\mathcal{M}(\mu,\lambda)$  as follows:
{\small\begin{eqnarray}
	\begin{array}{lll}
	\label{eqn:M_mu_lambda}
	\hspace{-0.4cm}\mathcal{M}(\mu,\lambda)\hspace{-0.3cm}&=& \hspace{-0.4cm} \displaystyle\int\limits_{0}^{\infty}\dfrac{[\ln(1+x)]^{2}e^{-{x}/\lambda}}{x+\mu} {\D}x,\\
	&=&\exp\left(\dfrac{\mu}{\lambda}\right)   \displaystyle\int\limits_{0}^{\infty}\dfrac{\ln(1+x)E_{1}\left(\frac{x+\mu}{\lambda}\right)}{1+x} {\D}x,
	\end{array}
	\end{eqnarray}} 
  where the second line is obtained using integration by parts. Now, the average delay is given by \cite{Zlatanov2013_1}:
{\small\begin{equation}\label{eqn:delay_expr}
\hspace{-0.2cm}\overline{T} \leq \dfrac{1}{2} \dfrac{1}{(\xi\E[(1-d)C_{s}])^2}\dfrac{\xi^2\E[(1-d)C_{s}^2]+(2\xi-1)\E[dC_{r}^2]}{\xi-1},
\end{equation}} 
where $\xi=\E[dC_{r}]/\E[(1-d)C_{s}]$ is greater than one when the buffer is starved. $\E[(1-d)C_{s}]$ and $\E[dC_{r}]$ are obtained from  (\ref{eqn:AR_CABR_SR}) and the reversibility relation  (\ref{eqn:reversibility}).  Further it can be seen using integration by parts that:
{\small\begin{eqnarray*}
	\begin{array}{lll}
		\E[(1-d)C_{s}^2]&=&-\displaystyle\int\limits_{0}^{\infty} [\log_{2}(1+x)]^2\,{\D}F^c_{d,\gamma_{s}}(0,x),\\
		&=& \dfrac{1}{\ln(2)}\displaystyle\int\limits_{0}^{\infty} \dfrac{\log_{2}(1+x)F^c_{d,\gamma_{s}}(0,x)}{1+x}\,{\D}x,
	\end{array}
\end{eqnarray*}} 
where the joint CCDF relation of (\ref{eqn:CCDF_SR}) is used in the above.  The expression for $\E[(1-d)C_s^2]$ is the same as that of (\ref{eqn:CCDF_SR}), with  $\mathcal{I}(\mu,\lambda)$ replaced by $\mathcal{J}(\mu,\lambda)/\ln(2)$, and all $\mathcal{J}(\mu,\lambda)$  replaced by $\mathcal{M}(\mu,\lambda)/\ln(2)$ (this equation is omitted for brevity). Unfortunately, ${\cal M}(\mu,\lambda)$ cannot be expressed in closed form, and needs to be evaluated numerically or using approximations to the exponential integral in $\mathcal{M}(\mu,\lambda)$. Various values  of average delay can be obtained by varying $\rho$ from $0$ to $\rho_{opt}$.
\subsection{Fixed-Rate Transmission}
In this subsection we analyze the delay performance assuming fixed-rate transmission. For ease of exposition, we assume that the fixed rate is unity (analysis can always be generalized for arbitrary rate $R$). We discuss a variant of threshold based transmission protocol presented in \cite{Islam2015} for the cooperative scenario, and analyze the tradeoffs between SER, delay, and throughput.
\begin{figure}[ht]
	\begin{center}
		\includegraphics[scale=.4]{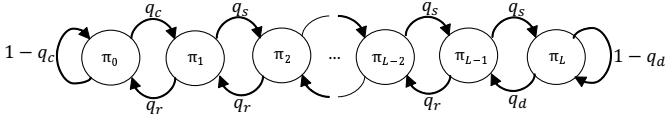}
		\caption{State transition diagram for the modified threshold based transmission protocol.}
		\label{fig:fig_1}
	\end{center}
	\vspace{-.75cm}
\end{figure}
\par Consider a FIFO buffer of size $L$.  As indicated in Fig.~\ref{fig:fig_1}, it has $L+1$ states  with state probabilities $\pi_{i}$, $i=0,1,2,\ldots,L$.  It can be seen that the state changes from $i$ to $i+1$ with probability $q_s$ ($\SSS-\SR$ link is chosen), and from $i+1$ to $i$ with probability $q_r$ ($\SR-\SD$ link chosen).
We use probabilities $q_c$ and $q_d$ to incorporate various options when the buffer is full or empty. The following special cases are of interest:
\begin{enumerate}
	\item Choosing $q_c = q_s$  ($q_d = q_r$) amounts to not taking cognizance of the buffer being empty (full). It results in poorer throughput, but better SER performance \cite{Islam2015} as compared to the case when $q_c=1$ ($q_d=1$).
	\item $q_c=1$ ($q_d=1$) amounts to  choosing the $\SSS-\SR$ ($\SR-\SD$) link when the buffer is empty (full). It results in better throughput, but poorer SER performance \cite{Islam2013_2} as compared to the case when $q_c=q_s$ ($q_d=q_r$). 
\end{enumerate}
Using $q_c$ ($q_d$) value in between these extreme values \footnote{Using $q_{c}<q_{s}$ decreases throughput considerably (especially when the buffer is starved or is limit in size). We do not consider this case here.} allows one to tradeoff SER and throughput. We discuss choice of $q_c$ ($q_d$) later in this subsection.
Here  $q_{c}=\Pr\{\rho_{c}\gamma_{s}>\gamma_{r}\}$, and $q_{d}=\Pr\{\rho_{d}\gamma_{s}>\gamma_{r}\}$, where $\rho_{c}$ and $\rho_d$ are the link-selection threshold parameters {\em when the buffer is empty and full respectively}.  
\par We also explore the relationship between delay, throughput and SER, and their dependence on choice of $q_{c}$. It is obvious that the queuing model of considered protocol given in Fig.~ \ref{fig:fig_1} models a  birth-death process, and is hence reversible. By applying the detailed balance equations, we get:
{\small\begin{eqnarray*}
\begin{array}{lll}
\hspace{0cm}q_{c}\,\pi_{0} = q_{r}\,\pi_{1},\quad
\pi_{i}=\xi\, \pi_{i+1},\,\, 1\leq i\leq L-2,\quad
q_{s}\,\pi_{L-1} = q_{d}\,\pi_{L},
\end{array}
\end{eqnarray*}}
where
{\small\begin{eqnarray}
		\begin{array}{lll}\label{eqn:xi_xic_xid}
			\hspace{0cm}\xi=\dfrac{q_{r}}{q_{s}}=\dfrac{1-q_{s}}{q_{s}},\quad\quad \xi_{c}=\dfrac{1-q_{c}}{q_{c}},\quad\xi_{d}=\dfrac{1-q_{d}}{q_{d}}.
		\end{array}
	\end{eqnarray}} 
Note that $\xi>1$ when the buffer is starved, whereas $\xi_{c}>1$ ($\xi_{d}>1$) when $q_{c}<0.5$ ($q_{d}<0.5$). After solving the local balance equations, the steady-state probabilities of buffer in empty and full states are:
{\small\begin{eqnarray}\label{eqn:trans_prob}
\begin{array}{lll}
\pi_{0} &=& \left( 1+\dfrac{q_{c}}{q_{d}}\xi^{1-L}+\dfrac{q_{c}}{q_{s}}\dfrac{\xi^{-1}-\xi^{-L}}{1-\xi^{-1}}\right)^{-1}\hspace{-0.4cm},\\
\pi_{L} &=& \dfrac{q_{c}}{q_{d}}\xi^{1-L}\pi_{0}= \left( 1+\dfrac{q_{d}}{q_{c}}\xi^{L-1}+\dfrac{q_{d}}{q_{r}}\dfrac{\xi^{L}-\xi}{\xi-1}\right)^{-1}\hspace{-0.4cm}.
\end{array}
\end{eqnarray}} 
It is observed that $\pi_{L}$ can also be obtained from $\pi_{0}$ using (\ref{eqn:reversibility}).  The throughput $\overline{\tau}$ is clearly equal to the arrival rate $\A$. The departure rate $\DD$ is greater than the arrival rate $\A$ when the buffer is starved i.e. $\xi>1$. The expressions for $\overline{\tau},\,\A$ and $\DD$ can be written from Fig.~ \ref{fig:fig_1}, as follows\cite{Islam2015}:  
{\small\begin{eqnarray}\label{eqn:A_Tau_D}
\begin{array}{lll}
\A \hspace{-0.25cm}&=&\hspace{-0.25cm} q_{c}\pi_{0}+q_{s}(1-\pi_{0}-\pi_{L})=q_{d} \pi_{L} + q_{r}(1-\pi_{0}-\pi_{L}),\\
\overline{\tau}\hspace{-0.25cm}&=&\hspace{-0.25cm}\A=1/2\{1-(1-q_{c})\pi_{0}-(1-q_{d}) \pi_{L}\},\\
\DD \hspace{-0.25cm}&=&\hspace{-0.25cm} q_{d} \pi_{L} + q_{r}(1-\pi_{0}-\pi_{L})+ (1-q_{c})\pi_{0}.
\end{array}
\end{eqnarray}}  
Using  (\ref{eqn:trans_prob}), (\ref{eqn:A_Tau_D}) and some simple manipulations, we get:
{\small\begin{eqnarray}\label{eqn:qc_pi0_tau}
\dfrac{q_{c}\pi_{0}}{\overline{\tau}}=\dfrac{1-\xi^{-1}}{1-\xi^{-L}},\quad \dfrac{q_{d}\pi_{L}}{\overline{\tau}}=\dfrac{\xi-1}{\xi^{L}-1}.
\end{eqnarray}
}
It is clear from the above equation that the ratio of influx from $\pi_{0}$ to $\pi_{1}$ ($\pi_{L}$ to $\pi_{L-1}$) to $\overline{\tau}$ i.e. $q_{c}\pi_{0}/\overline{\tau}$ ($q_{d}\pi_{L}/\overline{\tau}$) cannot be changed by varying $q_{c}$ ($q_{d}$) alone. This is a useful observation in evaluating the performance of this model. \par We now derive expressions for the SER, average delay, and throughput. It is evident that the SER $\overline{\mathcal{P}}'_{s}$ of $\SSS-\SR$ link with the protocol in Fig.~\ref{fig:fig_1} is computed as the ratio of packets in error to the total number of packets, and is given by: 
{\small
\begin{eqnarray} 
\begin{array}{lll}\label{eqn:ps_temp}
\hspace{-0.25cm}\overline{\mathcal{P}}'_{s}\hspace{-0.3cm} &=&\hspace{-0.3cm} \dfrac{q_{c}\pi_{0}\overline{\mathcal{P}}_{c}+q_{s}(1-\pi_{0}-\pi_{L})\overline{\mathcal{P}}_{s}}{q_{c}\pi_{0}+q_{s}(1-\pi_{0}-\pi_{L})},
\end{array}
\end{eqnarray}}
where {\small $\overline{\mathcal{P}}_{s}=\overline{\mathcal{P}}_{s}^{CABR}$} is given by (\ref{eqn:P_obs}). Please note that the superscript CABR has been omitted for convenience. Similarly {\small $\overline{\mathcal{P}}_{c}$} is given by (\ref{eqn:P_obs}) with $\rho_{c}$ and $q_{c}$ used in place of $\rho$ and $q_{s}$. After substituting  $q_{s}(1-\pi_{0}-\pi_{L})=\overline{\tau}-q_{c}\pi_{0}$ from (\ref{eqn:A_Tau_D}) and   $q_{c}\pi_{0}/\overline{\tau}$ from (\ref{eqn:qc_pi0_tau}) in (\ref{eqn:ps_temp}), we get:
{\small
	\begin{eqnarray} 
		\begin{array}{lll}\label{eqn:Ps_fin}
			\hspace{-0.25cm}\overline{\mathcal{P}}'_{s}\hspace{-0.3cm} &=&\hspace{-0.3cm}\left( \dfrac{1-\xi^{-1}}{1-\xi^{-L}}\right) \overline{\mathcal{P}}_{c}+\left( \dfrac{\xi^{-1}-\xi^{-L}}{1-\xi^{-L}}\right) \overline{\mathcal{P}}_{s}.
		\end{array}
	\end{eqnarray}}
The SER of the $\SR-\SD$ link i.e. $\overline{\mathcal{P}}'_{r}$, can be evaluated from $\overline{\mathcal{P}}'_{s}$ using (\ref{eqn:reversibility}) as:
{\small
	\begin{eqnarray} 
	\begin{array}{lll}\label{eqn:Pr_fin}
	\hspace{-0.25cm}\overline{\mathcal{P}}'_{r}\hspace{-0.3cm} &=&\hspace{-0.3cm}\left( \dfrac{\xi-1}{\xi^{L}-1}\right) \overline{\mathcal{P}}_{d}+\left( \dfrac{\xi^{L}-\xi}{\xi^{L}-1}\right) \overline{\mathcal{P}}_{r}.
	\end{array}
	\end{eqnarray}}
\par We now derive expressions for the delay. The average system delay $\overline{T}$ comprises of  two components: the average delay $\overline{T}_{q}$ due to data queuing, and the average delay $\overline{T}_{s}$ due to silent time-slots (i.e. $\overline{T} =\overline{T}_{q}+\overline{T}_{s} $). Note that  the silent time-slots arise due to either overflow ($\overline{T}_{o}$) or underflow ($\overline{T}_{u}$) so that $\overline{T}_{s} =\overline{T}_{o} + \overline{T}_{u}$. It can be seen from Fig.~ \ref{fig:fig_1} that:
{\small
\begin{eqnarray}\label{eqn:T_u}
\begin{array}{lll}
\overline{T}_{u} &=& \dfrac{(1-q_{c})\pi_{0}}{\A} \overset{m}{=} \dfrac{1-q_{c}}{q_{c}}\dfrac{1-\xi^{-1}}{1-\xi^{-L}}=\xi_{c}\dfrac{1-\xi^{-1}}{1-\xi^{-L}},\\ \overline{T}_{o} &=& \dfrac{(1-q_{d})\pi_{L}}{\A} \overset{n}{=} \dfrac{1-q_{d}}{q_{d}}\dfrac{1-\xi}{1-\xi^{L}}=\xi_{d}\dfrac{1-\xi}{1-\xi^{L}},
\end{array}
\end{eqnarray}
}
where equality $m$ and $n$  are obtained from (\ref{eqn:qc_pi0_tau}) using $\overline{\tau}=\mathcal{A}$ and some simple manipulations. It is observed that $\overline{T}_{o}$ can also be obtained from $\overline{T}_{u}$ using (\ref{eqn:reversibility}).  It is clear from the expression that $\overline{T}_{u}=\overline{T}_{o}=0$ if a transition is forced the when the buffer is empty or full (implying that $q_{c}=q_{d}=1$). When  starving the buffer so that $\xi>1$, $\overline{T}_{u}$ increases while $\overline{T}_{o}$  decreases.
\par Now, the average delay due to queuing is $\overline{T}_{q} = {\overline{Q}}/{\A}$, where {\small $\overline{Q}=\sum\limits_{i=0}^{L}i\,\pi_{i}$} is the average queue size and $\A$ is the arrival rate.\footnote{The equivalent queuing delay of LIFO buffer can be expressed as $\overline{T}_{e} = {(L-\overline{Q})}/{\A}$, which can be derived from the expression of $\overline{T}_{q}$ using (\ref{eqn:reversibility}).} After writing these delays in terms of $q_{c}\pi_{0}/\mathcal{A}$ given by (\ref{eqn:qc_pi0_tau}) and using some manipulations, we get:
{\small\begin{eqnarray}\label{eqn:T_q}
\begin{array}{lll}
\hspace{.3cm}\overline{T}_{q} &=& 1+\dfrac{2}{\xi-1}+\dfrac{L(\xi-1)}{\xi^{L}-1}\left(\xi_{d}-\dfrac{2}{\xi-1}\right).
\end{array}
\end{eqnarray}}
It is readily observed from (\ref{eqn:T_u}) and (\ref{eqn:T_q}) that increasing $\xi_{c}$ increases $\overline{T}_{u}$ but has no effect on $\overline{T}_{q}$ while $\xi_{d}$ increases both $\overline{T}_{o}$ and $\overline{T}_{q}$.  It can be observed that $\overline{T}_{q}=1+{2}/{(\xi-1)}={1}/{(1 - 2 q_{s})}$ when $L\rightarrow\infty$. Also for a finite-sized buffer, decreasing $q_{d}$ increases $\xi_{d}$, thereby increases $\overline{T}_{q}$ in (\ref{eqn:T_q}). When $\xi_{d}<{2}/{(\xi-1)}$, or equivalently when $q_{d}>1-2 q_{s}$, the second term in (\ref{eqn:T_q}) is negative, and the queuing delay is lower with finite-sized buffers than those with infinite size employing buffer starving. It is also observed that the minimum achievable queuing delay is unity which can be achieved when either $\xi\rightarrow\infty$ or when $L=1$ with $\xi_{d}=0\,(q_{d}=1)$. Another important observation is that when $\xi_d = 2/(\xi -1)$, the queuing delay is $1 + 2/(\xi -1)$, which is independent of the buffer size, and the same as that of the infinite-sized buffer with starving.
\par We can write the throughput of the system from  (\ref{eqn:A_Tau_D}) in terms of $\overline{T}_{u}$ and $\overline{T}_{o}$ using (\ref{eqn:T_u}) as follows:
{\small
 {\begin{eqnarray}\label{eqn:tau_Tu2}
	\hspace{-0.5cm}2\hspace{-0.3cm}&=&\hspace{-0.cm} \dfrac{1}{\overline{\tau}}-\dfrac{(1-q_{c})\pi_{0}}{\overline{\tau}}-\dfrac{(1-q_{d})\pi_{L}}{\overline{\tau}}=\dfrac{1}{\overline{\tau}}-\overline{T}_{u}-\overline{T}_{o},\\\label{eqn:tau_Tu}
	\hspace{-0.5cm}&\Rightarrow&\hspace{1.cm}\overline{\tau}=\dfrac{1}{2+(\overline{T}_{u}+\overline{T}_{o})}.
\end{eqnarray}} 
}
It is obvious that the average delay due to both underflow and overflow leads to loss in throughput. We note that $\overline{\tau}\leq 1/2$ as expected. \par In several applications, we might want to constraint the delay and throughput i.e. we might want to impose the constraints $\overline{T}\leq\overline{T}_{max}^{*}$ and $\overline{\tau}\geq\overline{\tau}_{min}^{*}$. In \cite{Islam2015}, it is mentioned that all choices of $\overline{\tau}_{min}^{*}$ and $\overline{T}_{max}^{*}$ are not feasible. Here, we bring out the constraints on the choices using (\ref{eqn:tau_Tu2}) as follows.
{\small\begin{eqnarray}\label{delay_throughput}
	\hspace{0cm}\overline{T}_{u}+\overline{T}_{o}\hspace{-0.3cm}&=&\hspace{-0.cm} \dfrac{1}{\overline{\tau}}-2\leq\dfrac{1}{\overline{\tau}_{min}^{*}}-2,\nonumber\\\label{eqn:sear_pace_const_tau2}
	\Rightarrow\overline{T}_{q}+\overline{T}_{u}
	+\overline{T}_{o}\hspace{-0.3cm}&\leq&
	\hspace{-0.cm}\overline{T}_{max}^{*}\Rightarrow\overline{T}_{q}\leq\overline{T}_{max}^{*}-\dfrac{1}{\overline{\tau}_{min}^{*}}+2.
	\end{eqnarray}}
Substituting $\overline{T}_{q}\geq1$\footnote{For starving buffer ($\xi\geq1$), $\dfrac{\xi^{L}-1}{\xi-1}=\sum_{i=0}^{L-1}\xi^i \geq L$. Substituting it in (\ref{eqn:T_q}), we get $\overline{T}_{q}\geq1+\xi_{d}$, where equality is achieved when $\xi_{d}=0\, (\text{i.e. } q_{d}=1)$ with either $\xi\rightarrow\infty\,(\text{i.e. } q_{s}\rightarrow0)$ or with $L=1$.} in (\ref{eqn:sear_pace_const_tau2}), we get:
{\small\begin{eqnarray}\label{eqn:sear_pace_const_tau}
	1\leq\overline{T}_{q}\leq\overline{T}_{max}^{*}-\dfrac{1}{\overline{\tau}_{min}^{*}}+2\Rightarrow\overline{\tau}_{min}^{*}(1+\overline{T}_{max}^{*})\geq 1.
	\end{eqnarray}}
It is clear that $\overline{\tau}_{min}^{*}(1+\overline{T}_{max}^{*})\geq 1$ together with $\overline{\tau}_{min}^{*}\leq 1/2$ and $\overline{T}_{max}^{*}\geq 1$ are the constraints on choice of $\overline{\tau}_{min}^{*}$ and $\overline{T}_{max}^{*}$. Now, to extract useful insights, we consider two alternatives to control the delay - use of limited-size buffers, and buffer starving.\\
\\\emph{Limiting buffer size at optimum point ($\rho=\rho_{opt}$):}
\par It is obvious that for $\rho=\rho_{opt}$, $\xi=1$. As described earlier that the choice of $q_{c}$ and $q_{d}$ are irrelevant for the infinite-sized buffer. However, limiting the size of the buffer makes $\pi_{0}$ and $\pi_{L}$ finite, thereby making the choice of $q_{c}$ and $q_{d}$ important in deciding the trade-offs between throughput, delay and SER. Using {\small $\dfrac{1-\xi^{-L}}{1-\xi^{-1}}=\sum\limits_{k=0}^{L-1}\xi^{-k}$} ($=L$ when $\xi=1$), the SER and delay at $\rho=\rho_{opt}$ are given by (\ref{eqn:Ps_fin})-(\ref{eqn:T_q}) as follows\footnote{$\overline{T}_{q}$ and $\overline{T}_{e}$ for $\xi=1$ are obtained from \ref{eqn:T_q} using some manipulations and using limit $\xi\rightarrow\ 1$.}:
{\small\begin{eqnarray} 
\label{eqn:p_fin_rho_opt}
\overline{\mathcal{P}}'_{s}\hspace{-0.25cm}&=&\hspace{-0.25cm} \dfrac{\overline{\mathcal{P}}_{c}}{L} +\left(1-\dfrac{1}{L}\right) \overline{\mathcal{P}}_{s},\quad
\overline{\mathcal{P}}'_{r} = \dfrac{\overline{\mathcal{P}}_{d}}{L} +\left(1-\dfrac{1}{L}\right) \overline{\mathcal{P}}_{r},\\\label{eqn:Delay_FinMark_L_fin_rho_opt}
\overline{T}_{q}\hspace{-0.25cm}&=&\hspace{-0.25cm} L(1+\overline{T}_{o}),\quad \overline{T}_{u} = \dfrac{\xi_{c}}{L},\quad\overline{T}_{o} =\dfrac{\xi_{d}}{L}.
\end{eqnarray}} 
It is clear from (\ref{eqn:Delay_FinMark_L_fin_rho_opt}) that when $q_{d}=1,\,\xi_{d}=0$ and $\overline{T}_{o}=0$, so that the minimum value of queuing delay $\overline{T}_{q}=L$ is achieved. However, it is clear that there is an increase in $\overline{\mathcal{P}}'_{r}$ (because of increase in $\overline{\mathcal{P}}_d$), which can be limited only by increasing $L$. \\
\\\emph{Starving the buffer}
\par It is clear that for starving buffer, $\xi>1$. First, we provide expressions for the SER and throughput assuming $L\rightarrow\infty$, and then discuss selection of $q_{c}$ (as function of $q_{s}$) to meet certain constraints. The SER and queuing delay for the infinite-sized buffer are given from (\ref{eqn:Ps_fin}),(\ref{eqn:T_u}) and (\ref{eqn:T_q}) as follows:
{\small\begin{eqnarray}\label{eqn:SER_SR_Markov} 
\overline{\mathcal{P}}'_{s}&=&\left(1-\xi^{-1}\right)\overline{\mathcal{P}}_{c}+\xi^{-1}\overline{\mathcal{P}}_{s},\quad \overline{\mathcal{P}}'_{r}=\overline{\mathcal{P}}_{r},\\\label{eqn:delay_inf_tq_tu} 
\overline{T}_{q} &=& 1+\dfrac{2}{\xi-1},\,\overline{T}_{u}=\xi_{c}(1-\xi^{-1}).
\end{eqnarray}}
The average system delay can be written from (\ref{eqn:delay_inf_tq_tu}) as:
{\small\begin{eqnarray} \label{eqn:delay_inf}
	\overline{T}  &=& \overline{T}_{q}+\overline{T}_{u}= 1+\dfrac{2}{\xi-1}+ \xi_{c}\xi^{-1}(\xi-1).
	\end{eqnarray}} 
We consider a scenario where constraints are placed on maximum delay and minimum throughput, and another when the desired throughput is specified. \\
  {\bf Maximum Delay and Minimum Throughput (MDMT) Constraint Scheme}\\ It is clear that for the choice of $\xi_{c}$, when $\xi_{c}\,\xi^{-1}=x^{*}$ such that $x^{*}$ is a constant, $\overline{T}_{u}$ is inversely proportional to ($\overline{T}_{q}-1$), and the overall delay turns out to be  convex w.r.t. $\xi$ as follows:
{\small\begin{eqnarray}\label{eqn:x_convex_delay}
	\overline{T}_{} &=&1+\dfrac{2}{\xi-1}+ x^{*}(\xi-1),\quad\text{where }\xi_{c}=\xi x^{*}\\\nonumber
	&=& 1+\sqrt{2x^{*}}\Bigg[\sqrt{\dfrac{2}{x^{*}}}  \dfrac{1}{\xi-1}+\sqrt{\dfrac{x^{*}}{2}}(\xi-1)  \Bigg],\\\nonumber 
	&=&  1 + \sqrt{2x^{*}}\left(\dfrac{1}{\psi}+\psi\right), 
	\end{eqnarray}} 
where for constant $x^{*}$, {\small$\psi= \sqrt{{x^{*}}/{2}}\ (\xi-1)$} is a monotonic increasing function over $\xi$. Hence the system delay is convex w.r.t. $\xi$ and hence $q_{s}$, and the minimum value $\overline{T}=1+2\sqrt{2x^{*}}$ which is attained at $\xi=1+\sqrt{2/x^{*}}$ (after solving for $\psi=1$).
\par Of interest is the feasible range of $\xi$ for specified $\overline{T}_{max}$. Using (\ref{eqn:x_convex_delay}), we can write: $\xi_{min}\leq\xi\leq\xi_{max}$ where:
{\small\begin{eqnarray*}
		\begin{array}{lll}
		 \xi_{min}&=&\,1+\dfrac{1}{2x^{*}}\left[\overline{T}_{max}^{*}-1-\sqrt{(\overline{T}_{max}^{*}-1)^2-8x^{*}}\right],\\
		 \xi_{max}&=&\,1+\dfrac{1}{2x^{*}}\left[\overline{T}_{max}^{*}-1+\sqrt{(\overline{T}_{max}^{*}-1)^2-8x^{*}}\right].
		\end{array}
	\end{eqnarray*}}
Also of interest is the feasible range of $\xi$ to ensure a minimum desired throughput $\overline{\tau}_{min}^{*}$. Imposing  $\overline{\tau}=1/(2+\overline{T}_{u})\geq\overline{\tau}_{min}^{*}$ and using $\overline{T}_{u}=x^{*}(\xi-1)$, we get:
{\small\begin{eqnarray*}
		\begin{array}{lll}
			\xi_{max\overline{\tau}}=\,1+\dfrac{1}{x^{*}}\left[\dfrac{1}{\overline{\tau}_{min}^{*}}-2\right].
		\end{array}
	\end{eqnarray*}}
To ensure maximum delay of $\overline{T}_{max}^{*}$ and minimum throughput $\overline{\tau}_{min}^{*}$, the constraint on $\xi$ becomes: $\xi_{min}\leq\xi\leq\min(\xi_{max},\xi_{max\overline{\tau}})$.   To obtain a  constraint on feasible values of $\overline{T}_{max}^{*}$ and $\overline{\tau}_{min}^{*}$  we solve $\xi_{min}\leq\xi_{max\overline{\tau}}$ to get:
{\small\begin{eqnarray*}
		\begin{array}{lll}
			(\overline{T}_{max}+3)\overline{\tau}_{min}\leq2+\overline{\tau}_{min}(\sqrt{(\overline{T}_{max}^{*}-1)^2-8x^{*}}).
		\end{array}
	\end{eqnarray*}}
\par Consider first the case  when $x^{*}=1$. In this case, $(\xi_{c}=\xi)$, $q_{c}=q_{s}$ and $\overline{T}=1+2\sqrt{2}$ at $\xi=1+\sqrt{2}$ as in \cite{Islam2015}. Note that when $x^{*}<1$ so that $\xi_{c}<\xi$ (i.e. $q_{c}>q_{s}$), $\overline{T}_{u}$ in (\ref{eqn:delay_inf}) decreases w.r.t. the case when $x^{*}=1$, thereby increasing throughput as is evident from (\ref{eqn:tau_Tu}).\\
{\bf Constant Throughput (CT) Scheme}:\\ In some situations, we might want to ensure a constant throughput $\overline{\tau}^{*}$. Substituting $\overline{T}_{u}=\xi_{c}(1-\xi^{-1})$ from (\ref{eqn:delay_inf_tq_tu}) in (\ref{eqn:tau_Tu}) and noting that  $\overline{T}_{o}=0$ ($L$ infinite), we have:
%
{\small\begin{eqnarray}
\begin{array}{lll}\label{eqn:x_const_tau}
2 = \dfrac{1}{\overline{\tau}^{*}}-\overline{T}_{u}&\Rightarrow&\xi_{c} = \dfrac{1}{1-\xi^{-1}} \dfrac{1-2{\overline{\tau}^{*}}}{\overline{\tau}^{*}}.\\
\end{array}
\end{eqnarray}}
To obtain the range of $\xi$ for maximum delay $\overline{T}_{max}$, we use (\ref{eqn:T_q}) and (\ref{delay_throughput}) to get:
{\small\begin{eqnarray*}
	\begin{array}{lll}
	\overline{T}_{q}\leq\overline{T}_{max}^{*}-\dfrac{1}{\overline{\tau}^{*}}+2\Rightarrow \xi\geq\,1+\dfrac{1}{2}\dfrac{\overline{\tau}^{*}}{\overline{\tau}^{*}(1+\overline{T}_{max}^{*})-1}.
	\end{array}
	\end{eqnarray*}}
It is apparent that since $\overline{T}_{u}$ is constant in CT scheme, the maximum delay constraint is met when we starve the buffer.\\

{\em SER Analysis}
\par Analysis of SER performance, and tradeoffs between SER, throughput and delay are of interest. However, the non-linear nature of the  LSP (evaluated using (\ref{eqn:CCDF_SR})) present in these expressions  makes it  difficult to understand the impact of $\rho$ on system delay and throughput. Also computer simulations are required to understand the choice of system parameters $q_c$, $q_d$ and $q_s$  that ensure a minimum SER while achieving the target delay and throughput.
  However, if we restrict our attention to the PIP case (i.e.  $p_{s}=p_{r}=1$), some useful insights can be obtained on choice of $q_c$ and $q_s$ from $\overline{\mathcal{P}}'_{s}$. Note that for an infinite-sized buffer, $q_{d}$ does not impact the SER. Expression for ${\small\overline{\mathcal{P}}_{s,asym}}$ can be found in Table~\ref{tab:CCDF_E_LSP_ASER_AR} to be: {\small$\overline{\mathcal{P}}_{s,asym}\,\approx  \dfrac{3 \varphi}{4\eta^2}\dfrac{\rho}{q_{s}\mu_{s}\mu_{r}}$}. An expression for ${\small\overline{\mathcal{P}}_{c,asym}}$ can be found by replacing $q_s$ by $q_c$ to be: {\small$\overline{\mathcal{P}}_{c,asym}\,\approx  \dfrac{3 \varphi}{4\eta^2}\dfrac{\rho}{q_{c}\mu_{s}\mu_{r}}$}. Using these expressions in  (\ref{eqn:SER_SR_Markov}), $\overline{\mathcal{P}}_{s,asym}^{\,'}$ can be written  as:
 {\small\begin{eqnarray}\label{eqn:SER_Markov_asym} 
 	\hspace{-0.1cm}\begin{array}{lll}
 	\overline{\mathcal{P}}_{s,asym}^{\,'} &\approx& \dfrac{3 \varphi}{4\eta^2}\dfrac{1}{\mu_{s}\mu_{r}}\left[\dfrac{\xi^{-1}\rho}{q_{s}}+\dfrac{(1-{\xi^{-1}})\rho_{c}}{q_{c}}\right],
 	\end{array}
 	\end{eqnarray}}
 where the LSP of $\SSS-\SR$ link in the PIP case is listed in Table~\ref{tab:CCDF_E_LSP_ASER_AR} as:
 {\small\begin{eqnarray}\label{eqn:ps_sr_asym} 
 \begin{array}{lll}
 q_{s}=\hspace{-0.1cm}\dfrac{-\rho\mu_{s}}{\mu_{r}-\rho \mu_{s}}-\dfrac{\rho\mu_{s}\mu_{r}}{(\mu_{r}-\rho \mu_{s})^2}\ln\dfrac{\rho \mu_{s}}{\mu_{r}}.
 \end{array}
 \end{eqnarray}}
 The nonlinear dependence makes determination of $\rho$ from the LSP $q_{s}$  very difficult\footnote{For a starving buffer, $\rho$ can be evaluated from the series expression of $q_{s}$ , which is given in Appendix C, using series reversion \cite[eq. (3.6.25)]{Abramowitz1965}.}. However, it is shown in Appendix C that for buffer starvation ($\xi\geq1$), $q_{s}$ (hence $\rho$) can be approximated as follows ($\rho_{c}$ is approximated in the similar fashion):
 \begin{eqnarray}\label{eqn:ps_sr_asym2} 
 \begin{array}{lll}
 q_{s} \hspace{-0.1cm}\approx\dfrac{\rho\mu_{s}}{\mu_{r}+\rho\mu_{s}} \Rightarrow \rho\approx\xi^{-1}\dfrac{\mu_{r}}{\mu_{s}},\text{ similarly }\, \rho_{c}\approx\xi_{c}^{-1}\dfrac{\mu_{r}}{\mu_{s}}.
 \end{array}
 \end{eqnarray}
 Substituting the approximations for $\rho$ and $\rho_{c}$ in (\ref{eqn:SER_Markov_asym}) and after some manipulations, we get:
 \small{\begin{eqnarray}\label{eqn:x_slope_control} 
 	\hspace{-0.2cm}\begin{array}{lll}
 	\overline{\mathcal{P}}_{s,asym}^{\,'}\hspace{-0.2cm} &\approx&\hspace{-0.2cm} \dfrac{3 \varphi}{4\eta^2}\dfrac{1}{\mu_{s}^{2}}\left[2-(1 - \xi^{-1})\varepsilon\right],\\
 	 \hspace{1cm}\varepsilon&=& 1+\xi^{-1}-\xi_{c}^{-1},\quad \text{or }\xi_{c}= (1+\xi^{-1}-\varepsilon)^{-1}
 	\end{array}
 	\end{eqnarray}}\normalsize
 where $\varepsilon$ is treated as the parameter which provides the choice of $\xi_{c}$ in controlling {\small$\overline{\mathcal{P}}_{s,asym}^{\,'}$}. Since $\xi>1$ in starving scenario, we need to use $\varepsilon\geq 0$ to minimize SER. A large value of $\varepsilon$ cause $\xi_c$ to be large, which decreases $q_c$, and causes loss in throughput. It is clear that for $\varepsilon=1$, {\small $\overline{\mathcal{P}}_{s,asym}^{\,'}=\overline{\mathcal{P}}_{s,asym}=\dfrac{3 \varphi}{4\eta^2\mu_{s}^2q_{r}}$} (where we have used the fact that $1/(1+\xi^{-1}) = q_r/(q_r + q_s) = q_r$). 
 \par For the MDMT scheme,   substituting $\xi_{c}=\xi x^{*}$ in $\varepsilon=1+\xi^{-1}-\xi_{c}^{-1}\geq 0$, we get $\frac{1-x^{*}}{x^{*}}\leq \xi < \infty$. Since  $\xi\geq1$ in a starving buffer,  $x^{*}\geq0.5$ for $\varepsilon\geq 0$, which ensures that ${\small \overline{\mathcal{P}}_{s,asym}^{\,'}}\leq \dfrac{3 \varphi}{2\eta^2}\dfrac{1}{\mu_{s}^{2}}$. Note that when $x^{*}<0.5$, the SER cannot be bounded similarly over the entire range of $\xi$. For the CT scheme,  substituting $\xi_{c}$ from (\ref{eqn:x_const_tau}) for $\varepsilon=1+\xi^{-1}-\xi_{c}^{-1}\geq 0
 $, we get {\small$1\leq \xi\leq\dfrac{1-\overline{\tau}*}{3\overline{\tau}*-1}$}. 
It is clear that $\overline{\tau}^{*}\leq1/3$ implies that  $\varepsilon\geq0$, which ensures that ${\small \overline{\mathcal{P}}_{s,asym}^{\,'}}\leq \dfrac{3 \varphi}{2\eta^2}\dfrac{1}{\mu_{s}^{2}}$ over the entire range of $1\leq \xi \leq \infty$. When $\overline{\tau}^{*}>1/3$, the range of $\xi$ is clearly limited.
\section{Simulation Results}\label{sec:SimRes}
In this section, we evaluate the average rate and SER performance by simulations, and compare the same with the derived analytical expressions.  We assume $\gamma_{p}=10$ dB.

\begin{figure*}
	\vspace{-0.2 in}
	\centering
	\begin{minipage}{.3\textwidth}
		\centering
		\includegraphics[height=2.2 in, width =2.3 in]{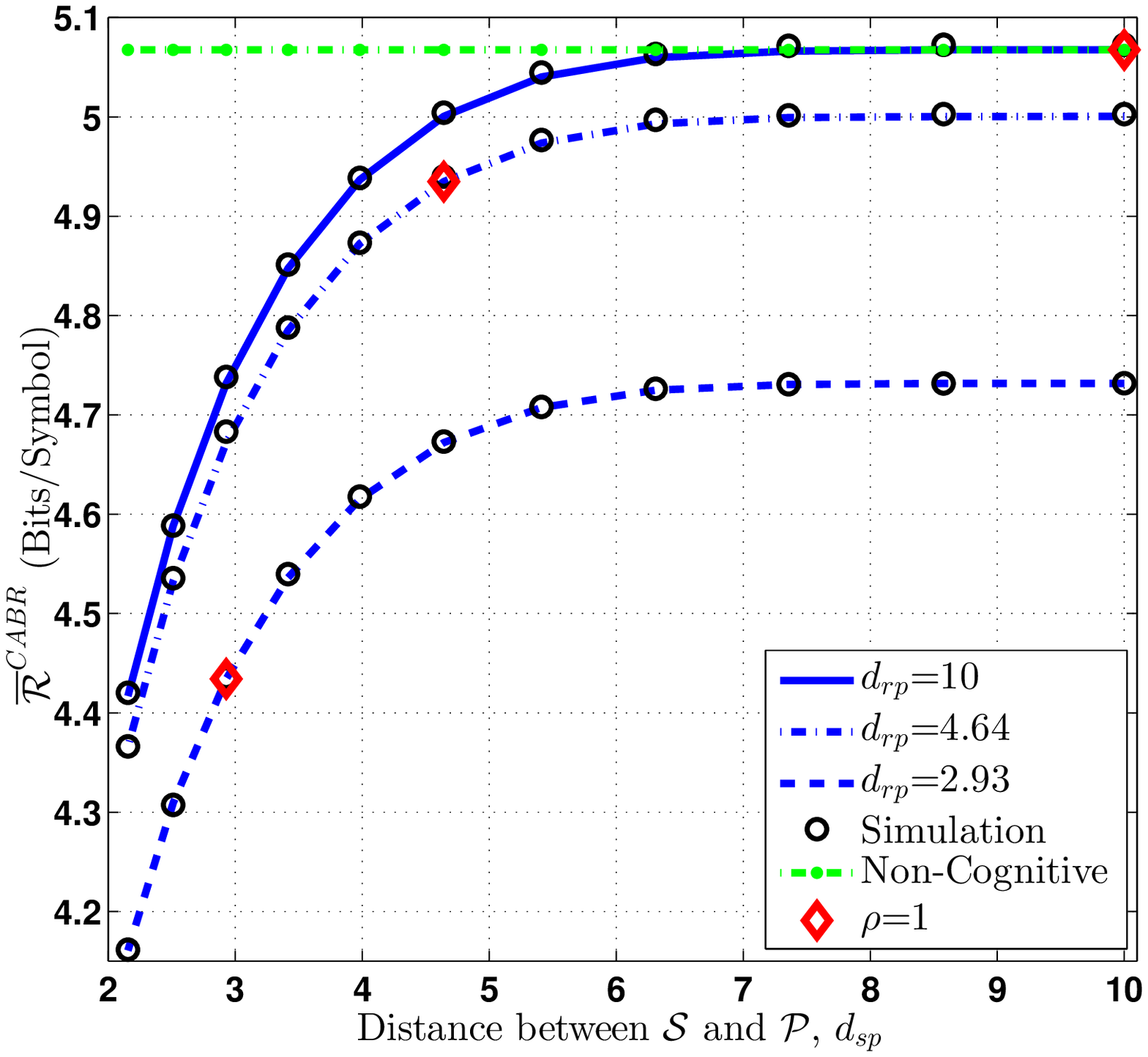}
		\caption{Achievable rate of CABR vs $d_{sp}$, cf. (\ref{eqn:rate_expr})
			$\gamma_{max}=30\,dB$, $\gamma_{p}=10\,dB\,,\,\Omega_{hs}=\Omega_{hr}=1$.}
		\label{fig:OptimumThroughput}
	\end{minipage}\hfill
	\begin{minipage}{.3\textwidth}
		\centering
		\includegraphics[height=2.2 in, width =2.3 in]{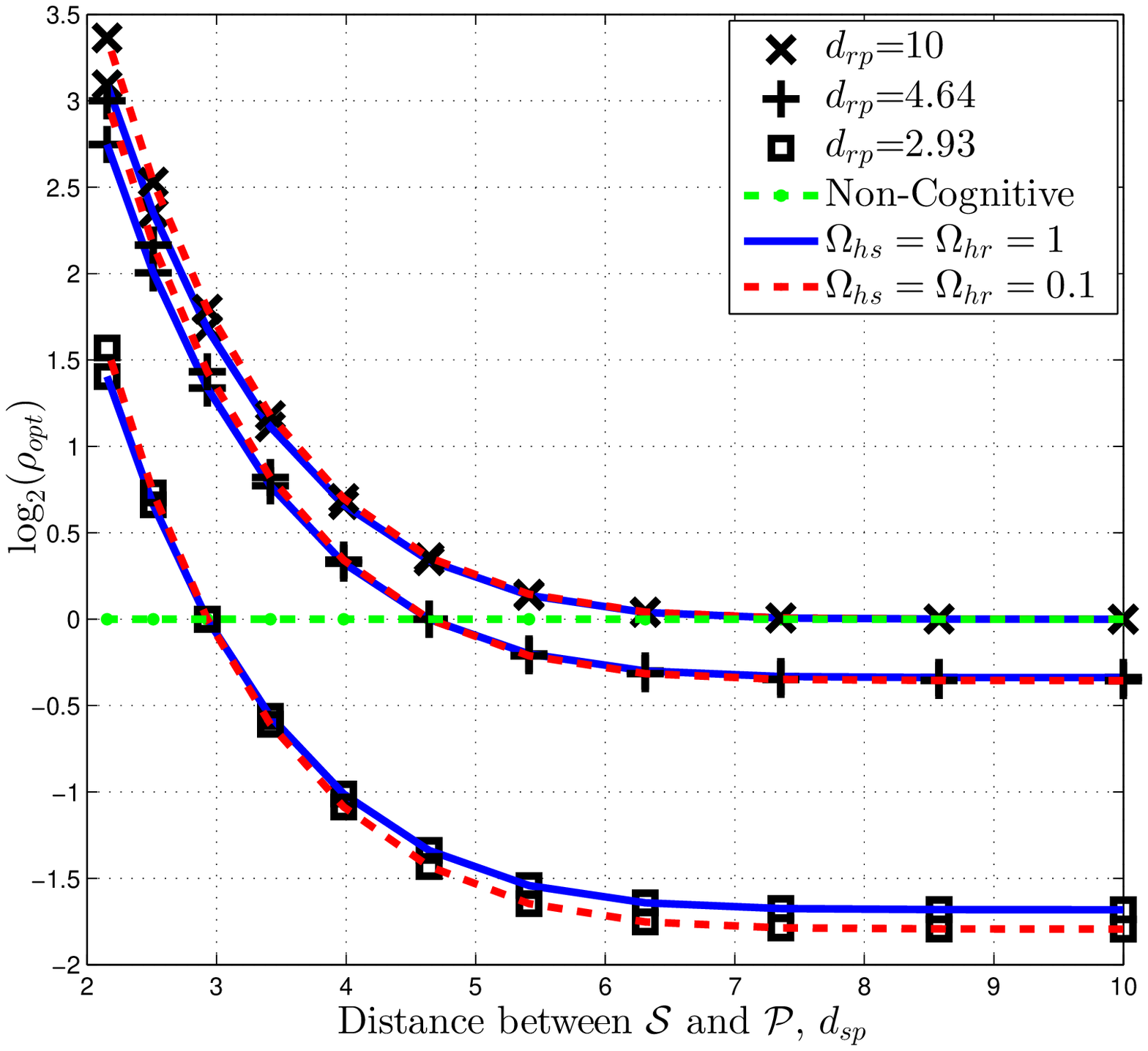}
		\caption{Link selection parameter $\rho$ of CABR vs $d_{sp}$, cf. (\ref{eqn:rate_expr}) $\gamma_{max}=30\, dB$, $\gamma_{p}=10\,dB$, $\alpha=3$.}
		\label{fig:OptimumRho} 
	\end{minipage}\hfill
	\begin{minipage}{.3\textwidth}
		\centering
		\includegraphics[height=2.2 in, width =2.3 in]{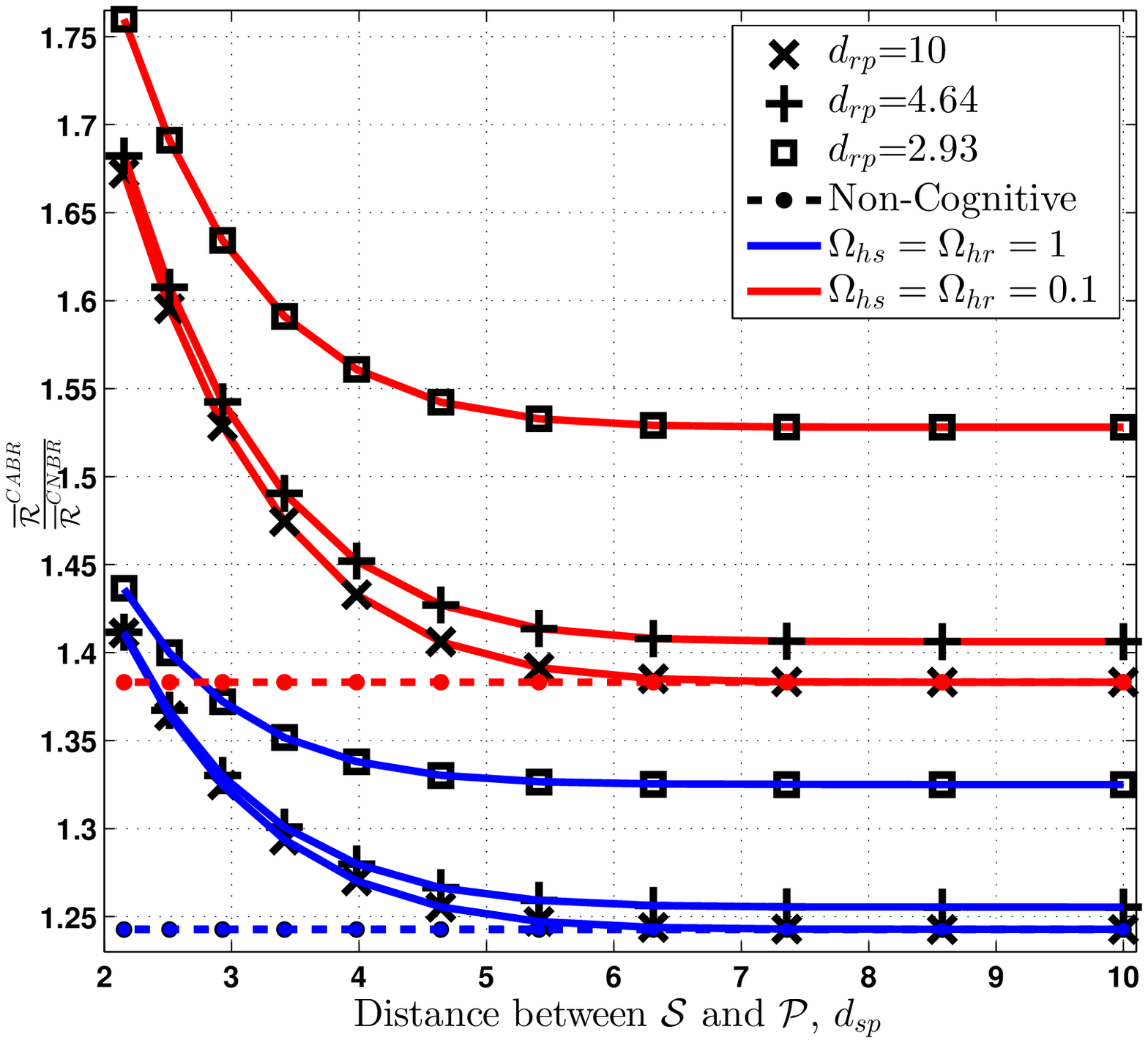}
		\caption{ Ratio of achievable rate of CABR w.r.t. CNBR, cf. (\ref{eqn:rate_expr}), (\ref{eqn:AR_CNBR1}), (\ref{eqn:AR_CNBR2}), $\gamma_{max}=30\,dB$, $\gamma_{p}=10\,dB$, $\alpha=3$.} 
		\label{fig:IncreaseInThroughput}
	\end{minipage}
	\begin{minipage}{.3\textwidth}
		\centering
		\includegraphics[height=2.2 in, width =2.3 in]{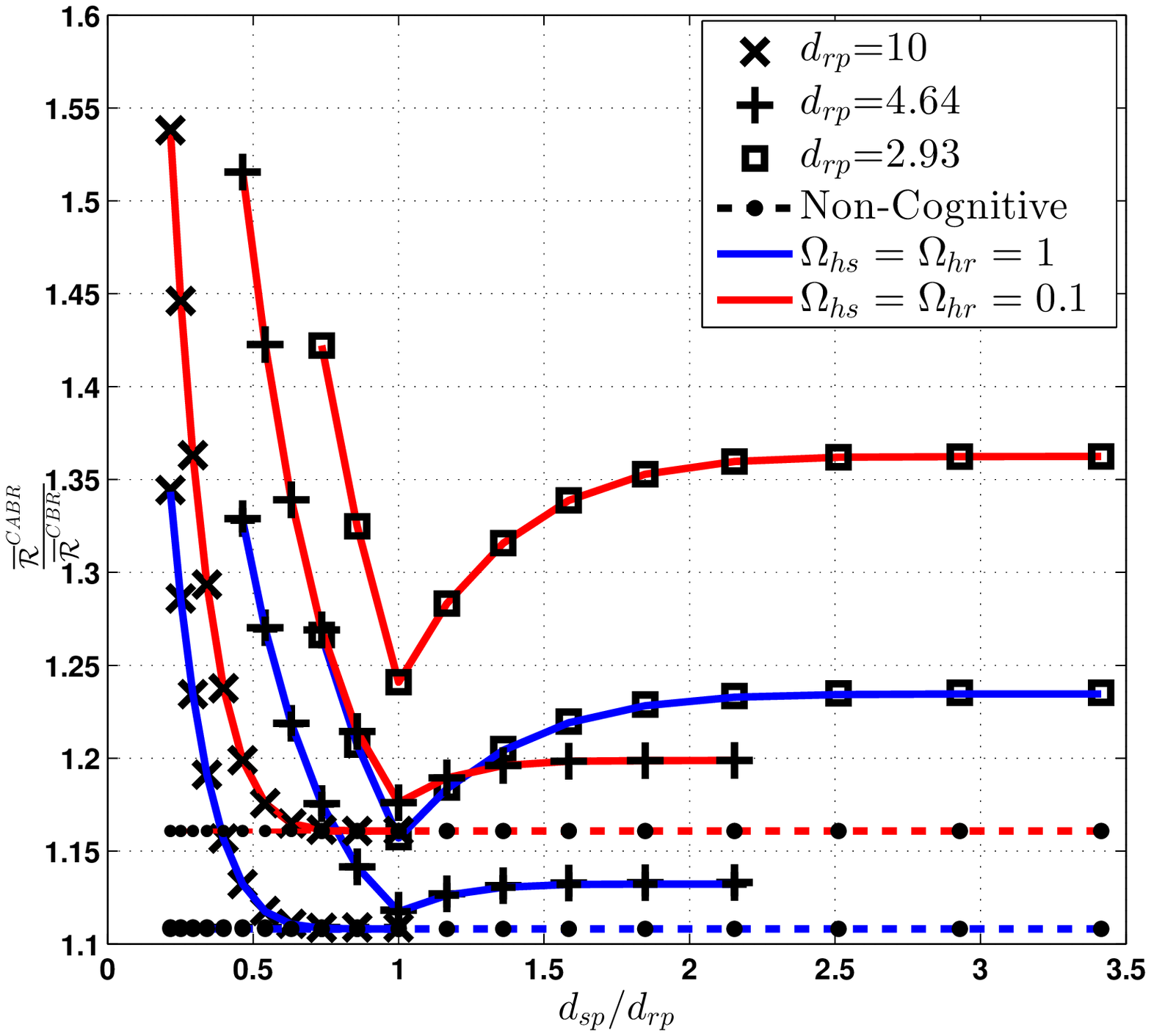}
		\caption{ Ratio of Achievable Rate of CABR w r t CBR, cf. (\ref{eqn:rate_expr}), (\ref{eqn:AR_CBR}) $\gamma_{max}=30\,dB$, $\gamma_{p}=10\,dB$, $\alpha=3$.} 
		\label{fig:IncreaseInThroughput2}
	\end{minipage}\hfill
	\begin{minipage}{.3\textwidth}
		\centering
		\includegraphics[height=2.2 in, width =2.3 in]{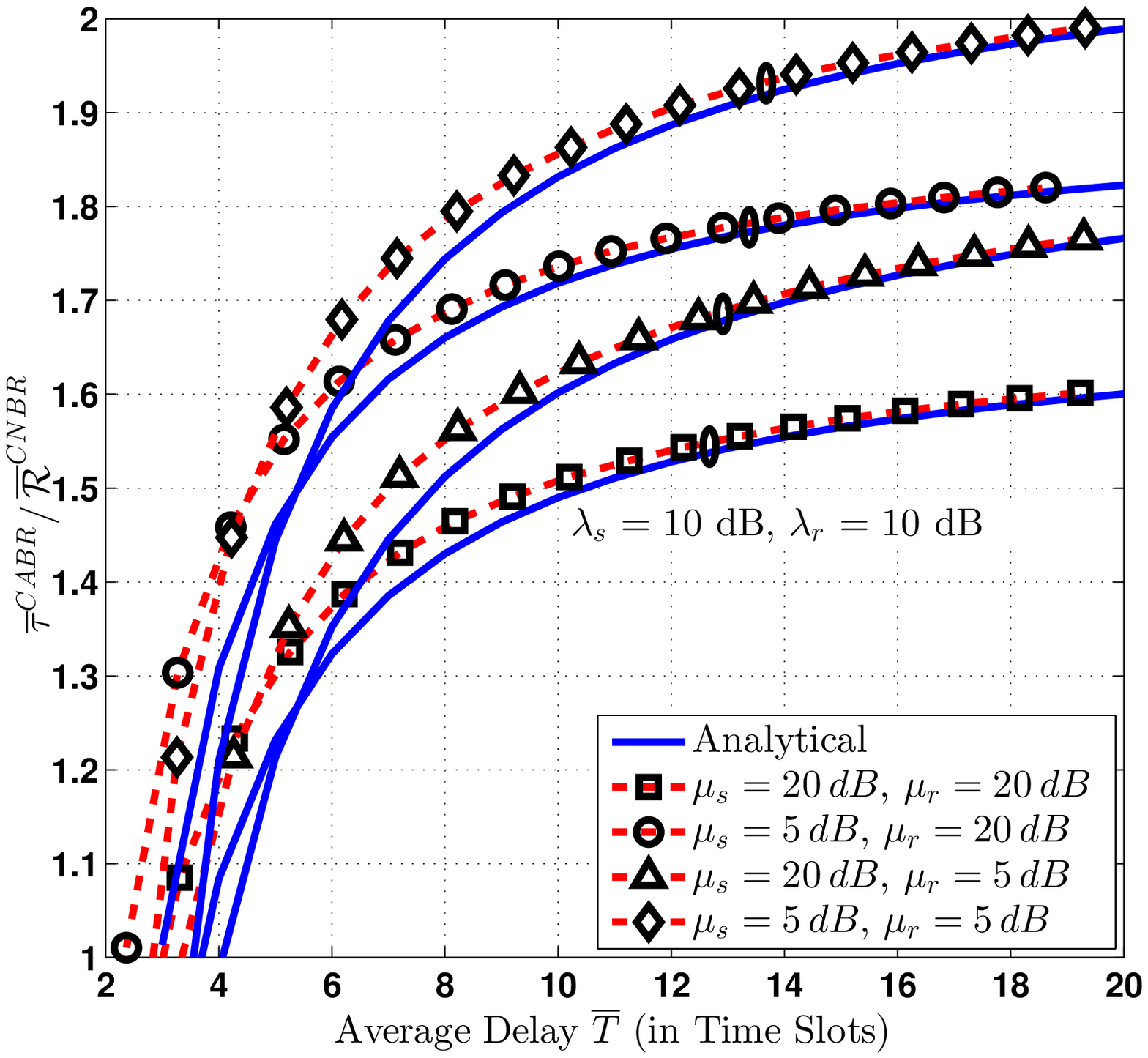}
		\caption{Ratio of throughput of CABR w.r.t.  CNBR, cf. (\ref{eqn:rate_expr}), (\ref{eqn:AR_CNBR1}), (\ref{eqn:AR_CNBR2}), where $\rho$ is chosen to satisfy delay constraint cf. (\ref{eqn:delay_expr}). }  
		\label{fig:PerGainPerTimeSlot} 
	\end{minipage}\hfill
	\begin{minipage}{.3\textwidth}
		\centering
		\includegraphics[height=2.2 in, width =2.3 in]{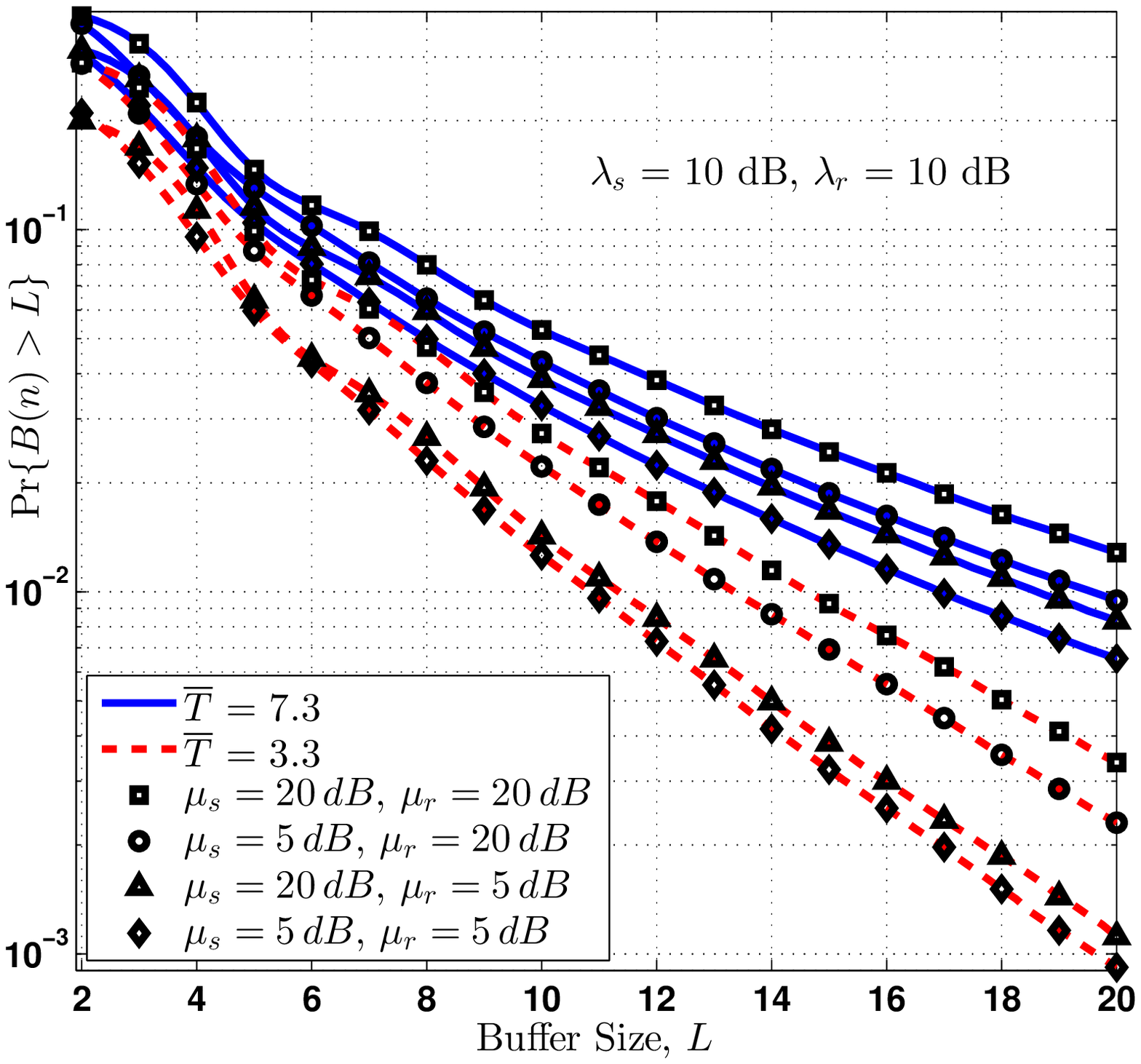}
		\caption{Probability of a buffer overflow vs  buffer size $L$, assuming delay is constrained cf. (\ref{eqn:delay_expr}) to $7.3$ and $3.3$.}  
		\label{fig:ProbDropBitsVsQueueSize}
	\end{minipage}
	\begin{minipage}{.3\textwidth}
		\centering
		\includegraphics[height=2.2 in, width =2.3 in]{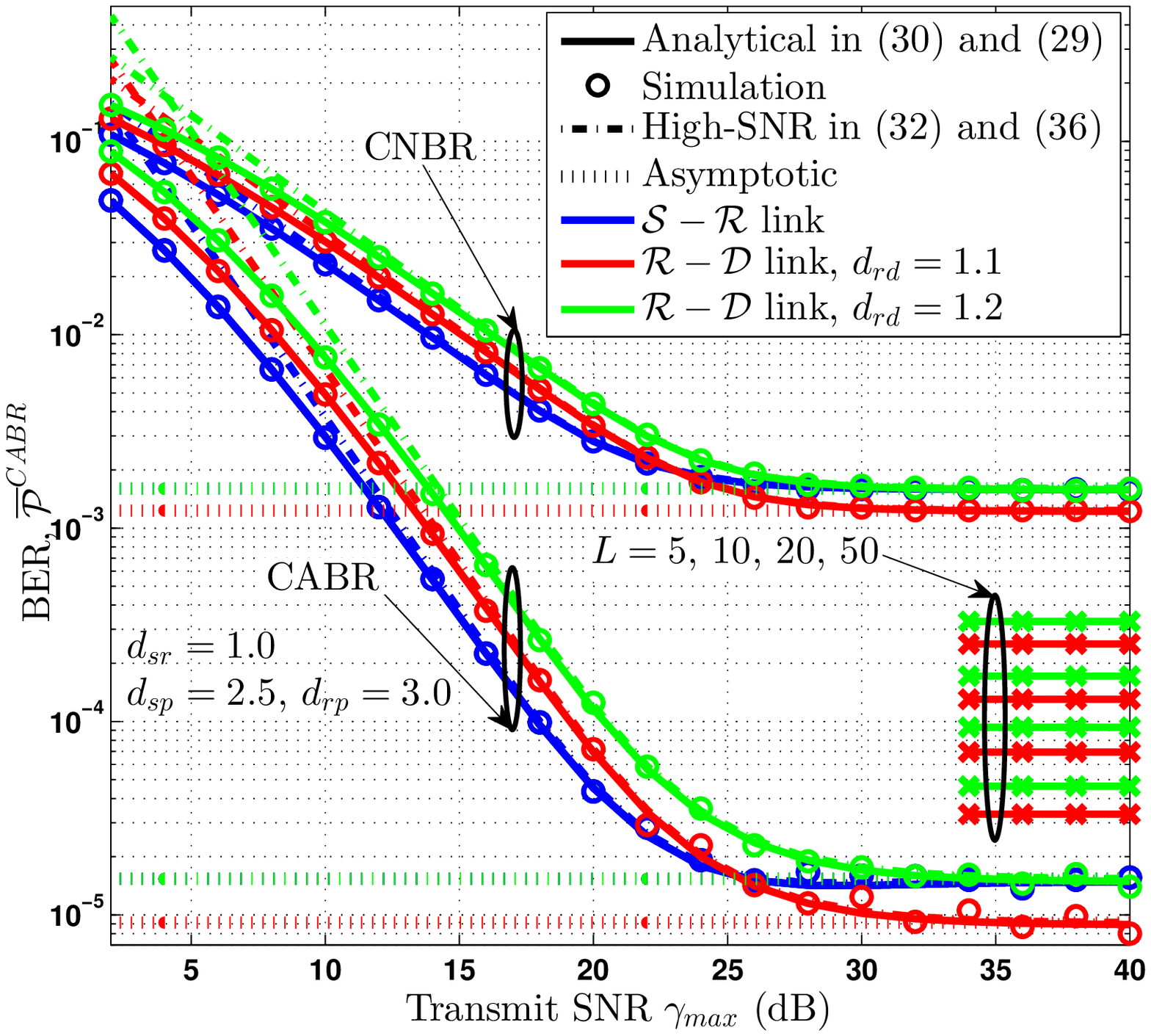}
		\caption{BER (versus transmit SNR) of $\SSS-\SR$ and $\SR-\SD$ link , as well as high-SNR expressions, and asymptotes.}
		\label{fig:ber_s_r_sym}
	\end{minipage}\hfill
	\begin{minipage}{.3\textwidth}
		\centering
		\includegraphics[height=2.2 in, width =2.3 in]{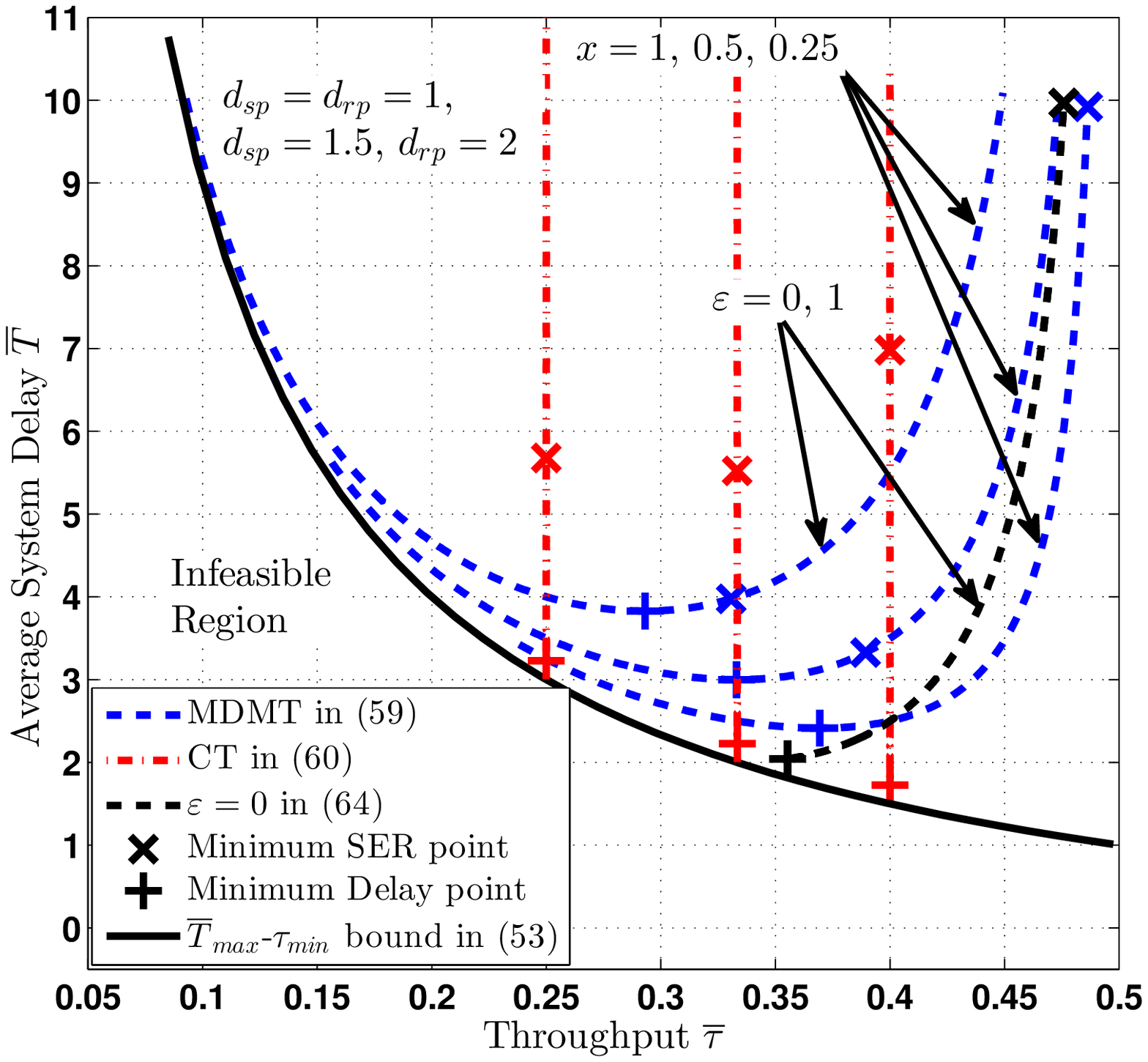}
		\caption{System Delay vs throughput, c.f. (\ref{eqn:delay_inf}) and (\ref{eqn:tau_Tu}) for PIP case when $L\rightarrow\infty$ c.f. (\ref{eqn:sear_pace_const_tau}), (\ref{eqn:x_convex_delay}), (\ref{eqn:x_const_tau}) and (\ref{eqn:x_slope_control}).} 
		\label{fig:DelayvsTauFixedRate}
	\end{minipage}\hfill
	\begin{minipage}{.3\textwidth}
		\centering
		\includegraphics[height=2.2 in, width =2.3 in]{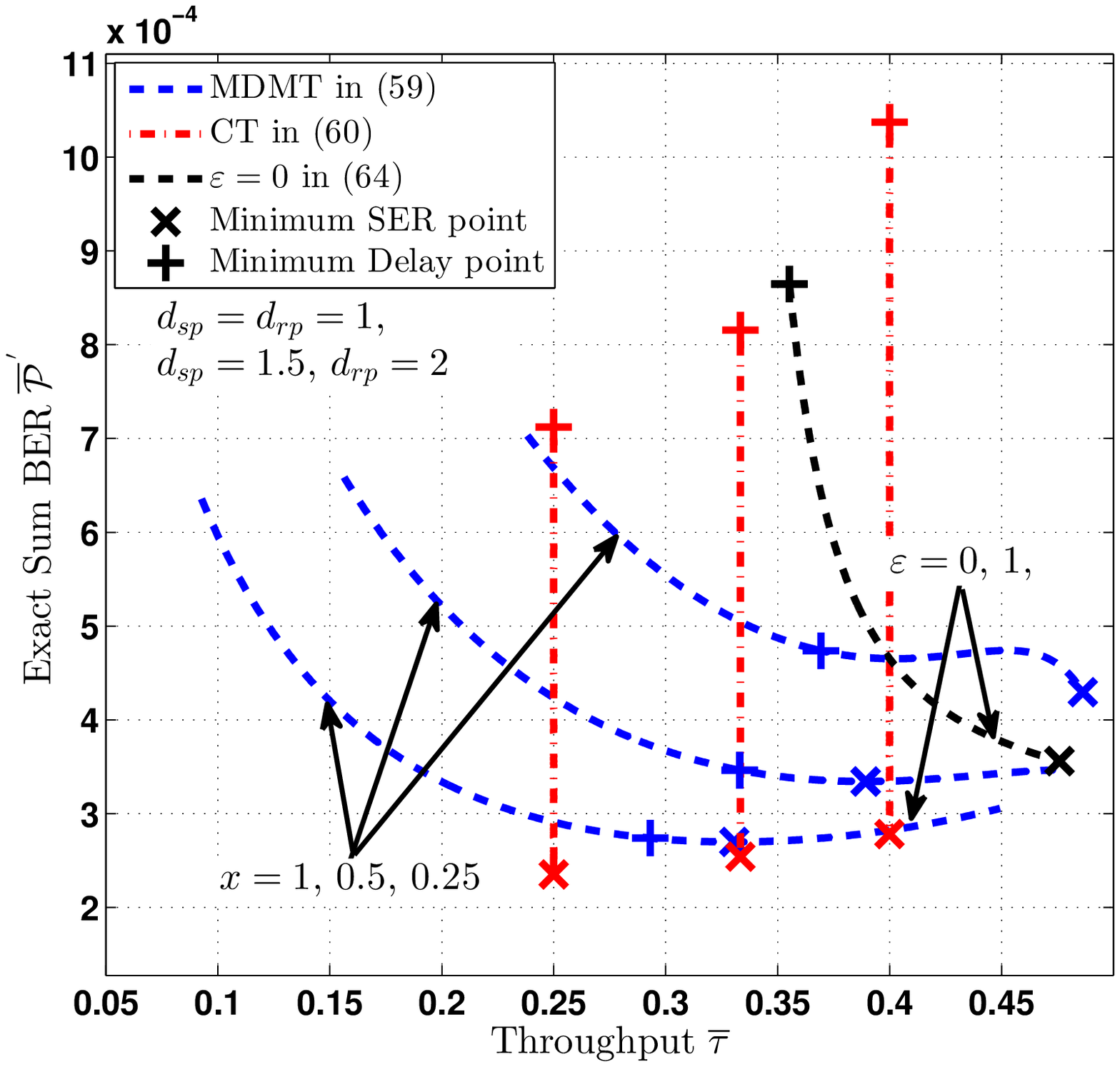}
		\caption{Exact Sum BER vs throughput, c.f. (\ref{eqn:SER_SR_Markov}) and (\ref{eqn:tau_Tu}) for PIP case when $L\rightarrow\infty$ c.f. (\ref{eqn:x_convex_delay}), (\ref{eqn:x_const_tau}) and (\ref{eqn:x_slope_control}).}
		\label{fig:BERvsTauFixedRate}
	\end{minipage}
	\vspace{-0.1 in}
\end{figure*}
\par  We first plot the average rate performance of the CABR scheme with an infinite-size buffer, and compare it with conventional schemes (CNBR and CBR). We assume $\gamma_{max}=30dB$ and $\Omega_{hs}=\Omega_{hr}=1$ for carrying out the simulations. Fig.~\ref{fig:OptimumThroughput} depicts the average rate of the CABR scheme versus $d_{sp}$ (the distance of $\SSS$ from $\PD$) for various $d_{rp}$ values. It can be seen that the average rate is larger for larger $d_{rp}$ for the same $d_{sp}$. For the same $d_{rp}$, the average rate saturates for higher $d_{sp}$ and does not improve further unless $d_{rp}$ is increased (thereby improving the second link performance). When $d_{sp}$ and $d_{rp}$ are both very large, the system essentially becomes non-cognitive (both $\SSS$ and $\SR$ are in the PTP regime), and corresponds to the system considered in \cite{Zlatanov2013_1}. The rate of this non-cognitive system is also plotted in the figure.

\par In Fig.~\ref{fig:OptimumRho}, $\log_{2}(\rho_{opt})$ is plotted  versus $d_{sp}$ for various $d_{rp}$. It is clear that when $d_{sp}=d_{rp}$, $\rho=1$. When $d_{rp}>d_{sp}$, $\rho>1$ and when $d_{rp}<d_{sp}$, $\rho<1$. When $d_{sp}$ and $d_{rp}$ are very large (corresponding to the non-cognitive system in (\cite{Zlatanov2013_1})), $\rho=1$. When the quality of $\SSS-\SR$ channel link is poor ($d_{sp}$ is small), the bottleneck link needs to be selected more often, thereby increasing $\rho_{opt}$. In the simulations, since we have assumed $\Omega_{hs}=\Omega_{hr}$, $\rho$ is not influenced by their value.
\par Fig.~\ref{fig:IncreaseInThroughput} depicts the rate improvement of CABR w.r.t. CNBR versus $d_{sp}$ (cf. (\ref{eqn:rate_expr}), (\ref{eqn:AR_CNBR1}) and (\ref{eqn:AR_CNBR2})). It can be seen that the improvement is highest when $d_{rp}$ is smallest. {\em This clearly demonstrates that adaptive link-selection is highly beneficial in the interference constrained scenarios typically encountered in cognitive radio systems.}  
\par In Fig.~\ref{fig:IncreaseInThroughput2}, the ratio of rates  of CABR to that of CBR is plotted versus $d_{sp}/d_{rp}$, cf. (\ref{eqn:rate_expr}) and (\ref{eqn:AR_CBR}). It is clear that the ratio saturates for larger $d_{sp}$, and has a minimum when $d_{sp}=d_{rp}$. Although the average rate itself decreases, the ratio always improves when the channel between $\SSS-\SR$ and $\SR-\SD$ degrades (for both CNBR and CBR schemes).
\par Fig.~\ref{fig:PerGainPerTimeSlot} shows the rate ratio of CABR and CNBR schemes when the average delay is bounded as in (\ref{eqn:delay_expr}) with an infinite-size buffer. It is evident from the figure that the when the average delay is constrained to be small, the gain with buffer use is also small, though it continues to be greater than one. It is also seen from the figure that performance of the asymmetric link is poor in the case  when $\SR-\SD$ link is poor. This is because of buffer starving (use of LIFO buffer alleviates this problem).
\par Fig.~\ref{fig:ProbDropBitsVsQueueSize} shows the probability of buffer overflow with buffer of size $L$  ($\Pr\{B(n)>L\}$) when the average delay $\overline{T}$ is fixed at $7.3$ and $3.3$ as per (\ref{eqn:delay_expr}). It is clear from the figure that a stronger channel to the primary receiver leads to smaller chance of overflow since the average size of the buffer also becomes smaller. 
\par  For SER analysis with fixed-rate transmission, we consider a BPSK modulation scheme so that $\eta=2$, $\varphi=1$, and $R_s=R_r=1$ bits per channel use (bpcu). We plot the SER performance of CABR and CNBR schemes, and compare the same with simulation results. We consider a "symmetric" channel (ratio of the distance of secondary transmitters from respective secondary receiver to that of primary receiver $\PD$ is constant), for which   $\mu_{s}=\mu_{r}$. To ensure symmetry, we choose $\Omega_{hs}=1$ and $\Omega_{hr}=0.5787\,(d_{rd} =1.2)$ (which makes $\mu_s=\mu_r=156.25$). We also consider an "asymmetric" channel  where we choose $\Omega_{hs}=1$ and $\Omega_{hr}=0.751\,(d_{rd} =1.1)$ ($\mu_s=156.25$ and $\mu_r=202.50$). Figs.~\ref{fig:ber_s_r_sym} depicts the SER of $\SSS-\SR$ and $\SR-\SD$ links for the CABR and CNBR schemes in the symmetric and asymmetric cases. It can be seen that the analytical results match perfectly with the simulation results. It is evident from Fig.~\ref{fig:ber_s_r_sym} that the CABR scheme outperforms the CNBR scheme even in the PIP regime. This validates the observation from  (\ref{eqn:SER_ASYM_CBR}) and (\ref{eqn:SER_ASYM_CABR}) that in the PIP regime, the SER of CNBR scheme depends on the inverse of average SNRs of $\SSS-\SR$ and $\SR-\SD$  links, whereas it depends on the inverse of the square of the corresponding SNRs in the case of the  CABR scheme. 
\par In Fig.~\ref{fig:ber_s_r_sym}, we have also shown and verified through analytical (using (\ref{eqn:p_fin_rho_opt}) and (\ref{eqn:Delay_FinMark_L_fin_rho_opt})) and simulation respectively the impact of finite buffer on SER performance in the PIP regime (note the marked improvement as buffer size increases). In these simulations, $\SSS$ transmits ($\SR$ transmits) whenever the buffer is empty (full) i.e. $q_c=1$ ($q_d=1$). It is clear from (\ref{eqn:Delay_FinMark_L_fin_rho_opt})) that $\overline{T}_{q}=L$ time-slots whereas $\overline{T}_{u}=\overline{T}_{o}=0$. Note also that in the low to medium SNR range, the SER of the $\SSS-\SR$ link channel is better than that of $\SR-\SD$  since $\Omega_{hs}>\Omega_{hr}$. It  is clear from (\ref{eqn:link_sel_pr_half}) that the value of $\rho=\lambda_{r}/\lambda_{s}$ is less than one in the PTP regime, which implies that the  $\SSS-\SR$ link is chosen less frequently.
As we move towards the PIP regime, the  SERs of both the links converge to the same floor in Fig.~\ref{fig:ber_s_r_sym}. Since we consider a symmetric case where  $\mu_{s}=\mu_{r}$, the value of $\rho=1$ in the PIP regime, and both links are chosen equally. On the contrary, the floor of  $\SR-\SD$ link is lower than that of the $\SSS-\SR$ link since $\mu_{r}>\mu_{s}$. Hence $\SR-\SD$ channel is better than $\SSS-\SR$ in the PIP regime and is chosen less frequently, which is contrary to the situation in the PTP regime (since $\Omega_{hs}>\Omega_{hr}$). The SERs crossover at one SNR, and at the intersection point, $\rho=1$.
\par In Fig.~\ref{fig:ber_s_r_sym} , it can be seen that the asymptotes at low to medium SNR match closely with the actual SER. It is also clear that SER of the CABR scheme exhibits a higher slope in this range. At high SNRs, the SER exhibits a floor, and the asymptotes of SER at high SNR match closely with simulations.  \par Fig.~\ref{fig:DelayvsTauFixedRate} and Fig.~\ref{fig:BERvsTauFixedRate} show the average system delay and gross BER vs. throughput for the modified threshold-based transmission protocol depicted in Fig.~\ref{fig:fig_1} when both $\SSS-\SR$ and $\SR-\SD$ links are in the PIP regime. However, to validate the analysis in high interference condition, we choose $d_{sr}=d_{rd}=1$ and $d_{sp}=1.5$ and $d_{rp}=2.0$ for the result (i.e $\mu_{s}=33.75$ and $\mu_{r}=80$). We choose various values of $\xi_{c}$ such that the required constraint for a particular scheme is satisfied (i.e. $\xi_{c}$ is chosen from (\ref{eqn:x_convex_delay}), (\ref{eqn:x_const_tau}) or (\ref{eqn:x_slope_control})). First we plot the inequality given in (\ref{eqn:sear_pace_const_tau}), which  constrains the throughput and delay. We also plot the delay and BER performance for $\varepsilon=0$ where BER of $\SSS-\SR$ link is independent of $\xi$ (see (\ref{eqn:x_slope_control})). We further plot the delay and BER performance of the MDMT scheme for $x^{*}=1, 1/2,1/4 $ and mark the points for minimum BER and minimum delay.  The minimum delay values are $ 1+2\sqrt(2),\,3,\,1+\sqrt{2} $   which occur at ${\small\overline{\tau}=\dfrac{1}{2+\sqrt{2}},\,\dfrac{1}{3},\,\dfrac{\sqrt{2}}{1+2\sqrt{2}}}$ respectively. It is clear that performance with  $x^{*}=0.5$ approaches that with $\varepsilon=0$ when  $\xi\rightarrow1$. For the CT scheme, we plot the BER and mark the point for minimum BER and delay. As we can see in the figure, that SER can be improved by choosing lower value of throughput. The minimum SER is always less than that of the MDMT scheme with $x^{*}=1$, whereas the delay is larger.  It is obvious that since throughput varies inversely w.r.t. to underflow delay, the delay in CT scheme is not convex and minimum is achieved at the bound implied by (\ref{eqn:sear_pace_const_tau}). Also, since $\xi_{c}$ decreases with increasing $\xi$, the SER degrades and is maximum at the throughput delay floor implied by (\ref{eqn:sear_pace_const_tau}).
As we can see that for $\overline{\tau}=2/5$ and $x=0.25$, $\varepsilon=0$ for $\xi=3$. Hence for $\overline{\tau}=2/5$ and $x=0.25$, the BER of $\SSS-\SR$ link is bounded w.r.t. $\xi$ for $\xi\leq3$ and $\xi\geq3$ respectively.
\par It is to be observed that  the average rate, sum BER and average delay remain the same for the LIFO buffer with the reversed channel (i.e. with channel statistics of $\SSS-\SR$ and $\SR-\SD$ link interchanged). For example, the same BER and delay performance with LIFO are obtained if  we choose $d_{rd}=d_{sr}=1$ and $d_{rp}=1.5$ and $d_{sp}=2.0$ (i.e $\mu_{r}=33.75$ and $\mu_{s}=80$) in Fig.~\ref{fig:DelayvsTauFixedRate} and Fig.~\ref{fig:BERvsTauFixedRate}. Please note that the performance of LIFO is better than FIFO for the above specified statistics.
\section{Conclusion}\label{sec:ConClu}
In this paper, expressions are derived for average rate and SER performance of an adaptive link-selection scheme  in a cognitive two-hop network based on a  buffer-aided decode-and-forward  relay.  Performance is  compared with that of conventional schemes.  It is shown that adaptive link-selection is of utmost importance in interference-constrained underlay cognitive radio scenarios. We analyze delay performance, and discuss trade-off between delay, symbol error rate and throughput. Performance of a threshold based transmission scheme is analysed.  An alternative to starving the buffer (to improve performance in some scenario) is discussed, that uses the buffer in an unconventional manner. 

{\small
\section*{Appendix A}\label{app:A}
The derivation of (\ref{eqn:CCDF_SR1}) and (\ref{eqn:CCDF_SR2}) is presented in this Appendix. We use (\ref{eqn:I_n_mu_lambda_x}) in the derivation extensively. Using integration by parts in (\ref{eqn:I_n_mu_lambda_x}), we obtain the following recursion relation:
{\small\begin{eqnarray*}
	\begin{array}{lll}
		\mathcal{I}_{n}(\mu,\lambda;x)\hspace{-0.025in}= \dfrac{1}{n-1}\left[e^{-{x}/{\lambda}}\left(\dfrac{\mu}{x+\mu}\right)^{n-1}\hspace{-0.15in}-\dfrac{\mu}{\lambda}\mathcal{I}_{n-1}(\mu,\lambda;x)\right].
	\end{array}{}
\end{eqnarray*}} 
We know that
$\hspace{-0.0in}F^c_{d,\gamma_{s}}(0,x)= F^c_{\gamma_{s}}(x)-F^c_{d,\gamma_{s}}(1,x)$ where CCDF $F^c_{\gamma_{s}}(x)$ is given by (\ref{eqn:F_Gi_s}). Now
{\small\vspace{-.25cm}
	\begin{eqnarray*}
		\begin{array}{lll}\vspace{-0.2cm}
			F^c_{d,\gamma_{s}}(1,x)=\Pr\{\frac{\gamma_r}{\rho}>\gamma_s>x\}=\displaystyle\int\limits_{x}^{\infty}F^c_{\gamma_{r}}(\rho s)f_{\gamma_{s}}(s) {\D}s.
		\end{array}
	\end{eqnarray*}
}
 Substituting from (\ref{eqn:F_Gi_s}), we get:
 {\small
 	\begin{eqnarray*}
 		\begin{array}{lll}
 			\hspace{-0.1 in}F^c_{d,\gamma_{s}}(1,x)\hspace{-0.1 in}&=&  \dfrac{1}{\lambda_{s}}\displaystyle\int\limits_{x}^{\infty} e^{-(\rho s)/\lambda_{r}}\left[1-p_{r}\left(1- \dfrac{\mu_{r}}{\rho s+\mu_{r}}\right) \right]\\
 			&&\hspace{-0.8 in} \times\
 			e^{-s/\lambda_{s}}\left[1-p_{s}\left(1- \dfrac{\mu_{s}}{s+\mu_{s}}- \dfrac{\lambda_{s}\mu_{s}}{(s+\mu_{s})^2}\right) \right]{\D}s\\
 			&&\hspace{-0.8 in} =\
 			\underbrace{ \dfrac{1}{\lambda_{s}}\int\limits_{x}^{\infty} e^{-(\rho\,s)/\lambda_{\rho}} {\D}s}_{T_{1}}-p_{r}   \underbrace{\dfrac{1}{\lambda_{s}}\int\limits_{x}^{\infty} e^{-(\rho\,s)/\lambda_{\rho}}\left(1- \dfrac{\mu_{r}}{\rho s+\mu_{r}}\right) {\D}s}_{T_{4}}
 		\end{array}
 	\end{eqnarray*}
 }
{\small
	\begin{eqnarray}
			&&\hspace{-0.3 in} -\
			p_{s} \underbrace{\dfrac{1}{\lambda_{s}}\int\limits_{x}^{\infty} e^{-(\rho\,s)/\lambda_{\rho}}\left(1- \dfrac{\mu_{s}}{s+\mu_{s}}- \dfrac{\lambda_{s}\mu_{s}}{(s+\mu_{s})^2}\right) {\D}s}_{T_{2}-T_{3}}+ \,p_{s}p_{r} \nonumber\\
			&&\hspace{-0.3 in} \times\
			\underbrace{\dfrac{1}{\lambda_{s}}\int\limits_{x}^{\infty} e^{-(\rho\,s)/\lambda_{\rho}}\left[1- \dfrac{\mu_{s}}{s+\mu_{s}}- \dfrac{\lambda_{s}\mu_{s}}{(s+\mu_{s})^2}\right]\left[1- \dfrac{\mu_{r}}{\rho s+\mu_{r}}\right] {\D}s}_{(T_{2}-T_{3})-T_{5}} \nonumber\\\label{appA_T_terms}
			&&\hspace{-0.3 in} =\
			T_{1} - p_{s} (T_{2}-T_{3})-p_{r} T_{4} + p_{s}p_{r}\{(T_{2}-T_{3})-T_{5}\}
	\end{eqnarray}
}
 {where the last line is obtained by collecting $p_{s}$, $p_{r}$ and $p_{s}p_{r}$ terms together. We now present expressions for each of the integrals $T_1$ - $T_5$. It can be shown that $T_1$ - $T_4$ are given by:}
{\small
	\begin{eqnarray*}
		\begin{array}{lll}
			T_{1}=\dfrac{1}{\lambda_{s}}\displaystyle\int\limits_{x}^{\infty} e^{-(\rho\,s)/\lambda_{\rho}} {\D}s =\dfrac{\lambda_{\rho}}{\rho\lambda_{s}} e^{-(\rho\,x)/\lambda_{\rho}}\\
			&&\hspace{-2.525in} T_{2}=\dfrac{\rho}{\lambda_{\rho}}\hspace{-0.1cm}\displaystyle\int\limits_{x}^{\infty}\hspace{-0.05cm} \left(1- \dfrac{\mu_{s}}{s+\mu_{s}}- \dfrac{\lambda_{\rho}\mu_{s}}{\rho(s+\mu_{s})^2}\right) e^{-(\rho\,s)/\lambda_{\rho}} {\D}s\overset{l}{=} \dfrac{x\,e^{-(\rho\,x)/\lambda_{\rho}}}{x+\mu_{s}}\\
			&&\hspace{-2.525in}T_{3}=\dfrac{\rho}{\lambda_{r}}\displaystyle\int\limits_{x}^{\infty} \left(1-\dfrac{\mu_{s}}{s+\mu_{s}}\right)  e^{-(\rho\,s)/\lambda_{\rho}} {\D}s\\
			&&\hspace{-2.35in} \overset{m}{=} \dfrac{\lambda_{\rho}}{\lambda_{r}}\left[e^{-(\rho\,x)/\lambda_{\rho}}-\dfrac{\rho\mu_{s}}{\lambda_{\rho}} \exp\left(\dfrac{\rho\mu_{s}}{\lambda_{\rho}}\right) E_{1}\left(\dfrac{\rho x+\rho\mu_{s}}{\lambda_{\rho}}\right)\right]\\
			&&\hspace{-2.525in}T_{4}=\dfrac{1}{\lambda_{s}}\displaystyle\int\limits_{x}^{\infty} \left(1-\dfrac{\mu_{r}}{\rho s+\mu_{r}}\right)  e^{-(\rho\,s)/\lambda_{\rho}} {\D}s\\
			&&\hspace{-2.35in} \overset{n}{=} \dfrac{\lambda_{\rho}}{\rho\lambda_{s}}\left[e^{-(\rho\,x)/\lambda_{\rho}}-\dfrac{\mu_{r}}{\lambda_{\rho}}\exp\left(\dfrac{\mu_{r}}{\lambda_{\rho}}\right)E_{1}\left(\dfrac{\rho x+\mu_{r}}{\lambda_{\rho}}\right)\right]. 
		\end{array}
	\end{eqnarray*}
}
 {Equality $l$ is derived using (\ref{eqn:I_n_mu_lambda_x}) and its recursion whereas equality $m$ and $n$ use only (\ref{eqn:I_n_mu_lambda_x}). Generally, $T_{5}$ is given as:}
{\small
	\begin{eqnarray*}
		\begin{array}{lll}
			\hspace{-.7in}T_{5}=\dfrac{1}{\lambda_{s}}\displaystyle\int\limits_{x}^{\infty} e^{-(\rho\,s)/\lambda_{\rho}}\left(1- \dfrac{\mu_{s}}{s+\mu_{s}}- \dfrac{\lambda_{s}\mu_{s}}{(s+\mu_{s})^2}\right)\dfrac{\mu_{r}}{\rho s+\mu_{r}} {\D}s\\
			&& \hspace{-3.35in}\overset{p}{=}\
			\dfrac{\mu_{r}}{\mu_{r}-\rho\mu_{s}}\left[(T_{2}-T_{3})-T_{4}-T_{6}\right], \text{where}\vspace{-0.3cm}
		\end{array}
	\end{eqnarray*}
	\begin{eqnarray*}
		\begin{array}{lll}\vspace{-.3cm}		
			\hspace{-.4in}T_{6}=\displaystyle\int\limits_{x}^{\infty}\dfrac{\rho\mu_{s}}{(s+\mu_{s})(\rho s+\mu_{r})} e^{-(\rho\,s)/\lambda_{\rho}} {\D}s\overset{q}{=}\dfrac{\rho\mu_{s}}{\mu_{r}-\rho\mu_{s}}\\
		\end{array}
	\end{eqnarray*}
	\begin{eqnarray*}
		\begin{array}{lll}
			&&\hspace{-.3in} \times\
			\left[\exp\left(\dfrac{\rho\mu_{s} }{\lambda_{\rho}}\right)E_{1}\left(\dfrac{\rho x+\rho\mu_{s}}{\lambda_{\rho}}\right)-\exp\left(\dfrac{\mu_{r} }{\lambda_{\rho}}\right)E_{1}\left(\dfrac{\rho x+\mu_{r}}{\lambda_{\rho}}\right)\right],
		\end{array}
	\end{eqnarray*}
}
 {In the above, equality $p$ and $q$ result from partial fraction expansion and some manipulation using (\ref{eqn:I_n_mu_lambda_x}).Under particular condition when $\mu_{r}=\rho\mu_{s}$, $T_{5}$ is given as:}\vspace{-0.25cm}
{\small
	\begin{eqnarray*}
		\begin{array}{lll}
			T_{5}\hspace{-.02in}=\dfrac{1}{\lambda_{s}}\displaystyle\int\limits_{x}^{\infty} e^{-(\rho\,s)/\lambda_{\rho}}\left(1- \dfrac{\mu_{s}}{s+\mu_{s}}- \dfrac{\lambda_{s}\mu_{s}}{(s+\mu_{s})^2}\right)\dfrac{\mu_{s}}{s+\mu_{s}} {\D}s\\
			&&\hspace{-3.31in}\overset{r}{=}\
			\dfrac{\mu_{s}}{\lambda_{s}}\exp\left(\dfrac{\mu_{r}}{\lambda_{\rho}}\right)E_{1}\left(\dfrac{\rho x+\mu_{r}}{\lambda_{\rho}}\right)-\dfrac{1}{2}e^{-(\rho\,x)/\lambda_{\rho}}\bigg[\left(\dfrac{\mu_{s}}{x+\mu_{s}}\right)^2\\
			&&\hspace{-3.31in}
			-\left(\dfrac{\mu_{r}}{\lambda_{r}}-\dfrac{\mu_{s}}{\lambda_{s}}\right)\left[\dfrac{\mu_{s}}{x+\mu_{s}}-\dfrac{\rho\mu_{s}}{\lambda_{\rho}}\exp\left(\dfrac{\mu_{r}}{\lambda_{\rho}}\right)E_{1}\left(\dfrac{\rho x+\mu_{r}}{\lambda_{\rho}}\right)\right]\bigg]
		\end{array}
	\end{eqnarray*}
}
 {
	Equality $r$ is established using (\ref{eqn:I_n_mu_lambda_x}) after some manipulation.
	After rearranging all the terms, we get  (\ref{eqn:CCDF_SR}).
}
\vspace{-.5cm}
\section*{Appendix B}\label{app:B}
The derivation of (\ref{eqn:PBUFASYM_SR}) is presented in this Appendix. We carry out the Taylor's series expansion of all the terms of (\ref{eqn:CCDF_SR1}) assuming moderate values of average SNRs ($\lambda_{i}$  and $\mu_{i}$ $i\in\{s,r\}$). We get the same result if (\ref{eqn:CCDF_SR2}) is used in place of (\ref{eqn:CCDF_SR1}). We first re-arrange $F^c_{d,\gamma_{s}}(0,w)$ given in (\ref{eqn:CCDF_SR1}) by collecting $p_{s},\,p_{r}$ and $p_{s}p_{r}$ terms using $F^c_{\gamma_{s}}(w)$ in (\ref{eqn:F_Gi_s}) and $F^c_{d,\gamma_{r}}(1,w)$ in (\ref{appA_T_terms}) in following way:
\begin{equation}\label{eqn:FCNDGS}
F^c_{d,\gamma_{s}}(0,w)=T_1^{'}-p_{s}T_2^{'}+p_{r} T_3^{'} - p_{s}p_{r}T_4^{'}
\end{equation}
Now for high SNR analysis, all these terms in above expression can be approximated by the second order terms as follows. $T_1^{'}$ is given by:
\begin{equation*}
T_1^{'}=e^{-{ w}/\lambda_{s}}-\dfrac{\lambda_{\rho}}{\rho\lambda_{s}} e^{-{\rho w}/\lambda_{\rho}}.
\end{equation*}
Assuming $1/\lambda_{s}$ and $\rho/\lambda_{\rho}$ to be small at high SNRs, and using Taylor's series around $w=0$, we get
{\small\begin{eqnarray}\label{eqn:PSN_PRN_Term}
	\begin{array}{lll}
T_1^{'} &\approx& 1-\dfrac{w}{\lambda_{s}}+\dfrac{w^2}{2\lambda_{s}^2}-\dfrac{\lambda_{\rho}}{\rho\lambda_{s}} \left(1-\dfrac{\rho w}{\lambda_{\rho}}+\dfrac{\rho^2 w^2}{2\lambda_{\rho}^2}\right),\\
&=& 1-\dfrac{\lambda_{\rho}}{\rho\lambda_{s}}-\dfrac{\rho\,w^2}{2 \lambda_{s}\lambda_{r}}
\end{array}
\end{eqnarray}} 
Similarly, applying the procedure for $p_{s}$ term $T_{2}^{'}$, we get:
{\small\begin{eqnarray*}
	\begin{array}{lll}
		\hspace{-0cm}T_2^{'}&=&\dfrac{w}{w+\mu_{s}}(e^{-{ w}/\lambda_{s}}-e^{-{\rho w}/\lambda_{\rho}})+\dfrac{\lambda_{\rho}}{\lambda_{r}}e^{-{\rho w}/\lambda_{\rho}}\\
		&&\hspace{-0.6cm}\times\
		\left[1-\dfrac{\rho\mu_{s}}{\lambda_{\rho}} \exp\left(\dfrac{\rho w+\rho\mu_{s}}{\lambda_{\rho}}\right) E_{1}\left(\dfrac{\rho w+\rho\mu_{s}}{\lambda_{\rho}}\right)\right],\\
		&&\hspace{-.25in} =\
		e^{-{ w}/\lambda_{s}}+\left(1+\dfrac{w}{\mu_{s}}\right)^{-1}( e^{-{\rho w}/\lambda_{\rho}}-e^{-{ w}/\lambda_{s}})\\
		&&\hspace{-.6cm}-\
		\dfrac{\lambda_{\rho}}{\rho\lambda_{s}}e^{-{\rho w}/\lambda_{\rho}}-\dfrac{\rho\mu_{s}}{\lambda_{r}} \exp\left(\dfrac{\rho\mu_{s}}{\lambda_{\rho}}\right) E_{1}\left(\dfrac{\rho w+\rho\mu_{s}}{\lambda_{\rho}}\right)
	\end{array}
\end{eqnarray*}} 
Although $\lambda_{\rho}>>\mu_{s}$ for high SNR, we also assume that cross interference link is not severe i.e. $\mu_{s}>>1$. After applying Taylor's series at $w=0$ and some manipulation, we get:
{\small\begin{eqnarray*}
	\begin{array}{lll}
		\hspace{-0.25cm}T_2^{'}&\approx& 1-\dfrac{\lambda_{\rho}}{\rho\lambda_{s}}-\dfrac{\rho\mu_{s}}{\lambda_{r}} \exp\left(\dfrac{\rho\mu_{s}}{\lambda_{\rho}}\right) E_{1}\left(\dfrac{\rho\mu_{s}}{\lambda_{\rho}}\right)\\
		&+&
		\dfrac{\rho w^2}{\lambda_{r}}\left(\dfrac{1}{\lambda_{s}}+\dfrac{1}{\mu_{s}}\right)-\dfrac{\mu_{s}\rho^{2} w^2}{\lambda_{r}\lambda_{\rho}^2}\exp\left(\dfrac{\rho\mu_{s}}{\lambda_{\rho}}\right)E_{-1}\left(\dfrac{\rho\mu_{s}}{\lambda_{\rho}}\right).
	\end{array}
\end{eqnarray*}} 
Substituting  {\small$\exp(x)E_{-1}(x)=\exp(x)\int\limits_{1}^{\infty}t e^{-xt}{\D}t = x^{-2}\exp(x)\,\Gamma(2,x) \approx x^{-1}+x^{-2}\,$}  in the above and using some simple manipulations, we get:
{\small\begin{eqnarray}\label{eqn:PS_Term}
\hspace{-0.2cm}T_2^{'} \hspace{-0.1cm}&\approx&\hspace{-0.1cm} \dfrac{\lambda_{\rho}}{\lambda_{r}}\left[1-\dfrac{\rho\mu_{s}}{\lambda_{\rho}} \exp\left(\dfrac{\rho\mu_{s}}{\lambda_{\rho}}\right) E_{1}\left(\dfrac{\rho\mu_{s}}{\lambda_{\rho}}\right)\right] + \dfrac{\rho \,w^2}{2\lambda_{r}\mu_{s}}.
\end{eqnarray}} 
The following terms are also approximated using similar approach (details omitted for brevity):
{\small\begin{eqnarray}\label{eqn:PR_Term}
\begin{array}{lll}
T_3^{'} \approx \dfrac{\lambda_{\rho}}{\rho\lambda_{s}}\left[1-\dfrac{\mu_{r}}{\lambda_{\rho}}\exp\left(\dfrac{\mu_{r} }{\lambda_{\rho}}\right)E_{1}\left(\dfrac{\mu_{r}}{\lambda_{\rho}}\right)\right] - \dfrac{\rho\,w^2}{2\lambda_{s}\mu_{r}}
\end{array}
\end{eqnarray}} 
{\small\begin{eqnarray}\label{eqn:PS_PR_Term}
\hspace{-0.25cm}T_4^{'} \hspace{-0.35cm}&\approx&\hspace{-0.35cm} -\dfrac{\rho\mu_{s}}{\mu_{r}-\rho\mu_{s}}\hspace{-0.1cm}\left[1- \dfrac{\lambda_{\rho}}{\lambda_{r}}\exp\left(\dfrac{\rho\mu_{s} }{\lambda_{\rho}}\right)E_{2}\left(\dfrac{\rho\mu_{s}}{\lambda_{\rho}}\right)\right]+ \dfrac{\mu_{r}}{\mu_{r}-\rho\mu_{s}}\nonumber\\
&& \hspace{-0.6cm}\times\
 \dfrac{\lambda_{\rho}}{\rho\lambda_{s}}\exp\left(\dfrac{\mu_{r} }{\lambda_{\rho}}\right)E_{2}\left(\dfrac{\mu_{r}}{\lambda_{\rho}}\right)- \dfrac{\rho\,w^2}{2\mu_{s}\mu_{r}}-\dfrac{\rho\mu_{s}\mu_{r}}{(\mu_{r}-\rho\mu_{s})^2}\nonumber\\
&& \hspace{-0.6cm}\times\
 \left[\exp\left(\dfrac{\rho\mu_{s} }{\lambda_{\rho}}\right)E_{1}\left(\dfrac{\rho\mu_{s}}{\lambda_{\rho}}\right)-\exp\left(\dfrac{\mu_{r} }{\lambda_{\rho}}\right)E_{1}\left(\dfrac{\mu_{r}}{\lambda_{\rho}}\right)\right] .
\end{eqnarray} 
Substituting equations (\ref{eqn:PSN_PRN_Term})-(\ref{eqn:PS_PR_Term})  in  (\ref{eqn:FCNDGS}), we get: 
{\small\begin{equation*}
F^c_{d,\gamma_{s}}(0,w)\approx F^c_{d,\gamma_{s}}(0,0)-\dfrac{ w^2}{2}\left(\dfrac{1}{\lambda_{s}}+\dfrac{1}{\mu_{s}}p_{s}\right)\left(\dfrac{\rho}{\lambda_{r}}+\dfrac{\rho}{\mu_{r}}p_{r}\right),
\end{equation*}} 
where $q_{s}=F^c_{d,\gamma_{s}}(0,0)$. Substituting $F^c_{d,\gamma_{s}}(0,w)$ in  (\ref{eqn:P_obs}), we get:
{\small\begin{equation}
\hspace{-.1cm}\overline{\mathcal{P}}^{CABR}_{s} = \dfrac{\varphi\,\mathbf{\E}_{w}[w^2]}{4\, q_{s}}\left(\dfrac{1}{\lambda_{s}}+\dfrac{1}{\mu_{s}}p_{s}\right)\left(\dfrac{\rho}{\lambda_{r}}+\dfrac{\rho}{\mu_{r}}p_{r}\right).
\end{equation}} 
Substituting $\mathbf{\E}_{w}[w^k]=(2k-1)!!/\eta^k$, we get  (\ref{eqn:PBUFASYM_SR}). The same equation can be obtained by proceeding with (\ref{eqn:CCDF_SR2}) instead of (\ref{eqn:CCDF_SR1}).

\section*{Appendix C}\label{app:C}
To prove the bound of LSP of $\SSS-\SR$ link $q_{s}$ in PIP case ($p_{s}=p_{r}=1$), first we write the expression from (\ref{eqn:ps_sr_asym}) as:
{\small\begin{eqnarray*}
	\begin{array}{lll}
	q_{s}=\hspace{-0.1cm}\dfrac{-z}{1-z}-\dfrac{z}{(1-z)^2}\ln z,\quad \text{where } z =\dfrac{\rho\mu_{s}}{\mu_{r}}<1 \text{ for } \xi>1.
	\end{array}
	\end{eqnarray*}}
After substituting the power series for $\ln z=-\sum\limits_{k=1}^{\infty}\dfrac{(1-z)^k}{k}$ in previous equation and some simple manipulations, we get:
{\small\begin{eqnarray*}
		\begin{array}{lll}
			q_{s}&=&\hspace{-0.1cm}z\sum\limits_{k=0}^{\infty}\dfrac{(1-z)^k}{k+2}\overset{m}{\approx}z\sum\limits_{k=0}^{\infty}\dfrac{(1-z)^k}{2^{k+1}}\\
			&=&\dfrac{z}{2}\sum\limits_{k=0}^{\infty}\left(\dfrac{1-z}{2}\right)^k\overset{n}{=}\dfrac{z}{2}\dfrac{2}{z+1}=\dfrac{z}{z+1},
		\end{array}
	\end{eqnarray*}}
where the approximation $m$ is obtained by exploiting the fact that in starving scenario when $z<1$, using $2^{k+1}$ in place of $k+2$ results in negligible error  (note that $2^{k+1}\geq k+2,\, \forall k\geq0$). Equality $n$ is achieved using sum of infinite geometric series.
\bibliographystyle{ieeetr}
\bibliography{biblogrpy}
\end{document}